\documentclass[10pt]{article}
\pdfoutput=1
\usepackage[utf8]{inputenc}
\usepackage{cite}
\usepackage{amsmath,amssymb,amsbsy,amstext,amsthm,simplewick,amsfonts}
\usepackage{graphicx}
\usepackage{wrapfig}
\usepackage{upgreek}
\usepackage{bm} 
\usepackage{framed}
\usepackage{bbm}
\usepackage{textcomp}
\usepackage{tikz}
\usepackage{pifont}
\usetikzlibrary{matrix,shapes,fit,tikzmark,calc}
\usepackage{adjustbox}
\usepackage{makecell}
\usepackage{tcolorbox}
\usepackage{physics}
\usepackage{empheq}
\usepackage[normalem]{ulem}
\usepackage{enumitem}
\usepackage{braket} 
\usepackage{array}
\usepackage{dsfont}
\usepackage{ulem}

\numberwithin{equation}{section}

\def\be{\begin{equation}}
\def\ee{\end{equation}}

\def\e{\epsilon}

\def\ba{\begin{eqnarray}}
\def\ea{\end{eqnarray}}
\def\k{\kappa}

\def\bfx{\textbf{x}}

\def\bfk{\textbf{k}}

\def\del{\partial}

\usepackage{caption}
\usepackage{subcaption}

\usepackage[framemethod=default]{mdframed}
\newmdenv[skipabove=7pt,
skipbelow=7pt,
rightline=false,
leftline=false,
topline=false,
bottomline=false,
backgroundcolor=gray!10,
linecolor=gray,
innerleftmargin=5pt,
innerrightmargin=5pt,
innertopmargin=5pt,
innerbottommargin=5pt,
leftmargin=0cm,
rightmargin=0cm,
linewidth=4pt]{eBox}
\newmdenv[skipabove=7pt,
skipbelow=7pt,
rightline=false,
leftline=false,
topline=false,
bottomline=false,
backgroundcolor=gray!10,
linecolor=gray,
innerleftmargin=5pt,
innerrightmargin=5pt,
innertopmargin=-5pt,
innerbottommargin=5pt,
leftmargin=0cm,
rightmargin=0cm,
linewidth=4pt]{eBox2}

\definecolor{blue3}{RGB}{31,119,180}
\definecolor{red3}{RGB}{214,39,40}
\definecolor{orange3}{RGB}{255,127,14}
\definecolor{green3}{RGB}{44,160,44}

\usepackage{colortbl}
\definecolor{lightgreen}{cmyk}{0.2, 0, 0.2, 0.2}
\definecolor{lightgray}{cmyk}{0.1,0.2,0,0.1}
\definecolor{lightgray2}{cmyk}{0.1,0.1,0,0.1}

\usepackage{hyperref}
\hypersetup{colorlinks=true,linkcolor=teal,citecolor=orange3,urlcolor=green3,pdfencoding=auto}

\makeatletter
\newlength{\apb@width}
\newcommand{\autoparbox}[2][c]{\settowidth{\apb@width}{#2}\parbox[#1]{\apb@width}{#2}}

\makeatother


\def\k{{\bm k}}

\def\disc{\text{Disc}}
\def\Mpl{M_{\text{P}}}

\def\nn{\nonumber}
\def\eps{\boldsymbol\epsilon}

\def\beq{\begin{equation}}
\def\eeq{\end{equation}}
\newcommand{\ex}[1]{\langle #1 \rangle}

\newcommand\EQ[1]{Eq.~\eqref{eq:#1}}

\allowdisplaybreaks[1]
\begin{document}


\begin{titlepage}
\setcounter{page}{1} \baselineskip=15.5pt 
\thispagestyle{empty}

\begin{center}
{\fontsize
{28}{28}\centering \bf Bootstrapping Large Graviton \\ \vspace{0.3cm} non-Gaussianities\;}\\
\end{center}

\vskip 18pt
\begin{center}
\noindent
{\fontsize{12}{18}\selectfont Giovanni Cabass\footnote{\tt gcabass@ias.edu}$^{,\star}$, Enrico Pajer\footnote{\tt enrico.pajer@gmail.com}$^{,\dagger}$, 
David Stefanyszyn\footnote{\tt dps56@cam.ac.uk}$^{,\dagger}$ and Jakub Supe{\l}\footnote{\tt js2154@cam.ac.uk}$^{,\dagger}$}
\end{center}

\begin{center}
\vskip 8pt
$\star$\textit{ School of Natural Sciences, Institute for Advanced Study, Princeton, NJ 08540, United States} \\ 
$\dagger$\textit{ Department of Applied Mathematics and Theoretical Physics, University of Cambridge, Wilberforce Road, Cambridge, CB3 0WA, UK}
\end{center}


\vspace{1.4cm}

\noindent Gravitational interferometers and cosmological observations of the cosmic microwave background offer us the prospect to probe the laws of gravity in the primordial universe. To study and interpret these datasets we need to know the possible graviton non-Gaussianities. To this end, we derive the most general tree-level three-point functions (bispectra) for a massless graviton \textit{to all orders in derivatives}, assuming scale invariance. Instead of working with explicit Lagrangians, we take a bootstrap approach and obtain our results using the recently derived constraints from unitarity, locality and the choice of vacuum. Since we make no assumptions about de Sitter boosts, our results capture the phenomenology of large classes of models such as the effective field theory of inflation and solid inflation. We present formulae for the infinite number of parity-even bispectra. Remarkably, for parity-odd bispectra, we show that unitarity allows for only a handful of possible shapes: three for graviton-graviton-graviton, three for scalar-graviton-graviton and one for scalar-scalar-graviton, which we bootstrap explicitly. These parity-odd non-Gaussianities can be large, for example in solid inflation, and therefore constitute a concrete and well-motivated target for future observations.


\end{titlepage} 


\newpage
\setcounter{tocdepth}{2}
{
\hypersetup{linkcolor=black}
\tableofcontents
}

\newpage


\section{Introduction}

\noindent Being the only force that stubbornly refuses to be described at arbitrarily high energies within the dominant framework of quantum field theory, gravity is a prominent testing ground for our understanding of fundamental physics. Ideas from string theory, the study of black holes and gauge-gravity duality suggest that the field-theoretic gravitons that appear to describe low-energy phenomena very well, most likely don't provide the right language to discuss non-perturbative and high-energy aspects of quantum gravity. Given how difficult it is to establish what gravity \textit{is}, a useful approach to the problem is to ask the related question: \textit{What can gravity be?} For example, given the framework of quantum mechanics as we know it, what different descriptions of gravity can be formulated that are mathematically and physically consistent? \\

\noindent Concrete and quantitative progress in this direction has been achieved for quantum fields on flat spacetime, e.g. via the derivation of positivity bounds that constrain effective field theories admitting standard and consistent UV-completions. To understand and model cosmology, and in particular inflation, dark energy and dark matter, we would like to use these bounds as a compass pointing us in the direction of the most promising consistent theories. However, the set of consistent theories of dynamical gravity is different in flat and cosmological spacetimes. Concrete examples of this difference include a theory of interacting massless spin 3/2 particles, which is given by supergravity in flat space, but is not known in de Sitter; or the theory of a scalar coupled to gravity with boost-breaking interactions, which is easily written down in cosmological spacetimes, as in many realistic models of inflation and dark energy, but which is inconsistent in flat spacetime as can be shown by examining amplitude factorization \cite{PSS}. At the same time, new probes of gravity have just become available through the observation of gravitational waves at interferometers, and there is a substantial international effort and a well-kindled hope to detect a cosmological background of gravitational waves from the primordial universe. In light of these considerations, it is highly desirable to study the consistency of effective field theories of gravity directly on the cosmological spacetimes where we want to use them for phenomenology. \\

\noindent In this work, we are interested in constraining the possible phenomenological descriptions of gravity around a (quasi) de Sitter spacetime, with an eye towards applications to inflation. To this end, we focus on the natural observables of this system: cosmological correlators, namely the expectation values of the product of fields in the space-like asymptotic future, which we will call the (conformal) boundary. Given a concrete model, such observables can be computed in perturbation theory using the in-in formalism. However, since we don't know what the ``right'' model is, we will follow a different approach, which is inspired by parallel progress in the study of amplitudes \cite{Benincasa:2013,Elvang:2013cua,TASI}. In particular, we aim to derive all possible correlators that are compatible with fundamental principles such as symmetry, unitarity and locality. This model-independent approach goes under the name of the cosmological bootstrap and has received growing attention in recent years \cite{Arkani-Hamed:2017fdk,CosmoBootstrap1,Baumann:2019oyu,Sleight:2019mgd,Sleight:2019hfp,Arkani-Hamed:2018bjr,Benincasa:2018ssx,Benincasa:2019vqr,Baumann:2020dch,COT,Sleight:2020obc,BBBB,MLT,Melville:2021lst,Goodhew:2021oqg,trispectrum,Baumann:2021fxj,DiPietro:2021sjt,Sleight:2021plv,Hogervorst:2021uvp,Meltzer:2021zin,Sleight:2021iix,Cespedes:2020xqq,Gomez:2021qfd,Green:2020ebl}. \\

\noindent We will focus on the simplest non-trivial correlators of massless spin-2 fields, a.k.a. gravitons, and massless scalars, namely three-point functions or bispectra. An important previous result is that of \cite{Maldacena:2011nz}, where, assuming invariance under the full isometry group of de Sitter, it was shown that for gravitons only three cubic cosmological wavefunctions are allowed, and of those only the two parity-even ones can lead to a non-vanishing bispectrum \cite{Soda:2011am}. Several additional results can be derived in this setup using conformal Ward identities, as done for example in \cite{Maldacena:2011nz,Mata:2012bx,Bzowski:2013sza,Kundu:2014gxa,Kundu:2015xta,Pajer:2016ieg}, and parity-odd correlators in CFT's were recently discussed in \cite{Jain:2021wyn,Jain:2021vrv}. While some of these results are remarkable because they are non-perturbative in nature, we are faced with the issue that de Sitter boosts are actually broken in all cosmological models and, in particular, during inflation. Unlike the breaking of scale-invariance, \textit{the breaking of boosts is in general not slow-roll suppressed and may be large}, as for example in so-called P-of-X models (a.k.a. ``k-inflation'' \cite{Armendariz-Picon:1999hyi}), where the Lagrangian is an arbitrary function of the kinetic term, or more general models captured by the effective field theory of inflation \cite{Creminelli:2006xe,Cheung:2007st}. In fact, as emphasized in \cite{Green:2020ebl}, the breaking of de Sitter boosts is a necessary condition to have phenomenologically large non-Gaussianities. \\

\noindent Therefore, to make contact with cosmological observations, in this work we will weaken the assumption of full de Sitter invariance and instead assume only the symmetries that have been observed in primordial perturbations, namely statistical homogeneity, isotropy and (approximate) scale invariance. In particular, we will allow for arbitrary breaking of de Sitter boosts. The price to pay for this smaller set of isometries is that we have to work in perturbation theory and we will restrict ourselves to tree-level. \\

\noindent Progress in understanding boost-breaking gravitational interactions has been achieved using effective field theories and the Lagrangian approach in a series of recent papers \cite{Creminelli:2014wna,Bordin:2017hal,CabassBordin,ZoologyNG,Bartolo:2020gsh,Bartolo:2017szm}. This approach is quite general and intuitive but its computational complexity grows quickly as one considers operators with an increasing number of derivatives. To overcome this difficulty, here we will instead follow the ``boostless'' cosmological bootstrap approach proposed in \cite{BBBB,MLT}, which partially builds upon results in \cite{Maldacena:2011nz,Arkani-Hamed:2017fdk,CosmoBootstrap1,PSS,COT} and is reviewed in Section \ref{sec:BootstrapTechniques}. Our approach leverages the powerful constraints of fundamental principles such as unitarity, locality and the choice of vacuum and allows us to bootstrap all tree-level graviton bispectra to \textit{any order in derivatives}, as well as all parity-odd mixed bispectra. At the end of our derivation we will see how the bootstrap results can be understood in the familiar Lagrangian language (see Section \ref{SymmetryBreaking}).\\

\noindent Our main results are summarized below:
\begin{itemize}
\item Unitarity and the choice of the Bunch-Davies vacuum highly restrict the allowed set of parity-odd correlators. In particular, for massless scalars and gravitons and to all orders in derivatives, there is only a finite number of tree-level correlators. In contrast, the number of possible wavefunction coefficients and Lagrangian interactions grows without bound as one increases the number of derivatives in the effective field theory expansion. In more detail, a contact parity-odd correlator can only arise when there is a logarithmic IR-divergence in the associated wavefunction coefficient. In turn this may only happen when $  2 n_{\partial_{\eta}} + n_{\partial_{i}} \leq 3 $, where $  n_{\partial_{\eta}} $ and $  n_{\partial_{i}} $ are respectively the number of time and space derivatives in the parity-odd interaction\footnote{This is valid for any contract $ n  $-point function and assumes that there is at most one time derivative per field. Interactions with more than one time derivative can always be re-written in terms of those with at most one time derivative using the equations of motion.}. This explains on general grounds why parity-odd correlators where found to vanish in the scale-invariant limit in a number of explicit calculations \cite{Soda:2011am,Shiraishi:2011st,Bartolo:2020gsh}.
\item We computed all tree-level graviton bispectra to any order in derivatives, assuming in particular scale-invariance and massless gravitons. There are infinitely many \textit{parity-even} graviton bispectra $  B_{3} $. For example, for the choice of all plus helicities these are given by the symmetrized products of three factors
\begin{equation}
e_{3}^{3} B^{+++}_3(\bfk_1,\bfk_2,\bfk_3) = \text{SH}_{+++}  \sum_{\rm permutations} h_{\alpha}(k_1, k_2, k_3) \psi^{\rm trimmed}_3(k_1,k_2,k_3)\,.
\end{equation}
The first factor SH$_{+++} $ includes the spinor helicities and  provides the correct little-group scaling. It is given by
\begin{align}
\text{SH}_{+++} &= \frac{[12]^2[23]^2[31]^2}{e^2_3}=- e^+_{ij}(\bfk_1) e^+_{jk}(\bfk_2) e^{+}_{ik}(\bfk_3)\\ 
&=-\frac{k_T^3 \left( 8 e_3 - 4 k_T e_2+ k_T^3 \right) }{16 \sqrt{2} e_3^2}\,,
\end{align}
where $  [ij] $ is the square-bracket product of helicity spinors, $  e_{ij} $ are polarization tensors, and $  k_{T} $, $  e_{2} $ and $  e_{3} $ are the elementary symmetric polynomials defined in \eqref{esp}. 
The second factor $  h_{\alpha} $ roughly accounts for the contractions between spatial derivatives and polarization tensors and can be any one of the following four possibilities
\begin{align} \label{h1a}
h_{0} = 1, \quad h_{2} = k_2 k_3, \quad h_{4} = I_1^2 I_2 I_3, \quad  h_{6} &= I^2_1I^2_2I^2_3 \,,
\end{align}
where 
\begin{align}
I_a &\equiv (k_T-2k_a) = k_{b}+k_{c}-k_{a} & a\neq b\neq c\,.
\end{align}
For parity-odd interactions there are a further five possibilities for $h_{\alpha}$. Finally, the third factor is the ``trimmed'' wavefunction $ \psi^{\rm trimmed}_3 $, which roughly accounts for the conformal time integrals of mode functions, time derivatives and spatial derivatives contracted with each other. This can be any of the infinitely-many rational-function solutions of the manifestly local test, $  \partial_{k_{a}} \psi^{\rm trimmed}_3 =0 $ at $  k_{a}=0 $ (see \eqref{MLT}), which are conveniently organized in terms of the increasing order of the polynomial in the numerator, roughly corresponding to the derivative expansion of an effective field theory. For concreteness, the first few explicit bispectra are given in \eqref{prima} through \eqref{ultima}. The bispectra corresponding to other helicity choices can be derived from the all-plus bispectrum as discussed in Section \ref{ToRule}.
\item Remarkably, \textit{there are only three parity-odd graviton bispectra at tree level to all orders in derivatives}. These are explicitly found to be
\begin{align}\nonumber
 B^{+++}_{3} &= \frac{g_{1,1}}{e_{3}^3} \text{SH}_{+++} k_T \left( k_T^2 - 2 e_2 \right) \,, &  B^{++-}_{3} &= \frac{g_{1,1}}{e_{3}^3} \text{SH}_{++-}  I_3 \left( k_T^2 - 2 e_2 \right)\,, \\ \nonumber
 B^{+++}_{3} &= \frac{g_{1,2} }{e_{3}^3}\text{SH}_{+++} \left(-3 e_3 + k_T e_2 \right)\,, &   B^{++-}_{3} &= \frac{g_{1,2}}{e_{3}^3} \text{SH}_{++-} \left[ \left(k_1 +k_{2} \right) \left(  k_{1}k_{2}+k_{3}^{2}\right)-(k_{1}^2 + k_{2}^2) k_{3}\right]\,,\\ \nonumber
 B^{+++}_{3} &= \frac{g_{3,3}}{e_{3}^3} \text{SH}_{+++} I_1 I_2 I_3 \,, &  B^{++-}_{3} & = \frac{g_{3,3}}{e_{3}^3} \text{SH}_{++-}  I_1 I_2 k_T\,,
\end{align}
where the $  g_{\alpha,p} $ are arbitrary real coupling constants whose indices denote respectively the number $  \alpha $ of spatial momenta contracted with polarization tensors and the total number of derivatives $  p $ in the associated interaction. The remaining two helicity configurations, namely $---$ and $--+$, can be obtained via a parity transformation, while keeping in mind the odd-parity of the above bispectra. In the effective field theory of inflation only one specific combination of these three shapes can appear and it must be accompanied by a parity-odd correction to the free theory. In this case, the final parity-odd graviton bispectrum must be small, and in particular much smaller than the standard parity-even contribution from General Relativity (GR) computed in \cite{Maldacena:2002vr}. By contrast, all three shapes above can appear in a general model of solid inflation \cite{SolidInflation}, without any modification to the free theory and with arbitrarily large amplitudes. Hence, these three parity-odd graviton bispectra are an important target for non-Gaussian searches in the graviton sector. Their shapes are plotted in Figure \ref{fig:B3Shapes}. In solid inflation they should be accompanied by correlated scalar-scalar-graviton and scalar-graviton-graviton bispectra with larger signal-to-noise ratios (see Section \ref{ssec:S2N}).
\item We show that \textit{there are only three parity-odd scalar-graviton-graviton bispectra and one scalar-scalar-graviton bispectrum at tree level to all orders in derivatives}, assuming scale invariance and manifest locality. These are given by 
\begin{align}
B_{3}^{00+} &=\frac{ h_{3,3}}{e_{3}^3} \frac{[13]^2 [23]^2}{k_{3}^2 [12]^2} I_{3}^2 k_{3}\,,\\
B_{3}^{0++} &= \frac{[23]^4}{k_{2}^2k_{3}^2 e_{3}^3}[q_{1,1}(k_{2}+k_{3})k_{1}^2 + q_{1,2,a}(k_{2}^3+k_{3}^3)+q_{1,2,b}(k_{2}k_{3}^2+k_{3}k_{2}^2)]\,,
\end{align}
where $  h_{3,3} $ and $  q_{\alpha,p} $ are arbitrary coupling constants. Notice, however, that for scalars non-manifestly local interactions do arise in GR. We show in Section \ref{SymmetryBreaking} that the above scalar-scalar-graviton bispectrum can be large in solid inflation, but not in the effective field theory of inflation, and can be the leading observational signal.
\end{itemize}

\noindent The rest of this work is organized as follows. In Section \ref{sec:BootstrapTechniques}, we review the framework and tools used to bootstrap correlators in general scale-invariant and boost-breaking theories, and in particular the boostless bootstrap rules, the constraints of unitarity in the form of the Cosmological Optical Theorem and associated cutting rules, the constraints from locality on massless fields in the form of the Manifestly Local Test, and finally the spinor helicity formalism for spinning cosmological correlators. The expert reader might skip directly to Section \ref{ParitySection}, where we derive a very general consequence of unitarity for tree-level contact correlators that implies that to all orders in derivatives there is only a small and finite number of non-vanishing parity-odd correlators. 
Then in Section \ref{sec:bispectra} we present formulae for all graviton bispectra to any order in the derivative expansion and show that there are only three non-vanishing parity-odd bispectra, and infinitely many parity-even ones. In Section \ref{SymmetryBreaking}, we show that the parity-odd bispectra can indeed arise in realistic models such as solid inflation, and study how they are constrained in the effective field theory of inflation. We also discuss their detectability by studying the associated signal-to-noise ratio. We conclude in Section \ref{summary} with an outlook on future research directions.


\paragraph{Notation and conventions} \label{Conventions}
Throughout we will work with the mostly positive metric signature $(-+++)$ and we define the three-dimensional Fourier transformation as 
\begin{align}
f(\bfx)&=\int \dfrac{d^3\bfk}{(2\pi)^3}{f}(\bfk)\exp(i\bfk\cdot\bfx)\equiv\int_{\bfk}{f}(\bfk)\exp(i\bfk\cdot\bfx) \label{FT} \,,\\
{f}(\bfk)&=\int d^3\bfx \,f(\bfx)\exp(-i\bfk\cdot\bfx)\equiv \int_{\bfx}  f(\bfx)\exp(-i\bfk\cdot\bfx)\,.
\end{align}
We use bold letters to refer to vectors, e.g. $\bf x$ for spatial co-ordinates and $\bf k$ for spatial momenta, and we write the magnitude of a vector as $k\equiv |\bfk|$. We will sometimes refer to these objects as ``energies" even though there is no time-translation symmetry in cosmology. We will use $i,j,k,\ldots = 1,2,3$ to label the components of $SO(3)$ vectors, and $a,b,c=1,\dots,n$ to label the $n$ external fields. For wavefunction coefficients and cosmological correlators we use $\psi_{n}$ and $B_n$ respectively:
\begin{align}
\psi_n(\bfk_1,\dots ,\bfk_n)& \equiv \psi_n'({\bfk_1,\dots ,\bfk_n})(2\pi)^3 \delta^3 \left(\sum \bfk_a \right)\,, \\ \nonumber
\langle {\cal{O}} {(\bfk_1)\dots {\cal{O}}(\bfk_n)} \rangle &\equiv \langle {\cal{O}} {(\bfk_1)\dots {\cal{O} }({\bfk_n}) }\rangle' (2\pi)^3\delta^3 \left(\sum \bfk_a \right)\\
&\equiv B_n ({\bfk_1,...,\bfk_n})\,  (2\pi)^3\delta^3\left(\sum \bfk_a \right)\,,
\end{align}
and we will drop the primes on $\psi_n$ when no confusion arises. We will also use a prime to denote a derivative with respect to the conformal time e.g. $\phi'=\del_\eta \phi$. We will often encounter polynomials that are symmetric in three variables, for example, for the $+++$ correlator. We write these in terms of the elementary symmetric polynomials (ESP):
\begin{align}\label{esp}
k_{T} &= k_{1}+k_{2}+k_{3}\,, \\
e_{2} &= k_{1}k_{2} + k_{1}k_{3}+k_{2}k_{3}\,, \\
e_{3} &= k_{1}k_{2}k_{3}\,.
\end{align}

\section{Bootstrap techniques from symmetries, locality and unitarity} \label{sec:BootstrapTechniques}

In this section, we define the objects that we will be bootstrapping, namely wavefunction coefficients appearing in the wavefunction of the universe and the associated cosmological correlators. In this part of the paper we also review bootstrap techniques that have been recently developed in the context of boost-breaking interactions. We outline how symmetries, locality and unitarity can be directly imposed on cosmological observables thanks to a set of Boostless Bootstrap Rules \cite{BBBB}, a Manifestly Local Test \cite{MLT} and the Cosmological Optical Theorem \cite{COT,Cespedes:2020xqq,Melville:2021lst,Goodhew:2021oqg}. Finally, we review the cosmological spinor helicity formalism that we will use to succinctly present graviton bispectra.  


\subsection{The wavefunction of the universe and cosmological correlators}
\noindent Let's start by reviewing the computation of the wavefunction of the universe $\Psi$ and defining wavefunction coefficients $\psi_{n}$ which will be our primary objects of interest. We will also remind the reader how correlation functions are extracted from knowledge of the wavefunction. \\ 

\noindent We take the background geometry to be that of rigid de Sitter (dS) spacetime which we write as\footnote{These are the so-called Poincar\'e or flat-slicing coordinates and cover half of the maximally extended de Sitter spacetime. This spacetime is the one relevant for the discussion of cosmological observations.}
\begin{align}
ds^2 = a^2(\eta) (- d \eta^2 + d {\bf x}^2)\,, \qquad a(\eta) =  -\frac{1}{\eta H}\,,
\end{align}
where the conformal time coordinate $\eta \in (-\infty,0)$ and $H$ is the constant Hubble parameter which we will often set to unity. This background geometry is an excellent approximation of an inflationary solution, and considering quantum fields fluctuating on this rigid background allows us to compute the leading contributions to inflationary non-Gaussianities, up to small slow-roll corrections \cite{Maldacena:2002vr}. Our methods in this paper will apply to general quantum field theories, but we will primarily be interested in the two massless modes that appear in \textit{all} inflationary models: a massless scalar $\phi(\eta, \bf x)$ and the transverse, traceless massless graviton $\gamma_{ij}(\eta, \bf x)$. When our results apply to both scalars and graviton, especially in Section \ref{ParitySection}, we will use $\varphi(\eta, \bf x)$ with any $SO(3)$ indices suppressed.  \\

\noindent The free action of a massless scalar is 
\begin{align}
S_{\phi, \text{free}} = \int d \eta d^3 {\bf x} \,a^2(\eta)\frac{1}{2} \left[\phi'^2 - c_{s}^2 \partial_{i}\phi \partial_{i}\phi \right]\,,
\end{align}
where we have allowed for an arbitrary, constant speed of sound $c_{s}$ which signals the fact we are allowing for dS boosts to be spontaneously or explicitly broken\footnote{When the speed of sound differs from the speed of light appearing in the metric, $  c_{s}\neq 1 $, the sound cone is not invariant under de Sitter boosts, a fact which can be simply seen in the flat-space limit, where de Sitter boosts reduce to Lorentz boosts.}. Working in momentum space, we write the quantum free field operator as 
\begin{align} \label{FreeField}
\hat{\phi}(\eta, {\bf k}) = \phi^{-}(\eta, k) a_\bfk + \phi^{+}(\eta,k) a_{-\bfk}^\dagger\,,
\end{align}
where the mode functions $\phi^{\pm} (\eta, k)$ correspond to solutions of the free classical equation of motion and are given by
\begin{align}
\phi^{\pm}(\eta, k) &= \frac{H}{\sqrt{2 c_{s}^3 k^3}} (1 \mp i c_{s} k \eta)e^{\pm i c_{s} k \eta}\,.\label{ModeFunctionsMassless}
\end{align}
The mode functions for graviton fluctuations take the same form as \eqref{ModeFunctionsMassless} (with $c_{s}=1$) with the addition of polarisation tensors $e^{h}_{ij}({\bf k})$, with $h = \pm 2$, as required by little group scaling. This is because for each polarisation mode the equation of motion is that of a massless scalar. The polarisation tensors satisfy the following conditions:
\begin{align}\label{pol1}
e_{ii}^{h}({\bf k})&=k^{i}e^{h}_{ij}({\bf k})=0 & \text{(transverse and traceless)}\,,\\
e_{ij}^{h}({\bf k})&=e_{ji}^{h}({\bf k})& \text{(symmetric)} \,, \\
e_{ij}^{h}({\bf k})e_{jk}^{h}({\bf k})&=0 & \text{(lightlike)} \,, \\
e^{h}_{ij}({\bf k})e^{h'}_{ij}({\bf k})^{\ast}&=2\delta_{hh'} & \text{(normalization)}\,, \label{normeps}  \\
e_{ij}^{h}({\bf k})^{\ast}&=e_{ij}^{h}(-{\bf k}) &\text{($  \gamma_{ij}(x) $ is real)}  \,.\label{poln}
\end{align} 

\noindent As we  explained in the introduction, we are interested in scenarios where dS boosts are broken since we know that these symmetries could not have been exact in the early universe, and large non-Gaussianities are associated with a large breaking of boosts \cite{Green:2020ebl}. We take the remaining symmetries of the dS group to be exact: spatial translations, spatial rotations and dilations. A general interaction vertex with $n$ fields, scalars and gravitons, therefore takes the schematic form
\begin{align}
S_{\text{int}} =  \int d \eta d^3 {\bf x} \, a(\eta)^{4 - N_{\text{deriv}}} \partial^{N_{\text{deriv}}} \varphi^{n}\,,
\end{align}
where $\partial$ stands for either time derivatives $\partial_{\eta}$ or spatial derivatives $\partial_{i}$, and $N_{\text{deriv}}$ is the total number of derivatives. Spatial derivatives and the graviton's indices are contracted with the $SO(3)$ invariant objects $\delta_{ij}$ and $\epsilon_{ijk}$ and the overall number of scale factors is dictated by scale invariance. \\

\noindent We now turn to the wavefunction of the universe which we denote as $\Psi$. We are interested in this wavefunction evaluated at the end of inflation or alternatively on the late-time boundary of dS space, at a conformal time which we denote as $\eta_{0}$. Ultimately we will take $\eta_{0} \rightarrow 0$. To illustrate the wavefunction of the universe method, let us focus on a single massless scalar $\phi$. The generalisation to gravitons simply requires the addition of $SO(3)$ indices where appropriate. We refer the reader to \cite{Guven:1987bx,WFCtoCorrelators1,COT,WFCtoCorrelators2,Baumann:2020dch} for further details. At late-times, the wavefunction has an expansion in the late-time value of the scalar, $\phi({\bf k}) \equiv \phi(\eta_{0}, {\bf k})$, given by
\begin{align}
\Psi[\eta_{0},\phi({\bf k})] = \text{exp}\left[-\sum_{n=2}^{\infty} \frac{1}{n!} \int_{{\bf k}_{1}, \ldots, {\bf k}_{n}} \psi_{n}({\bf k}_{1} \ldots {\bf k}_{n})\phi({\bf k}_{1}) \ldots \phi({\bf k}_{n}) \right],
\end{align}
where we have written the exponent as an expansion in powers of the field multiplied by the \textit{wavefunction coefficients} $ \psi_{n}({\bf k}_{1} \ldots {\bf k}_{n})$ which contain the dynamical information about the bulk processes. Invariance of the theory under spatial translations ensures that the $ \psi_{n}({\bf k}_{1} \ldots {\bf k}_{n})$ always contain a momentum conserving delta function and so we can write
\begin{align}
 \psi_{n}({\bf k}_{1} \ldots {\bf k}_{n}) =  \psi'_{n}({\bf k}_{1} \ldots {\bf k}_{n}) (2 \pi)^3 \delta^{3}({\bf k}_{1}+ \ldots+ {\bf k}_{n}).
\end{align}
We will often drop the prime even when we do not explicitly include the delta function. At weak coupling, we can compute the leading contribution to the wavefunction using the saddle-point approximation where the wavefunction is completely fixed by the value of the action evaluated on classical solutions:
\begin{align}
\Psi[\eta_{0},\phi({\bf k})] \approx e^{i S_{\text{cl}}[\phi({\bf k})]}.
\end{align}
Traditionally, one computes $S_{\text{cl}}[\phi({\bf k})]$ in perturbation theory using Feynman diagrams which involve bulk interaction vertices, bulk-boundary propagators $K(\eta,k)$ and bulk-bulk propagators $G(\eta, \eta', k)$. If we denote the scalar's free equation of motion as $\mathcal{O}(\eta,k) \phi$ = 0, then these propagators satisfy
\begin{align}
&\mathcal{O}(\eta,k) K(\eta, k) = 0, \\
&\mathcal{O}(\eta,k) G(\eta, \eta', k) = -\delta(\eta - \eta'), 
\end{align}
with boundary conditions
\begin{align}
\lim_{\eta \rightarrow \eta_{0}} K(\eta, k) &= 1, & \lim_{\eta \rightarrow -\infty(1 - i \epsilon)} K(\eta, k) &= 0 \\
\lim_{\eta, \eta' \rightarrow \eta_{0}} G(\eta, \eta', k) &= 0, & \lim_{\eta, \eta' \rightarrow -\infty(1 - i \epsilon)} G(\eta, \eta', k) &= 0.
\end{align}
We can then write both propagators in terms of the positive and negative frequency mode functions as
\begin{align}
K(k,\eta)&=\frac{\phi^+_k(\eta)}{\phi^+_k(\eta_0)}\,,\\
G ( p,\eta, \eta') &= i\left[ \theta(\eta-\eta')\left(\phi^+_p(\eta')\phi^-_p(\eta)-\frac{\phi^-_p(\eta_0)}{\phi^+_p(\eta_0)}\phi^+_p(\eta)\phi^+_p(\eta')\right)+(\eta \leftrightarrow \eta')\right] \nonumber \\
&=iP(p)\left[\theta(\eta-\eta')\frac{\phi^+_p(\eta')}{\phi^+_p(\eta_0)}\left(\frac{\phi^-_p(\eta)}{\phi^-_p(\eta_0)}-\frac{\phi^+_p(\eta)}{\phi^+_p(\eta_0)}\right)+(\eta \leftrightarrow \eta')\right],
\end{align}
where $P(p)$ is the power spectrum of $\phi$ and we have introduced the notation $\phi_{k}(\eta) \equiv \phi(\eta,k)$ to shorten the expressions. In deriving these expressions we have imposed the Bunch-Davies vacuum state as an initial condition which is the assumption that at very early times the mode functions are those of the flat-space theory. Physically this is because at very high energies the modes do not feel the expansion of the universe. \\
\begin{figure}
  \centering
  \includegraphics[width=.8\linewidth]{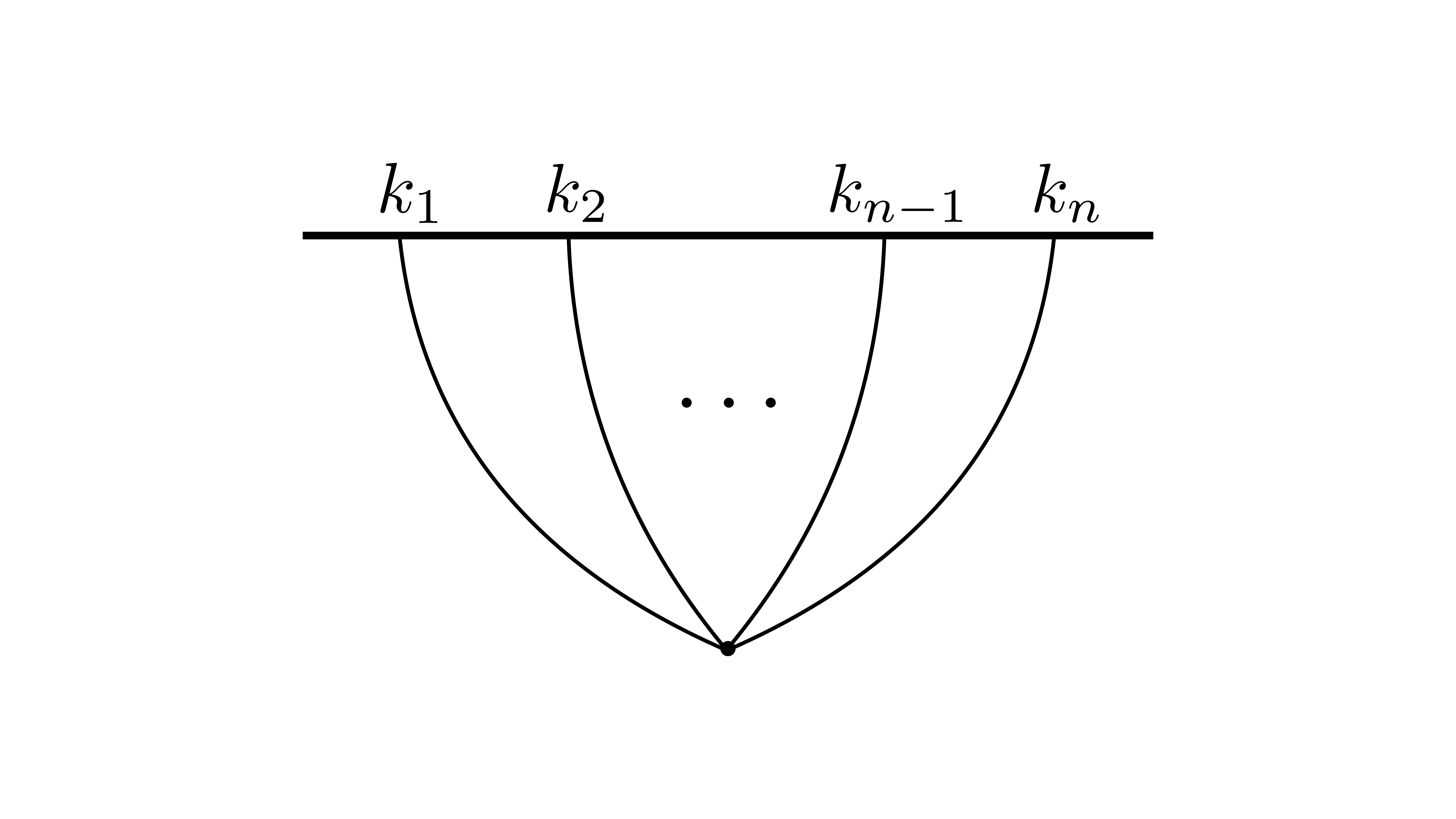}  
  \caption{Contact diagram for $n$ external fields}
  \label{fig:contactdiagram}
\end{figure}

\begin{figure}
  \centering
  \includegraphics[width=.8\linewidth]{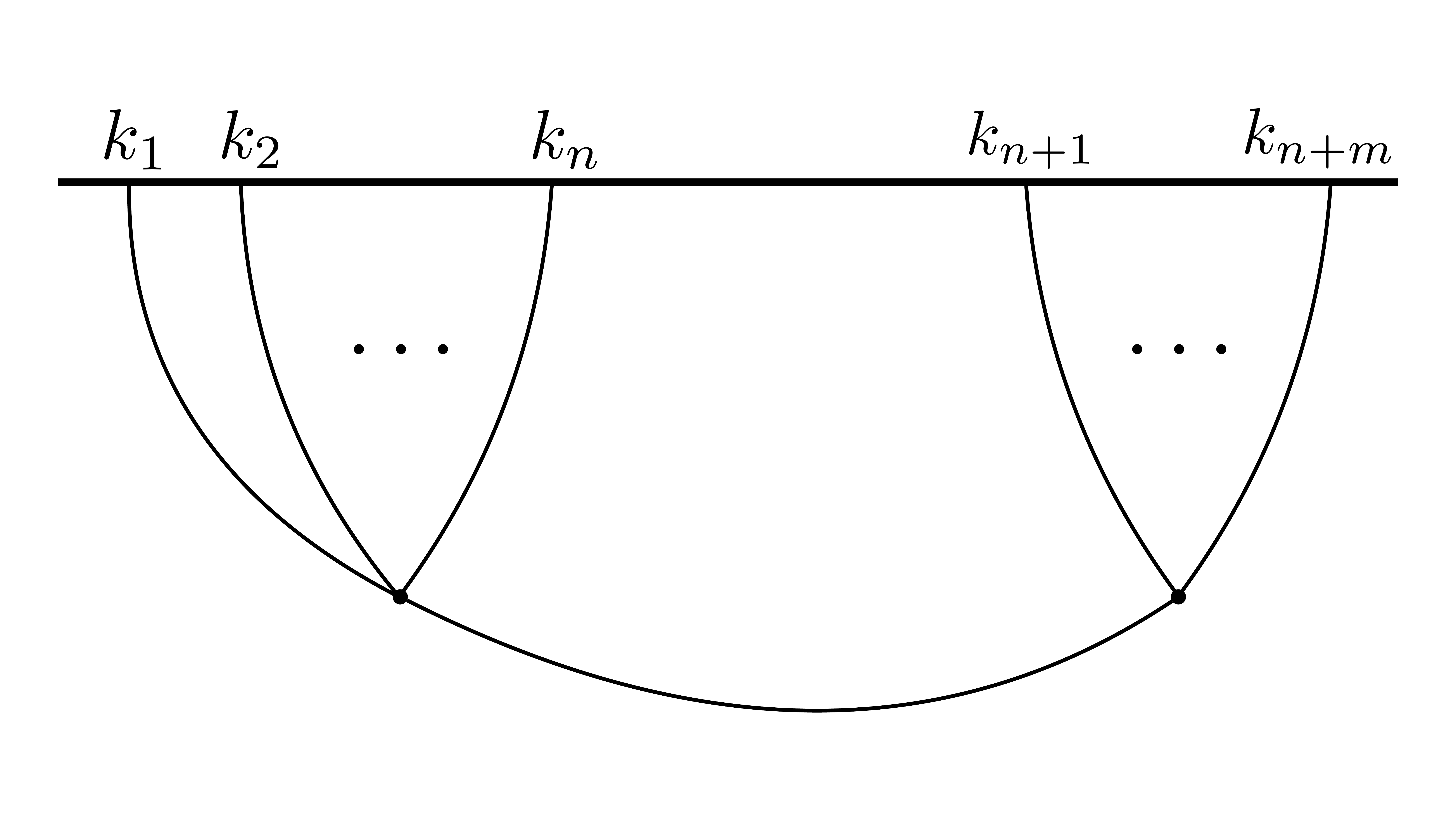}  
  \caption{Single exchange diagram for $n+m$ external fields}
  \label{fig:exchangediagram}
\end{figure}

\noindent Now to extract the wavefunction coefficients one follows the following Feynman rules. For a contact diagram like the one shown in Figure \ref{fig:contactdiagram}, we insert an overall factor of $(-i)$ and perform a single time integral where the integrand is a product of the coupling parameter, the $n$ bulk-boundary propagators and their derivatives (as dictated by the interaction vertex), and an appropriate number of scale factors (as dictated by scale invariance). Time derivatives act on the bulk-boundary propagators whereas spatial derivatives simply bring down a factor of $i k_{i}$, as is the case for scattering amplitudes. We integrate from the far past at $\eta = -\infty(1 - i \epsilon)$ to the future boundary at $\eta = \eta_{0}$. This $i \epsilon$ prescription ensures that there is a short period of evolution in Euclidean time rather than Lorentzian time that dampens the exponential factors appearing in the integral, thereby projecting the theory onto the vacuum state \cite{Hartle:1983ai, Maldacena:2002vr}. In analogy to scattering amplitudes, we finally sum over all possible permutations. For an exchange diagram like the one shown in Figure \ref{fig:exchangediagram} we now have two time integrals, one for each vertex. The vertices contribute $n$ and $m$ powers of the bulk-boundary propagators, possibly time-differentiated as dictated by the interaction vertices, while the internal line requires us to include one bulk-bulk propagator, which may also be differentiated with respect to time. The number of scale factors is fixed by scale invariance and as for contact diagrams we sum over all possible permutations. The generalisation of these rules to more complicated tree diagrams is simple, with a time integral for each local vertex. See Appendix A of \cite{MLT} for more details and examples. \\

\noindent As an example, for a massless scalar with a $\frac{a(\eta)}{3 !}\phi'^{3}$ self-interaction in the bulk, the three-point wavefunction coefficient is given by 
\begin{align}
\psi_{\phi'^3}(k_1,k_2,k_3)&=-i \int d\eta\,a(\eta)\,K'(k_1,\eta)\,K'(k_2,\eta)\,K'(k_3,\eta)\,,
\end{align}
while the $s$-channel four-point exchange diagram is given by
\begin{align}
\psi^s_{\phi'^3\times \phi'^3}=-i \int d\eta'\,d\eta\, \left(a(\eta) K'(k_1,\eta)K'(k_2,\eta)\right)\, \partial_\eta\partial_\eta' G(s,\eta,\eta')\,\left(a(\eta') K'(k_3,\eta')K'(k_4,\eta')\right),
\end{align}
where $s=|\bfk_1+\bfk_2|$ is the ``energy" of the internal line and we have suppressed the integration limits. This traditional computational process can be complicated due to the (nested) time integrals that have to be performed, which may obscure the origin of analytic properties of the final answer. In this paper we will usually avoid computing time integrals altogether and instead fix the final form of the wavefunction coefficients using symmetries, locality and unitarity, only computing explicit time integrals to verify that all parity-odd bispectra can be generated in solid inflation (Section \ref{SolidInflation}). In general, the wavefunction is a complex function of the kinematics and $\eta_{0}$, since we are evaluating the action on complex field configurations, and we will use our bootstrap methods to construct both the real and imaginary parts.  \\

\noindent With the wavefunction in hand, one can extract equal-time (late-time) expectation values using the usual quantum mechanics formula. We have 
\begin{align}
\label{eq:to_recall_correlators}
\langle \phi(\bfk_1) \ldots \phi(\bfk_n)  \rangle = \frac{\int \mathcal{D} \phi ~ \Psi \Psi^{\ast} ~ \phi(\bfk_1) \ldots \phi(\bfk_n)}{\int \mathcal{D} \phi ~ \Psi \Psi^{\ast}}\,,
\end{align}
for an $n$-point function of scalars. Here $\mathcal{D} \phi$ is the functional measure on a fixed time slice. Correlators are therefore fixed via the bulk dynamics through the probability distribution $ \Psi \Psi^{\ast}$. We will use this equation in Section \ref{ParitySection} to derive some general results for cosmological correlators arising from unitary time evolution in the bulk.
\subsection{Boostless Bootstrap Rules}
We now turn to reviewing bootstrap techniques for efficient computation of late-time wavefunctions/correlators. In \cite{BBBB} a set of \textit{Boostless Bootstrap Rules} was introduced that enables one to write down general structures for the three-point functions of massless scalars and gravitons without assuming full dS symmetries. In total, six rules were introduced, each based on the following principles:
\begin{itemize}
\item Rule 1: Spatial translations, spatial rotations and scale invariance, 
\item Rule 2: Tree-level approximation for wavefunctions and correlators in dS, 
\item Rule 3: High-energy boundary condition in the form of an amplitude limit, 
\item Rule 4: Bose statistics for wavefunctions/correlators of external bosons,
\item Rule 5: Bunch-Davies initial vacuum state,
\item Rule 6: Soft theorems.
\end{itemize}
For the curvature perturbation in inflation each of these six rules are necessary to bootstrap the bispectrum \cite{BBBB}, however for gravitons and spectator scalars that are the primary interest in this paper, rules $3$ and $6$ are not required and are replaced by the \textit{Manifestly Local Test} of \cite{MLT} which we will review in the following subsection. Before doing so let us first review the other rules ($1,2,4,5$) and refer the reader to \cite{BBBB} for further details on all rules.
\begin{itemize}
\item Rule 1: Spatial translations, spatial rotations and scale invariance. These symmetries ensure that wavefunction coefficients can be written as a product of a \textit{polarisation factor}, which is an $SO(3)$ invariant function of polarisation tensors and spatial momenta, multiplied by a \textit{trimmed wavefunction coefficient} which is only a function of the energies:
\begin{align} \label{GeneralWFC}
\psi_n = \sum_\text{contractions} \text{(polarization factor)} \times \text{(trimmed wavefunction coefficient)}\,.
\end{align}
We take all coefficients appearing in the polarisation factor to be real and therefore include any factors of $i$ that might appear when converting to momentum space, or simply as part of the Feynman rules, in the trimmed part which we will denote as $\tilde{\psi}_{n}$. We denote the total number of spatial momenta appearing in the polarisation factor as $\alpha$. For the bispectrum of massless gravitons which is our primary interest in this paper, we have 
\begin{align}\label{alpha}
\psi_3 &= \sum_\text{contractions} \left[e^{h_1}(\bfk_1) e^{h_2}(\bfk_2) e^{h_3}(\bfk_3)\bfk_1^{\alpha_1}\bfk_2^{\alpha_2}\bfk_3^{\alpha_3}\right] \psi^{\rm trimmed}_3,
\end{align}
with $\alpha_{1}+\alpha_{2}+\alpha_{3} = \alpha$. Here we have already stripped off the ever-present momentum conserving delta function that is a consequence of spatial momentum conservation. Furthermore, scale invariance ensures that for all $n$ we have $\psi_n \sim k^3$ which cancels the scaling of the three-dimensional delta function thereby ensuring invariance of $\Psi$. If one also includes dS boosts as a symmetry, the trimmed wavefunction coefficients for gravitons are very constrained \cite{Maldacena:2011nz}. In this paper we are interested in boost-breaking scenarios and so will not impose invariance under dS boosts.
    \item Rule 2: Tree-level approximation for wavefunctions/correlators in dS. This rule simply imposes that the bispectrum is a rational function of the external kinematics up to possible logarithmic terms. Such logs will indeed be captured by our bootstrap analysis. Our focus in this paper will be at tree-level but progress is now also being made on using bootstrap techniques at loop-level \cite{Melville:2021lst,DiPietro:2021sjt,Hogervorst:2021uvp}.
    \item Rule 4: Bose statistics for wavefunctions/correlators of external bosons. This rule enforces invariance under permutations of the momenta of identical fields.
    \item Rule 5: Bunch-Davies initial vacuum state. The assumption of a Bunch-Davies initial state enforces that the only allowed poles for contact diagrams are in the total energy $k_{T}  = \sum_{a =1 }^{n} k_{a}$. The degree of the leading $k_{T}$ pole is given by $p=1+\sum_A (\Delta_A-4)$ where the sum is over all vertices appearing in a given diagram and $\Delta_A$ is their mass dimension \cite{BBBB}. We only have one type of pole since the integrands appearing in the bulk formalism only depend on the positive frequency modes. For excited initial states both positive and negative frequency modes can contribute leading to so-called flattened singularities, see e.g. \cite{HolmanTolley,Signals} for the phenomenology of such poles. It is also interesting to note that the residue of the leading order $k_{T}$ poles contain the flat-space scattering amplitude for the same process \cite{Maldacena:2011nz,COT,Raju:2012zr}.
\end{itemize}
These four rules will play an important role in our ability to bootstrap graviton bispectra in Section \ref{sec:bispectra}.


\subsection{Manifestly Local Test}
In \cite{MLT} a condition, referred to as the \textit{Manifestly Local Test} (MLT), was introduced that must be satisfied by both contact and exchange $n$-point wavefunction coefficients of massless scalars and gravitons with manifestly local interactions. Manifestly local interactions are those with only positive powers of derivatives, i.e. without inverse Laplacians; this is a natural locality condition for gravitons and spectator scalars in dS at cubic order in perturbations \cite{BBBB}. Manifest locality can be violated upon integrating out the non-dynamical modes in a gravitational theory, so such a violation is a feature of the self-interactions of the inflationary curvature perturbation \cite{Maldacena:2002vr} as well as gravitons at quartic and higher order in the fields. The MLT was used in \cite{MLT} to bootstrap bispectra of the Goldstone mode in the Effective Field Theory of Inflation \cite{Cheung:2007st} to all orders in derivatives, and used in conjunction with partial energy recursion relations to bootstrap inflationary trispectra (see also \cite{Arkani-Hamed:2017fdk} for a use of energy shifts for the flat-space wavefunction). The MLT was also recently employed in \cite{Jain:2021vrv}. The MLT offers a conceptually simple yet very powerful bootstrap technique and will be a central feature of this work. \\

\noindent The MLT takes the form
 \begin{align}  \label{MLT}
    \frac{\partial }{\partial k_{c}} \psi_{n}(k_1,...,k_n;\{p \}; \{\bfk\})\Big|_{k_{c}=0}=0\,,\qquad \forall\, c=1,\dots,n\,, 
    \end{align}
where $k_{a}$ are the energies of the external fields, $\{p \}$ collectively denotes the energies of possible exchange fields while $\{\bfk\}$ collectively denotes a possible dependence of $n$-point functions on spatial momenta and polarisation tensors. We will also often also use $\{k \}$ to collectively denote the external energies. The derivative with respect to one of the external energies is taken while keeping all other variables fixed and this condition must be met for all external energies if they are those of a massless scalar or a graviton in de Sitter. Two complementary derivations of the MLT were given in \cite{MLT}. The first arises from demanding that exchange diagrams have the appropriate singularities while the second comes directly from the bulk representation of such $n$-point functions. We refer the reader to \cite{MLT} for details of the first method while reviewing the second here. \\
    
 \noindent The computation of tree-level diagrams in the bulk formalism reduces to nested time integrals of the following schematic form
\begin{align}
\psi_{n}(\{k\};\{p\};\{\bfk\}) \sim \int \left(\prod_A^V d \eta_A F_A\right) \left( \prod_a^n \partial_\eta^{\#} K_{\phi}(k_{a}) \right) \left(\prod_m^I \partial_\eta^\# G(p_m)\right) \,,
\end{align}
where the $F_A$'s denote the momentum dependence due to the spatial derivatives and polarisation tensors in the $V$ vertices, each vertex representing a contact interaction placed at the conformal time $\eta_A$. We have included a bulk-boundary propagator for each external field and have allowed for an arbitrary number of time derivatives acting on these propagators. Finally, we have allowed for $I$ internal bulk-bulk propagators $G$, possibly with time derivatives. Now we differentiate the above expression with respect to one of the external energies. This derivative acts only on the bulk-boundary propagator associated to this energy, because $F_A$ depend only on the spatial momenta and polarisations while $G(p_m)$ depend only on energies of internal legs. Assuming that $\eta$ integrals and $\frac{\partial}{\partial k_c}$ commute, we have
\begin{align} \label{eq:MLTproof2}
  \frac{\partial }{\partial k_{c}} \psi_{n}\Big|_{k_{c}=0} \sim \int \left(\prod_A^V d \eta_A F_A\right) \left( \prod_{a \neq c} \partial_\eta^{\#} K_{\phi}(k_{a}) \right) \left( \partial_\eta^{\#} \left( \frac{\partial}{\partial k_c} K_{\phi}(k_{c}) \right)\Big|_{k_{c}=0} \right) \left(\prod_m^I \partial_\eta^\# G(p_m)\right) \,.
\end{align}
The bulk-boundary propagator for a massless graviton is the same as for a massless scalar up to the presence of a polarisation tensor. In both cases, we have
\begin{align}
\frac{d}{d k} K(\eta,k)= \frac{d}{d k}  \left(  (1 - i k \eta )e^{ik\eta} \right) = k \eta^2 e^{i k \eta}\,,
\end{align}
which vanishes at $k=0$. It follows that (\ref{eq:MLTproof2}) must vanish. We emphasise that we have not assumed anything about the form of the $\psi_{n}$, so the MLT holds for contact and exchange diagrams, even those with IR-divergences: it follows from a simple property of the bulk-boundary propagators, namely that $\frac{d}{d k} K(\eta,k)$ vanishes at $k=0$. In fact, this property also holds in slow roll inflation, for both massless gravitons and massless scalars, and therefore the MLT (\ref{MLT}) is applicable in that case as well. The main obstacle to extending all of our results beyond exact scale invariance is therefore not the MLT itself, but the assumption of scale invariance (Rule 1), which allows us to write down a simple ansatz for the wavefunction coefficient before applying the MLT (as will be shown in detail in Section \ref{sec:bispectra}). We will return to the prospect of employing the MLT to construct slow-roll corrections in the future.  \\

\noindent The MLT, in conjuction with the bootstrap rules from the previous section, can be used to find \textit{all} consistent, tree-level, contact wavefunction coefficients for massless scalars and gravitons in de Sitter. Let us present a constructive proof of this claim. As a first step, we find an exhaustive list of polarization factors (see \eqref{GeneralWFC}), which covers all possible contractions of tensor indices. Then we write down an ansatz for $\psi^{\rm trimmed}_n$, consistent with rules 2 and 5 (rule 4 is automatically satisfied once we sum over the permutations). Any such ansatz can be written in the form of a bulk integral
\begin{align} \label{eq:MLTcompleteness}
\psi^{\rm trimmed}_{n} \sim \int d \eta  f(k_a , \bfk_a . \bfk_b; \eta) e^{i k_T \eta} ,
\end{align}
where $f(k_a , \bfk_a . \bfk_b; \eta)$ is a polynomial in the energies $k_a$ and the scalar products $\bfk_a . \bfk_b$, with appropriate factors of $\eta$ as required by scale invariance. The exponential factor contributes the needed poles in $k_T$, and these are the only possible poles, as dictated by rules 2 and 5. The IR divergences, which are of the form $\eta_0^{-m}$ or $\log(-k_T \eta_0)$, are fully accounted for by those terms in $f$ that have negative powers of $\eta$. \\

\noindent The final ingredient is the MLT, which imposes the following constraints on $f$:
\begin{equation} \label{eq:fconstraint}
\frac{\partial f}{\partial k_a} \Big|_{k_a = 0} + i \eta f|_{k_a=0} = 0.
\end{equation}
It is easy to see that any such polynomial (assuming scale invariance) can be written as
\begin{equation}
\label{eq:MLTcompletenessAnsatz}
f(k_a , \k_a . \k_b; \eta) = (1 - i k_1 \eta) g(k_2, \ldots , k_n, \k_a . \k_b; \eta) + k_1^2 h(k_a, \k_a . \k_b;\eta) \,,
\end{equation}
where $g, h$ are polynomials satisfying
\begin{align}
\frac{\partial g}{\partial k_a} \Big|_{k_a = 0} + i \eta g|_{k_a=0} &= 0, \quad a \neq 1, \\
\frac{\partial h}{\partial k_a} \Big|_{k_a = 0} + i \eta h|_{k_a=0} &= 0, \quad a \neq 1. 
\end{align}
Then, we can repeat the decomposition (\ref{eq:MLTcompletenessAnsatz}), albeit now for $g$ and $h$. By iterating over $a = 1, 2, \ldots, n$, we can arrive at a general form of $f(k_a , \k_a . \k_b; \eta)$:
\begin{align}
f(k_a , \k_a . \k_b; \eta) = \sum\limits_{S \subset \mathbb{Z}_n} \left( \prod\limits_{j \notin S} \left( 1 - i k_j \eta \right)  \prod_{j \in S} \left( k_j^2 \right) h_S \left( k_{a \in S}, \k_a . \k_b ; \eta \right) \right),
\end{align}
where $h_S$ are polynomials in the $k_a \in S$ and the scalar products $\bfk_a . \bfk_b$. The sum is taken over all subsets $S$ of the set $\mathbb{Z}_n := \{ 1, 2, \ldots, n\}$. It will now be sufficient to show that any term of the above sum can be produced by some linear combination of functions constructed from bulk-boundary propagators. In fact, we can focus on the case where $h_S$ is a monomial, since any polynomial is just a linear combination of those. If this monomial includes factors of $\bfk_a . \bfk_b$, we can generate them from the Lagrangian by writing pairs of spatial derivatives contracted with each other, so from now on, let us assume for simplicity that $h_S$ is a monomial that does not include such factors. Reinstating powers of $\eta$ as required by scale invariance, we are thus looking for a functional of bulk propagators that would generate
\begin{align}
\psi^{\rm trimmed}_{n} \sim \int d \eta \prod\limits_{j \notin S} \left( 1 - i k_j \eta \right)  \prod_{j \in S} \left( k_j^{2+n_j} \right) \eta^{\alpha + \sum_{j \in S} n_j + 2 |S| - 4}  e^{i k_T \eta} \,,
\end{align}
for some arbitrary $n_j \geqslant 0$; $\alpha$ is the energy dimension of the polarization factor. The linear combination we are looking for is, up to an overall constant,
\begin{align}
 \eta^{\alpha - 4} \prod\limits_{j \notin S} K(k_j, \eta) \prod_{j \in S} \left( K_{2+n_j}(k_j, \eta) \right) \,,
\end{align}
where $K(k_j, \eta)$ is the usual bulk-boundary propagator, and
\begin{align}
K_{2}(k, \eta) &\equiv \eta \partial_{\eta} K(k, \eta) = k^2 \eta^2 e^{i k \eta}, \\
K_{3}(k, \eta) &\equiv -i \left( \eta^{2} \partial^2_{\eta} K(k, \eta) - \eta \partial_{\eta} K(k, \eta) \right) = k^3 \eta^3 e^{i k \eta}, \\
K_{n+2}(k, \eta) &\equiv k^2 \eta^2  K_n(k, \eta) \quad \text{for} \ n \geqslant 2.
\end{align}
Each of these functions can be obtained from the massless bulk-boundary propagators by applying time derivatives, Laplacians $(k^2 \leftrightarrow -\nabla^2)$ and taking linear combinations. Recall that we can introduce the dependence on $\bfk_a . \bfk_b$ by introducing pairs of spatial derivatives, followed by taking linear combinations again to account for terms with distinct dependencies on $\bfk_a . \bfk_b$. Therefore, any integral of the form (\ref{eq:MLTcompletenessAnsatz}) can be generated by a linear combination of products of bulk-boundary propagators, their time derivatives, factors of $a(\eta)^2 k_a^2$ and by pairs of spatial derivatives contracted with each another. This entails that any solution to the MLT corresponds to a combination of some manifestly local operators.


\subsection{Cosmological Optical Theorem}
The final bootstrap tool we are going to review is the Cosmological Optical Theorem (COT) \cite{COT} which is a consequence of unitary time evolution in the bulk. It was shown in \cite{COT} that if the wavefunction of the universe is normalised at time $\eta$ then it only remains normalised at time $\eta'$ if contact and exchange wavefunction coefficients satisfy some simple yet powerful relations. Assuming a Bunch-Davies initial condition, the bulk-boundary propagator of fields of general mass and spin on any FLRW spacetime satisfies (see \cite{Goodhew:2021oqg} for a proof and a discussion of the related technical assumptions)
\begin{align} \label{BulkBoundaryCOT}
    K^*(-k^*,\eta)=K(k,\eta)\,,\qquad k \in \mathbb{C}\,,
\end{align}
from which one can derive the COT for contact diagrams \cite{COT}
\begin{align}\label{COT}
    \disc \left[ i \psi_{n}(k_1,...,k_n; \{\bfk\} ) \right] = i \left[ \psi_n(k_1,...,k_n;  \{\bfk\}) + \psi^*_n(-k^*_1,...,-k^*_n; \{ - \bfk\} ) \right] = 0 \,,
\end{align}    
which must be satisfied by any contact $n$-point function arising from unitary evolution in the bulk spacetime. Note that all spatial momenta in the second term get a minus sign, $ {\bfk} \to -\bfk$, and all energies are analytically continued. One is usually interested in real values of the energies $  k $, and so in the following we will drop the complex conjugation. This notation is unambiguous as long as one adopts the prescription that \textit{all negative energies are approached from the lower-half complex plane}. For scalars it is clear from \eqref{COT} how the second term should be computed but for spinning fields the presence of polarisation tensors introduces slight complications which were addressed in \cite{Goodhew:2021oqg}. Ultimately any polarisation factors appear as a common factor in this contact COT since e.g. $e_{ij}^{h}({\bf k})^{\ast} =e_{ij}^{h}(-{\bf k}) $. The COT is therefore not constraining the polarisation factor (which is constrained by symmetry), rather it is constraining the trimmed part of the wavefunction that in the bulk representation arises from performing the bulk time integrals. This of course makes sense as the COT is indeed a consequence of unitary time evolution. For our purposes in this paper the COT for contact diagrams is enough and we will use it in Section \ref{ParitySection} to derive some general results about cosmological correlators, but the consequences of unitarity for exchange diagrams are also known \cite{COT,Goodhew:2021oqg} and were used extensively in \cite{MLT} to bootstrap inflationary trispectra. The COT for exchange diagrams relates the discontinuity of an exchange diagram to products of the contributing sub-diagrams, multiplied by the power spectrum of the exchanged field. It is reminiscent of the factorisation theorem for scattering amplitudes. A complementary derivation of the COT was given in \cite{Cespedes:2020xqq} where the consequences of excited initial states were also considered. The COT was also extended to general FLRW spacetimes in \cite{Goodhew:2021oqg} and to loop level in the form of cutting rules in \cite{Melville:2021lst}, see also \cite{Baumann:2021fxj} for a recent discussion of cosmological cuts. Unitarity constraints on cosmological observables were also recently studied in \cite{DiPietro:2021sjt,Meltzer:2021zin,Hogervorst:2021uvp}. See \cite{Aharony:2016dwx,Meltzer:2019nbs,Meltzer:2020qbr} for analogous statements in anti-de Sitter (AdS) space. 

\subsection{Cosmological spinor helicity formalism} \label{SHF}
In this paper we are primarily concerned with bootstrapping graviton bispectra and just as is the case for scattering amplitudes, wavefunctions/correlators of spinning fields are most compactly presented using spinors rather than polarisation tensors. We end this section by reviewing the cosmological spinor helicity formalism and refer to the reader to \cite{Maldacena:2011nz,Baumann:2020dch} for other presentations. \\

\noindent The spinor helicity formalism is most useful when we have null momenta as is the case for massless on-shell particles in flat-space and has been used extensively in that setting. In our cosmological setting the spatial momentum ${\bf k}$ is not null, but we can define a null four-component object $k_{\mu} = (k, \bf{k} )$, with $k=|\bf{k}|$, which we can express as the outer product of two spinors via
\begin{equation}
\label{eq:spinor_review-1-A}
k_{\alpha\dot{\alpha}} = \sigma^{\mu}_{\alpha\dot{\alpha}}k_\mu = \lambda_{\alpha}\tilde{\lambda}_{\dot{\alpha}}\,, 
\end{equation} 
where $\sigma^\mu = (\mathds{1},\bm{\sigma})$ and $\bm{\sigma}$ are the Pauli matrices. Using the relation 
$\sigma^\mu_{\alpha\dot{\alpha}}\bar{\sigma}^{\dot{\beta}\beta}_\mu = 2\delta^\beta_\alpha\delta^{\dot{\beta}}_{\dot{\alpha}}$ (we follow the conventions used in \cite{Dreiner}), where $\bar{\sigma}^\mu = (\mathds{1},{-\bm{\sigma}})$, the inverse of \eqref{eq:spinor_review-1-A} is 
\begin{equation}
\label{eq:spinor_review-1-B}
k_\mu = \frac{1}{2}\bar{\sigma}^{\dot{\alpha}\alpha}_\mu k_{\alpha\dot{\alpha}}\,. 
\end{equation}
A little group transformation by definition should leave this four-momentum invariant, so we can model this transformation as $\lambda \rightarrow t \lambda$, $\tilde{\lambda} \rightarrow t^{-1} \tilde{\lambda}$ where each external field transforms with a different constant $t \in \mathbb{C}$. These very simple helicity transformations allow us to easily extract an overall dependence of a wavefunction/correlator on the spinors given some helicity configuration for the external fields, and is one of the primary virtues of the spinor helicity formalism. As usual, dotted and un-dotted indices are raised and lowered by $\epsilon_{\dot{\alpha}\dot{\beta}}$ and $\epsilon_{\alpha\beta}$ respectively e.g. $\smash{\tilde{\lambda}_{\dot{\alpha}} = \epsilon_{\dot{\alpha} \dot{\beta}} \tilde{\lambda}^{\dot{\beta}}}$, $\smash{\lambda_{\alpha} = \epsilon_{\alpha \beta} \lambda^{\beta}}$. 
\\
\\
Now for objects with three external fields, conservation of spatial momentum $\bfk_1+\bfk_2+\bfk_3=0$ leads to 
\begin{equation}
\label{eq:spinor_review-2}
\text{$\sum_{a=1}^3\lambda_{\alpha}^{(a)}\tilde{\lambda}^{(a)}_{\dot{\alpha}} = 
k_T(\sigma_0)_{\alpha\dot{\alpha}}$\quad and \quad $\braket{ab}[ab] = k_T(k_T-2k_c)\equiv k_TI_c$} \quad \text{for} \quad a \neq b \neq c\,,
\end{equation} 
where we have introduced 
\begin{align}
I_a \equiv (k_T-2k_a)\,,
\end{align}
and we recall that 
\begin{equation}
\label{eq:spinor_review-3}
\braket{ab} = \epsilon^{\alpha\beta}\lambda_{\alpha}^{(a)}{\lambda}^{(b)}_{\beta}\,\,, 
\quad[ab] = \epsilon^{\dot{\alpha}\dot{\beta}}\tilde{\lambda}_{\dot{\alpha}}^{(a)}\tilde{\lambda}^{(b)}_{\dot{\beta}}\,.
\end{equation} 
We remind the reader that the above spinors are not Grassmanian, so these angle and square brackets are anti-symmetric due to the anti-symmetric nature of the epsilon tensors. For scattering amplitudes one also has time translation invariance, which implies $k_{T} = 0$. In this case the above relations reduce to the usual flat-space ones, see e.g. \cite{Cheung:2017pzi}.
Now to construct $SO(3)$ invariant objects we can use \eqref{eq:spinor_review-3} but can also contract dotted and un-dotted indices using $\sigma^0_{\alpha\dot{\alpha}}$ \cite{Maldacena:2011nz,PSS}:
\begin{equation}
\label{eq:spinor_review-4}
(ab) = (\sigma^0)^{\alpha\dot{\alpha}}\lambda_{\alpha}^{(a)}\tilde{\lambda}^{(b)}_{\dot{\alpha}}\,, 
\end{equation}
with $(aa)=2k_a$. We can use \eqref{eq:spinor_review-2} to obtain an expression for $(ab)$ with $a \neq b$ i.e. the off-diagonal components. We have 
\begin{align}
(ab)[ac] &= I_b[bc]  \quad \text{for} \quad a \neq b \neq c\,, \label{eq:spinor_review-5-1} \\
(ab)\braket{bc} &= I_a \braket{ac}  \quad \text{for} \quad a \neq b \neq c\,, \label{eq:spinor_review-5-2}
\end{align}
and therefore a general three-point function is a function of the angle brackets, the square brackets and the energies. \\ 

\noindent For spinning fields, we will find it necessary to write polarisation tensors in terms of spinors. The transverse and traceless graviton polarisation tensors $e^{\pm}_{\mu\nu} $ are given by $e^{\pm}_{\mu}e^{\pm}_{\nu}$, where $e^{\pm}_{\mu}$ is the polarisation vector for a spin-$1$ particle of the same momentum. We therefore only need an expression for $e^{\pm}_{\mu}$ in the spinor helicity formalism. The form of the polarisation vectors follows from the fact that they must be lightlike, orthogonal to the corresponding momentum, and carry the appropriate helicity weight. We have (see e.g. \cite{Cheung:2017pzi,PSS})
\begin{equation}
\label{eq:spinor_review-6}
e^+_{\alpha\dot{\alpha}} = 2 \sqrt{2} \frac{\mu_\alpha\tilde{\lambda}_{\dot{\alpha}}}{\braket{\mu\lambda}}\,,\qquad 
e^-_{\alpha\dot{\alpha}} = 2 \sqrt{2} \frac{\lambda_\alpha\tilde{\mu}_{\dot{\alpha}}}{[\mu\lambda]}\,,
\end{equation} 
for generic reference spinors $\mu_\alpha$ and $\tilde{\mu}_{\dot{\alpha}}$. For scattering amplitudes in flat-space these reference spinors represent the redundancy in defining massless spinning fields as a representation of the Lorentz group, but for cosmology we can make a choice to eliminate this redundancy \cite{Maldacena:2011nz}. Indeed, we can use our freedom to mix dotted and undotted indices to choose 
\begin{equation}
\label{eq:spinor_review-7}
\mu_\alpha = (\sigma_0)_{\alpha\dot{\alpha}}\tilde{\lambda}^{\dot{\alpha}}\,, \qquad 
\tilde{\mu}_{\dot{\alpha}} = (\sigma_0)_{\alpha\dot{\alpha}}\lambda^{\alpha}\,, 
\end{equation} 
which makes the zero component of the polarisation vectors vanish. We can therefore write
\begin{equation}
\label{eq:spinor_review-8}
e^+_{\alpha\dot{\alpha}} = \sqrt{2} \frac{(\sigma_0)_{\alpha\dot{\beta}}\tilde{\lambda}^{\dot{\beta}}\tilde{\lambda}_{\dot{\alpha}}}{k}\,,\qquad 
e^-_{\alpha\dot{\alpha}} = \sqrt{2} \frac{(\sigma_0)_{\beta\dot{\alpha}}\lambda^\beta\lambda_{\alpha}}{k}\,,
\end{equation}
which has the correct normalisation. Under a helicity transformation we have $e^{+} \rightarrow t^{-2} e^{+}$ and $e^{-} \rightarrow t^{2} e^{-}$, as expected.  \\

\noindent With these relations at hand, we can easily convert any $SO(3)$ invariant object containing spatial momenta and polarisation vectors into the spinor helicity formalism using the necessary $\sigma$ and $\epsilon$ identities which are given in \cite{Dreiner}. We present a complete list of distinct contractions of $SO(3)$ indices for a massless graviton in Appendix~\ref{app:contractions}. We will use these relations extensively in Section \ref{subsec:tensor_structures} where we study the tensor structures for the graviton bispectrum. 


\section{Unitarity constraints on $n$-point cosmological correlators} \label{ParitySection}

In this section we are going to use the Cosmological Optical Theorem (COT) for contact diagrams to derive some general results about the form of cosmological correlators. Recall that with the wavefunction of the universe at hand, one can compute expectation values via Eq.~\eqref{eq:to_recall_correlators}, i.e.
\begin{align}
\langle \varphi(\bfk_1) \ldots \varphi(\bfk_n)  \rangle = \frac{\int \mathcal{D} \varphi ~ \Psi \Psi^{\ast} ~ \varphi(\bfk_1) \ldots \varphi(\bfk_n)}{\int \mathcal{D} \varphi ~ \Psi \Psi^{\ast}}\,,
\end{align}
where in the weak coupling approximation we are using here, the late-time wavefunction is given by 
\begin{align}
\Psi[\eta_{0},\varphi({\bf k})] = \text{exp}\left[-\sum_{n=2}^{\infty} \frac{1}{n!} \int_{{\bfk}_{1}, \ldots, {\bf{k}}_{n}} \psi_{n}(\{k \}; \{\bfk\})\varphi({\bf k}_{1}) \ldots \varphi({\bf k}_{n}) \right]\,.
\end{align}
Here we have made a distinction between the dependence of the wavefunction coefficients on the set of spatial momenta $\{\bfk\}$ and their norms $\{ k \}$, since in general we will work away from the physical configuration and treat $\{\bfk\}$ and $\{ k \}$ as independent objects, for reasons that will become clear. We have not included a possible dependence on internal energies $\{p \}$ since our focus in this section is on \textit{contact diagrams}. We are going to use the COT to constrain the form of the probability distribution $ \Psi \Psi^{\ast}$. Here and throughout this section we use $\varphi(\bfk)$ to schematically denote scalars and gravitons, with $SO(3)$ indices suppressed, and each of these fields satisfies $\varphi(\bfk) = \varphi(-\bfk)^{\ast}$ which follows directly from \eqref{FreeField}, \eqref{ModeFunctionsMassless} and \eqref{poln}. Now from this perturbative expression for the wavefunction, we have
\begin{align}
 - \log(\Psi \Psi^{\ast}) & =  \left( \sum_{n=2}^{\infty} \frac{1}{n!} \int_{{\bfk}_{1}, \ldots, {\bf{k}}_{n}} \psi_{n}(\{k \}; \{\bfk\})\varphi({\bf k}_{1}) \ldots \varphi({\bf k}_{n}) \right) \nonumber  \\ &+ \left( \sum_{n=2}^{\infty} \frac{1}{n!} \int_{{\bfk}_{1}, \ldots, {\bf{k}}_{n}} \psi_{n}(\{k \}; \{\bfk\})\varphi({\bf k}_{1}) \ldots \varphi({\bf k}_{n}) \right)^{\ast} \\
& =    \left( \sum_{n=2}^{\infty} \frac{1}{n!} \int_{{\bfk}_{1}, \ldots, {\bf{k}}_{n}} \psi_{n}(\{k \}; \{\bfk\})\varphi({\bf k}_{1}) \ldots \varphi({\bf k}_{n}) \right) \nonumber  \\ &+ \left( \sum_{n=2}^{\infty} \frac{1}{n!} \int_{{\bfk}_{1}, \ldots, {\bf{k}}_{n}} \psi^{\ast}_{n}(\{k \}; \{\bfk\})\varphi(-{\bf k}_{1}) \ldots \varphi(-{\bf k}_{n}) \right)\,.
\end{align}
If we  change the integration variables on the final line by sending $\{ \bfk \} \rightarrow  \{- \bfk \}$ we have 
\begin{align}
 - \log(\Psi \Psi^{\ast}) = \sum_{n=2}^{\infty} \frac{1}{n!} \int_{{\bfk}_{1}, \ldots, {\bf{k}}_{n}} [\psi_{n}(\{k \}; \{\bfk\})+ \psi^{\ast}_{n}(\{k \}; \{ - \bfk\})]\varphi({\bf k}_{1}) \ldots \varphi({\bf k}_{n})\,.
\end{align}
It follows from Gaussian integral formulae that the resulting correlators arising from these contact diagrams, in perturbation theory, are given by
\begin{align} \label{WFtoCorrelator}
B_{n}^{\text{contact}}(\{k \}; \{\bfk\}) =  -\frac{\psi'_{n}(\{k \}; \{\bfk\})+ \psi'^{\ast}_{n}(\{k \}; -\{\bfk\})}{\prod_{a=1}^{n} 2 \ \text{Re} ~ \psi'_{2}(k_{a})}\,,
\end{align}
where in deriving this expression we kept only terms linear in the coupling constants. For parity-even interactions of scalars and gravitons, the numerator is simply $2 \text{Re} ~ \psi'_{n}$ in which case our expression matches the one that usually appears in the literature. \\


\noindent Let's now use the contact COT to constrain $B_{n}^{\text{contact}}$. As we reviewed above, unitary time evolution in the bulk inflationary spacetime and the choice of the Bunch-Davies vacuum imply that \cite{COT}
\begin{align}\label{COT2}
\psi_n(\{ k \};  \{\bfk\}) + \psi^*_n(\{ -k\}; - \{\bfk\} ) = 0 \,.
\end{align}  
By directly comparing \eqref{WFtoCorrelator} and \eqref{COT2}, we conclude that 
\begin{center}
\textit{Any contribution to the wavefunction of the universe that is invariant under $\{ k \} \rightarrow \{ -k \}$, which is a flip in the sign of all external energies, does not contribute to the contact correlator.}
\end{center}

\noindent What are the implications of this observation? To answer this question we need to look more closely at the form of $  \psi_{n} $. After stripping away the polarization factor in $  \psi $, see \eqref{alpha}, the remaining trimmed wavefunction $  \psi^{\rm trimmed} $ for a contact interaction can have the following structures:
\begin{enumerate}
\item The trimmed wavefunction may be a rational functions of $\{ k\}$,
\begin{align} \label{eq:3dimensionsPsiTrimmed}
\psi^{\rm trimmed}_{n} \supset  \frac{\text{Poly}_{3 - \alpha + q}(\{ k \})}{\text{Poly}_{q}(\{ k \})} \,,
\end{align}
where the subscripts indicate the degrees of the polynomials and the combination $3 - \alpha + q$ is fixed by scale invariance such that $\psi_{n} \sim k^3$. If we further impose locality and the Bunch-Davies vacuum as in the bootstrap Rule 5 then the denominator must be $  k_{T} $ to some power, but we will not use this fact in the following. \\

\noindent If $\alpha$ is even, this trimmed wavefunction contains an overall odd number of energies and therefore is not invariant under $\{ k \} \rightarrow \{ -k \}$, whereas if $\alpha$ is odd, the trimmed wavefunction contains an overall even number of energies and so is invariant under $\{ k \} \rightarrow \{ -k \}$. So rational terms in the wavefunction can only contribute to the correlator if the polarisation factor has an even number of spatial momenta, which for scalars and gravitons implies parity-even. Conversely, parity-odd interactions of scalars and gravitons have an odd number of derivatives, which are contracted with a Levi-Civita tensor, and the contribution of their rational part to the correlator must vanish. This observation explains why $k_{T}$ poles were never found in the in-in computation of parity-odd graviton bispectra in the effective theory of inflation performed in \cite{CabassBordin}: they are simply incompatible with unitarity. 
\item The trimmed wavefunction may have logarithmic IR-divergences,
\begin{align}
\psi^{\rm trimmed}_{n} \supset \text{Poly}_{3 - \alpha}(\{ k \}) \log(- k_{T} \eta_{0}) \qquad 3-\alpha \geq 0\,,
\end{align}
where again the degree of the polynomial that multiplies the log is fixed by scale invariance. We cannot have any poles multiplying the log and so we need $3 - \alpha \geq 0$\footnote{This fact can be quite easily seen from the bulk representation and the corresponding time integrals one must perform. We don't have a better ``bootstrap" reason but it would be interesting to find one. We note that if the interactions violate manifest locality, there can be poles multiplying the log as they can come from inverse Laplacians.}. Such logs can arise from relevant operators in the bulk at tree-level but are also a common feature of loop corrections \cite{Gorbenko:2019rza,Cohen:2020php}. \\

\noindent These logs break the $\{ k \} \rightarrow \{ -k \}$ symmetry for both even and odd $\alpha$, so they can in principle contribute to the correlator. Unitarity in the form of the contact COT tells us that these logs do not appear on their own but rather always appear in the combination \cite{COT}
\begin{align}
\log(-k_{T} \eta_{0}) + \frac{i \pi}{2}\,,
\end{align}
multiplied by a \textit{real} function of $\{ k \}$, and possibly a polarisation factor (which also has real coefficients). Indeed, if we consider a wavefunction coefficient of the schematic form 
\begin{align} 
\psi^{\rm trimmed}_{n} \sim  \bfk^{\alpha} e^{\beta}(\bfk) [A \log(-k_{T} \eta_{0}) + B]\,,
\end{align}
where we have allowed for $\beta$ polarisation structures, a complex polynomial $A$ and a complex rational function $B$, then the COT \eqref{COT2} tells us that (recall that the polarisation factor becomes a common factor on the LHS of the COT)
\begin{align} \label{eq:ABequation}
A \log(-k_{T} \eta_{0}) + B - A^{\ast}[\log(-k_{T}\eta_{0}) + i \pi] - B^{\ast} = 0\,.
\end{align}
We therefore conclude that $\text{Im}(A) = 0$, $\text{Im}(B) = \frac{A \pi}{2}$ while $\text{Re}(B)$ is unconstrained and would actually contribute to the rational part of the wavefunction covered above in point 1. It then follows from \eqref{WFtoCorrelator} that for even $\alpha$ only the log contributes to the correlator and not the $ i \pi$ piece, whereas for odd $\alpha$ the $i \pi$ piece contributes to the correlator but the log does not. For parity-odd interactions of scalars and gravitons, which necessarily have an odd $\alpha$, we therefore conclude again that the singular part of the wavefunction does not contribute to the correlator. Indeed the parity-odd contributions to the graviton bispectrum computed in \cite{CabassBordin} come from this $\frac{i \pi}{2}$ part of the wavefunction.

\item The trimmed wavefunction may have a polynomial IR-divergence $  1/\eta_{0}^{q} $ with $  q\geqslant 1 $ as $\eta_{0} \to 0$. These terms may not have any singularity as $  k_{T}\to 0 $ because there we recover scattering amplitudes which, by time translation invariance, must be time independent. Scale invariance then tells us that 
\begin{align}
\psi^{\rm trimmed}_{n} \supset \sum_{q=1}^{3} \frac{\text{Poly}_{3 - \alpha - q}(\{ k \})}{\eta_{0}^q} \qquad 3-\alpha -q \geq 0\,.
\end{align}

\noindent Now we observe that we need $\alpha + q$ to be even in order to break the $\{ k \} \rightarrow \{ -k \}$ symmetry, while the MLT can only be satisfied if $3 - \alpha - q \geqslant 2$ or $3 - \alpha - q = 0$. These two conditions imply that $3 - \alpha - q \geqslant 3$, which contradicts the fact that $q \geqslant 1$. Thus, a combination of the COT and MLT leads us to conclude that $\eta_{0} = 0$ poles cannot contribute to cosmological correlators arising from manifestly-local bulk interactions\footnote{Although here our proof was outlined in $D=4$ spacetime dimensions, a generalised version of the MLT \cite{EnricoHarry} applies in all other dimensions and with this generalised MLT and the COT, one can show that $\eta_{0} = 0$ poles never appear in correlators. We thank Harry Goodhew for discussions on this point.}. 

\end{enumerate}
We have therefore seen that \textit{parity-odd contact correlators of scalars and gravitons do not contain any total-energy singularities}: the only part of the trimmed wavefunction that survives when we compute parity-odd correlators is finite or vanishing as $  k_{T} \to 0 $. These contributions arise from the polynomial function of $\{ k\}$ that multiplies $\log(-k_{T} \eta_{0}) + i \pi /2$ in the wavefunction and can only appear when the overall number of derivatives in bulk interactions is relatively small, which we will make precise in Section \ref{sec:bispectra}. This is consistent with the observation that the parity-odd Weyl-cubed vertex yields a vanishing bispectrum in dS space \cite{Maldacena:2011nz,Soda:2011am,Shiraishi:2011st}. In this case there are too many derivatives for a logarithm to appear in the wavefunction. Related observations about the consequences of unitarity cuts were recently made in \cite{Baumann:2021fxj}. We summarise these results in Table \ref{BootstrapResults} and remind the reader that the above discussion applies to contact diagrams, as relevant for this work. In Section \ref{sec:bispectra} we provide a full analysis of the form of the wavefunction for graviton cubic interactions and one can then use the results of this section to extract the contributions to the bispectra. \\

\noindent Before proceeding we would like to comment on what happens for tree-level contributions to the wavefunction that are not contact but include some exchange interaction (a bulk-bulk propagator in the  bulk representation). In that case, two things change: (i) the expression for the correlator in terms of wavefunction coefficients in \eqref{WFtoCorrelator} acquires additional contributions and (ii) the right-hand side of the Cosmological Optical Theorem (COT) does not vanish anymore \cite{COT}. Notice that both of these additional contributions are not singular as $  k_{T} \to 0 $. Hence, one can still conclude that any term in the wavefunction that is invariant under $\{ k \} \rightarrow \{ -k \}$ cannot contribute to the part of the correlator that is singular as $  k_{T}\to 0 $. Unfortunately, the wavefunction coefficients can become quite complicated for general exchange diagrams and we did not find a simple rule to establish when $  \psi^{\rm trimmed}_{n} $ is invariant under $\{ k \} \rightarrow \{ -k \}$.
\begin{table}[h!]
\begin{center}
\begin{tabular}{c c c c c}
  &  {$k_{T}$ poles}  & {$\log(-k_{T}\eta_{0}) + \frac{i \pi}{2}$}  & $\eta_{0}$ poles \\
\hline
even $\alpha$  & \tikzmark{a23}\checkmark & \tikzmark{a23}\checkmark ~ (only the log)  & \tikzmark{a22}\ding{55}   \\

odd $\alpha$  & \tikzmark{a22}\ding{55} & \tikzmark{a23}\checkmark ~ (only the $i \pi$) &  \tikzmark{a22}\ding{55}  \\
\end{tabular}
\caption{In this table we indicate which parts of the trimmed wavefunction, arising from contact diagrams, can contribute to cosmological correlators and which cannot. Here $  \alpha $ is the number of spatial derivatives contracted with polarizations tensors, as defined in \eqref{alpha}, and these results apply for three spatial dimensions, $d=3$.}
\label{BootstrapResults}
\end{center}
\end{table}

\section{Bootstrapping \textit{all} graviton bispectra}
\label{sec:bispectra}

In this section we bootstrap boost-breaking graviton bispectra at tree-level. We detail the general method that allows one to extract bispectra for any helicity configuration, and up to any desired order in derivatives. Throughout we employ the Boostless Bootstrap Rules and Manifestly Local Test, which were both reviewed in Section \ref{sec:BootstrapTechniques}. 

\subsection{Polarisation factors} \label{subsec:tensor_structures} 
\noindent It is the presence of spin-$2$ polarization tensors that distinguishes graviton bispectra from any other. As we reviewed in Section \ref{sec:BootstrapTechniques}, we write a general three-point wavefunction coefficient in terms of a polarisation factor multiplied by a ``trimmed'' wavefunction coefficient $\psi^{\rm trimmed}_3$ which is an $SO(3)$ scalar. We have \cite{BBBB}
\begin{equation}
\label{eq:trimmed_psi_3}
\psi^{h_1,h_2,h_3}_3(\bfk_1,\bfk_2,\bfk_3) = \sum_{\rm contractions} \Big[e^{h_1}(\bfk_1)e^{h_2}(\bfk_2)e^{h_3}(\bfk_3)\bfk_1^{\alpha_1}\bfk_2^{\alpha_2}\bfk_3^{\alpha_3}\Big] 
\psi^{\rm trimmed}_3(k_1,k_2,k_3)\,, 
\end{equation} 
where $h_{a} = \pm 2$ are the helicities of the external fields, and we remind the reader that we define the total number of \textit{spatial} momenta as $\alpha = \alpha_{1}+\alpha_{2}+\alpha_{3}$. Here index contractions between the momenta and polarization tensors are left implicit, and indeed our first goal is to construct all of the possible polarisation factors. As we explained in Section \ref{sec:BootstrapTechniques}, the trimmed wavefunction is constrained by the Manifestly Local Test (MLT) \cite{MLT} and the Cosmological Optical Theorem (COT) \cite{COT}, and so with the polarisation factors at hand, we will solve the MLT and obtain the complete three-point functions. \\

\noindent We first note that we can restrict our attention to $\alpha \leqslant 7$. This is because in order to construct an $SO(3)$-invariant object, we need to contract momenta with one of 
\begin{equation}
\label{eq:max_alpha}
\text{$e^{h_1}_{i_{1}i_{2}}e^{h_2}_{i_{3}i_{4}}e^{h_3}_{i_{5}i_{6}}$\quad or\quad $\epsilon_{i_{1}i_{2}i_{3}}e^{h_1}_{i_{4}i_{5}}e^{h_2}_{i_{6}i_{7}}e^{h_3}_{i_{8}i_{9}}$}\,,
\end{equation}
where the presence of a Levi-Civita tensor tells us that the resulting graviton bispectrum will violate parity. All remaining contractions are made with $\delta_{ij}$ and from now on we omit the dependence of polarization tensors on momenta for simplicity of notation. Now, it is straightforward to see that $\alpha$ can be at most $6$ in the parity-even case, with all six polarisation indices contracted with momenta, and $7$ in the parity-odd case since we can have at most two spatial momenta contracted with the Levi-Civita tensor due to momentum conservation. We will deal with the parity-even and parity-odd cases separately. \\

\noindent As is the case for scattering amplitudes, graviton bispectra are most compactly presented using the spinor helicity formalism rather than polarisation tensors. Indeed, this was the view advocated in \cite{Maldacena:2011nz} and is the route we will follow in this paper. A virtue of the spinor helicity formalism is that it can easily highlight possible degeneracies that could be hidden when using polarization tensors. Unfortunately, we do not have the means to construct the full structure of all allowed polarisation factors directly using spinors, so the approach we will take is to write down all possible polarisation factors in terms of polarisation tensors, with potential degeneracies still present, and to then convert these expressions into the spinor helicity formalism, where all degeneracies are manifest and can be easily eliminated.\\


\noindent We initially focus on the $+++$ helicity configuration, and in the following subsection we will show how to easily obtain the polarisation factors for all the other helicity configurations ($++-$, $--+$ and $---$) from this $+++$ building block. The helicity scaling of the external fields tells us that all $+++$ polarisation factors must contain
\begin{equation}
\label{eq:SH-1}
[12]^2[23]^2[31]^2\,,
\end{equation}
as an overall factor. This is the same factor that appears in three-point scattering amplitudes of massless gravitons \cite{Cheung:2017pzi,PSS} and is unique for this helicity configuration. The symmetries of the wavefunction then ensure that this can only be multiplied by $SO(3)$ invariant quantities that are simply functions of the three external energies. As explained in Section \ref{SHF}, whenever we convert a polarisation tensor into an expression with spinor brackets, we gain two powers of the corresponding energy in the denominator of the wavefunction. It is therefore not merely \eqref{eq:SH-1} that appears as an overall factor, but actually the \textit{dimensionless} quantity\begin{equation}
\label{eq:SH-2}
\text{SH}_{+++} = \frac{[12]^2[23]^2[31]^2}{e^2_3}\,,
\end{equation}
where $e_3=k_1k_2k_3$ is the third elementary symmetric polynomial. The above factor is ever-present. The information about the specific contraction is contained in an additional factor which is a function of the energies and which we denote as $h_{\alpha}(k_1, k_2, k_3)$. This is always a polynomial of degree $\alpha$. Finally, this product can be multiplied by the trimmed wavefunction, which in the bulk representation arises from bulk time integrals. This general form is true before we sum over all possible permutations, so the final form of the three-point function is 
\begin{equation}
\label{eq:psi+++}
\psi^{+++}_3(\bfk_1,\bfk_2,\bfk_3) = \frac{[12]^2[23]^2[31]^2}{e^2_3} \sum_{\rm permutations} h_{\alpha}(k_1, k_2, k_3) \psi^{\rm trimmed}_3(k_1,k_2,k_3)\,,
\end{equation}
where the sum over permutations ensures that the final expression is invariant under the exchange of any two external fields and their momenta, as dictated by Bose symmetry. In Appendix \ref{app:contractions} we construct all possible polarisation factors using polarisation tensors. 
With repeated use of \eqref{eq:spinor_review-5-1}, 
and recalling the definition of $I_a = k_T - 2k_a$, we find the following general structures for the $+++$ polarisation factors:
\begin{align}
h_{0} &= 1\,, \label{h1}\\
h_{1} &= i k_1 ~ \text{and perms}\,, \\
h_{2} &= k_1^2 ~ \text{and perms}, ~ k_1 k_2 ~ \text{and perms} \,,\\
h_{3} &= i k_1^3 ~ \text{and perms}, ~ i k_1^2 k_2 ~ \text{and perms}\,, ~ i k_1 k_2 k_3, \\
h_{4} &= I_1^2 I_2 I_3 ~ \text{and perms}\,,  \\
h_{5} &= i I_1^3 I_2 I_3 ~ \text{and perms}, ~  i I^2_1 I^2_2 I_3 ~ \text{and perms}\,,  \\
h_{6} &= I^2_1I^2_2I^2_3\,, \\
h_{7} &= i I^3_1I^2_2I^2_3 ~ \text{and perms}\,, \label{h7} 
\end{align}
where in some cases we have only presented one of the possible permutations, but we should keep in mind that one needs to sum over permutations in the final expression. For odd $\alpha$ we have included overall factors of $i$ which arise from the Levi-Civita tensor as shown in Appendix \ref{app:contractions}. Note that, if we only use spinor helicity variables, we do not have the means to derive the full form of the polarisation factors: for example, we did not find a good reason why a term like $I_{1}^7$ would be prohibited in the case of $\alpha = 7$. This was the main rationale for invoking polarization tensors in our argument, although it would be very interesting to derive the above list of structures, and to understand why some terms are not permitted, directly using spinors. \\ 


\noindent As we have explained in Sections \ref{sec:BootstrapTechniques} and \ref{ParitySection}, the general form of the trimmed wavefunction can be fixed by a set of Boostless Bootstrap Rules \cite{BBBB}. A combination of symmetries (including scale invariance), a weak-coupling approximation and Bunch-Davies initial conditions, ensures that the trimmed part of the wavefunction takes the form
\begin{align}
\label{eq:trimmedansatz}
\psi^{\rm trimmed}_3(k_1,k_2,k_3) &= \frac{\text{Poly}_{3+p-\alpha}(k_1, k_2, k_3)}{k_T^p} + \text{Poly}_{3-\alpha}(k_1, k_2, k_3) \log \left( -k_T \eta_0 \right) \nonumber \\ 
&\;\;\;\;+ \frac{\text{Poly}_{2 - \alpha}(k_1, k_2, k_3)}{\eta_{0}} + \frac{\text{Poly}_{1 - \alpha}(k_1, k_2, k_3)}{\eta_{0}^2} + \frac{\text{Poly}_{ - \alpha}(k_1, k_2, k_3)}{\eta_{0}^3}\,,
\end{align}
where we remind the reader that the degree of these complex polynomials is indicated by the subscripts. For those terms that diverge as $\eta_{0} \to 0$, we have strong restrictions on the allowed values of $\alpha$: a $1 / \eta_{0}^q$ singularity can only arise for $\alpha \leqslant 3-q$, which also justifies truncating the expansion at $q=3$. The above general form of the trimmed wavefunction is then further constrained by the MLT, which must be satisfied for all external energies. Note that we impose the MLT \textit{before} we sum over permutations in \eqref{eq:psi+++}, and so in that formula, each $\psi^{\rm trimmed}_3(k_1,k_2,k_3)$ is a solution to the MLT. The general recipe for constructing a $+++$ wavefunction coefficient is therefore the following:
\begin{enumerate}
\item Write down the spinor helicity factor $\text{SH}_{+++}$ and multiply it by one of the above choices for $h_{\alpha}(k_{1},k_{2},k_{3})$. 
\item Multiply this polarisation factor by a trimmed wavefunction coefficient of the form \eqref{eq:trimmedansatz} where the polynomials in this ansatz have been constrained by the MLT \eqref{MLT}. Note that for computational purposes it is useful to choose the permutation symmetry of this trimmed part to be the same as that of the polarisation factor. For example, if the polarisation factor is symmetric in the exchange of $k_{2}$ and $k_{3}$ then the trimmed part should be too, while if the polarisation factor has no symmetry then the trimmed part shouldn't either. 
\item Use the COT \eqref{COT} to deduce if unitarity demands real or imaginary coefficients.
\item Finally, sum over the remaining permutations such that the final wavefunction coefficient is fully symmetric, as dictated by Bose symmetry (Rule 4 of \cite{BBBB}).
\item To extract the corresponding three-point correlators, we use the results of Section \ref{ParitySection}. For even $\alpha$ we take the rational and log terms, with real coefficients, and divide by the appropriate powers of the power spectrum. For odd $\alpha$, we take the log part and simply replace the log with $i \pi /2$ such that we have some polynomial multiplied by a polarisation factor. Finally, we divide by the appropriate powers of the power spectrum. In both cases the result is real since for even $\alpha$ the polarisation factor is real, and is multiplied by a real function of the energies, while for odd $\alpha$ the polarisation factor is imaginary but it is multiplied by an imaginary function of the energies.
\end{enumerate}




\subsection{$+++$ to rule them all}\label{ToRule}
Before we constrain these $+++$ wavefunctions further, let us first show how we can obtain the $++-$, $--+$ and $---$ helicity configurations if $h_{\alpha}(k_{1},k_{2},k_{3})$ and $\psi^{\rm trimmed}_3(k_1,k_2,k_3)$ are known. It might be tempting to go back to the beginning, i.e. to the polarization tensors, and derive the spinor helicity form of tensor structures independently for each configuration. However, this is not necessary as the spinor variables can do most of the work for us. Let us first construct the $++-$ tensor structures in spinor helicity variables directly from the $+++$ ones. Flipping the helicity of the third graviton is equivalent to sending its energy from $k_3$ to $-k_3$ while keeping its momentum fixed. Under this transformation, the spinors transform according to \cite{PSS} \begin{equation}
\label{eq:trick-2-bis}
\tilde{\lambda}\mapsto i (\lambda_2,{-\lambda_1})\,\,,\quad\lambda\mapsto i ({-\tilde{\lambda}_2},\tilde{\lambda}_1)\,\,. 
\end{equation}
Using the definitions of the various brackets given in Section \ref{SHF}, we then have
\begin{equation}
\label{eq:trick-3}
[13] \mapsto -i (31)\, ,
\end{equation} 
\begin{equation}
\label{eq:trick-3}
[23] \mapsto -i (32)\,\,,
\end{equation} 
from which it follows that
\begin{equation}
\label{eq:trick-4}
\text{SH}_{+++}\mapsto \frac{[12]^2}{e_3^2}\, (31)^2 (32)^2 = \frac{[12]^6}{[23]^2 [31]^2 } \frac{I_1^2 I_2^2}{e_3^2} \equiv \text{SH}_{++-}\,\,. 
\end{equation}
So \textit{all} $++-$ wavefunction coefficients are multiplied by this common factor of $\text{SH}_{++-}$. Note the square brackets are completely fixed by the helicities of the external fields and are the same as for amplitudes \cite{PSS,Cheung:2017pzi}, while the ever-present $I_{1}^2 I_{2}^2$ factor in the numerator is required for the absence of divergences. Indeed, consider the following argument: with the help of (\ref{eq:spinor_review-2}), the spinor helicity factor $\frac{[12]^6}{[23]^{2} [31]^{2}}$ can be rewritten as
\begin{align}
\frac{[12]^6 \langle 23 \rangle^2 \langle 31 \rangle^2}{k_T^4 I_1^2 I_2^2}.
\end{align}
If the momenta are allowed to be complex, then $I_1$ can be taken to zero while keeping $k_T, I_2$ and the numerator finite. Such a divergence is forbidden and therefore we should include two factors of $I_{1}$ in the numerator to cancel it out. The absence of a divergence as $I_{2}$ is taken to zero similarly demands that we should include two factors of $I_{2}$. This argument can be easily generalised to other helicities to show that in general the bispectrum of any three fields with helicities $h_{a}$ for $  a=1,2,3 $ has to contain the following factor (for $H \equiv h_{1}+h_{2}+h_{3} \geq 0  $)
\begin{align}
\text{SH}_{h_{1},h_{2},h_{3}} = \frac{[12]^{d_3}[23]^{d_1}[31]^{d_2}}{\prod_{a=1}^{3} k_{a}^{|h_{a}|}} \prod^3_{b=1} I_{b}^{\text{max}[0,-d_b]} \,,
\end{align}
where
\begin{align}
d_a \equiv h_b + h_c - h_a = H - 2 h_a \quad (a\neq b \neq c) \,.
\end{align}
The scaling dimension of the spinor helicity factor $\text{SH}_{h_{1},h_{2},h_{3}}$ is $ \text{max}\{ 0, -d_1, -d_2, -d_3 \}$. The wavefunction coefficient then takes the form
\begin{align}
\psi_{3}^{h_{1},h_{2},h_{3}}(\bfk_1,\bfk_2,\bfk_3) = \text{SH}_{h_{1},h_{2},h_{3}} \times P_{m}(k_1, k_2, k_3)
\end{align}
where $P_m$ is a rational function of the energies (possibly also including $\log(-k_T \eta_0)$ multiplied by a polynomial) and $m$ is its scaling dimension. \\




\noindent To extract the $++-$ wavefunction, then, we take $\text{SH}_{++-}$ and multiply it by $h_{\alpha}(k_1,k_2,-k_3)$ and by $\psi^{\rm trimmed}_3(k_1,k_2,k_3)$. Note that only in $h_{\alpha}$ is the sign of $k_{3}$ flipped. Indeed, the structure of $h_{\alpha}$ is fixed by the form of the polarisation factor which certainly depends on the helicity configuration, whereas $\psi^{\rm trimmed}_3(k_1,k_2,k_3)$ is a product of time integrals in the bulk formalism and is therefore independent of the helicity configuration of the external fields. Therefore, the $++-$ wavefunction coefficients are given by
\begin{equation}
\label{eq:psi++-}
\psi^{++-}_3(\bfk_1,\bfk_2,\bfk_3) = \frac{[12]^6}{[23]^2 [31]^2 } \frac{I_1^2 I_2^2}{e_3^2} \sum_{\rm permutations} h_{\alpha}(k_1, k_2, -k_3) \psi^{\rm trimmed}_3(k_1,k_2,k_3)\,.
\end{equation}
The recipe we outlined above for the $+++$ configuration is then easily applied to this $++-$ case, with the symmetries of $\psi^{\rm trimmed}_3(k_1,k_2,k_3)$ fixed by $h_{\alpha}(k_1,k_2,-k_3)$ and with the final sum over permutations ensuring that the final wavefunction is symmetric under the exchange of $k_{1}$ and $k_{2}$, as dictated by Bose symmetry. \\

\noindent Finally, the $--+$ and $---$ wavefunction coefficients are then obtained directly from the $++-$ and $+++$ ones respectively, by sending $k_{a} \mapsto -k_{a}$ for $a=1,2,3$. This corresponds to all square brackets changing into (minus) angle brackets, such that 
\begin{subequations}
\label{eq:trick-6}
\begin{align}
\text{SH}_{+++}&\mapsto\frac{\langle 12 \rangle^2 \langle 23 \rangle^2 \langle 31 \rangle^2}{e^2_3} \equiv \text{SH}_{---} \,, \label{eq:trick-6-1} \\
\text{SH}_{++-}&\mapsto\frac{\langle 12 \rangle^6}{\langle 23 \rangle^2 \langle 31 \rangle^2} \frac{I^2_1I^2_2}{e_3^2} \equiv \text{SH}_{--+}  \,. \label{eq:trick-6-2}
\end{align}
\end{subequations}
Under $k_{a} \mapsto -k_{a}$, we have $h_{\alpha}(k_1,k_2,k_3) \mapsto (-1)^{\alpha} h(k_1,k_2,k_3)$, while $\psi^{\rm trimmed}_3(k_1,k_2,k_3)$ is again taken to be unchanged.  \\

\noindent  In conclusion, with knowledge of the building blocks of the $+++$ wavefunction coefficients, one can easily compute wavefunction coefficients for other helicity configurations. We note that our ability to do this is due to fact that time translations are no longer a symmetry in cosmology and therefore square, angle and round brackets are related as shown in Section \ref{SHF}. For scattering amplitudes, where time translations are a symmetry, one cannot simply map between different configurations in this way. As a very non-trivial check of this procedure, we verified that the $+++$ wavefunction coefficient arising from a parity-even $\text{Weyl}^3$ vertex in the bulk gives rise to a vanishing $++-$ coefficient, as it should \cite{Maldacena:2011nz}. 

\subsection{A further simplification of the polarisation factors}
\noindent Now given that $h_{\alpha}(k_{1},k_{2},k_{3})$ must be multiplied by a solution to the MLT, we can actually further simplify the structures given in \eqref{h1} to \eqref{h7}. The general $h_{\alpha}$ in \eqref{eq:psi+++} is given by an arbitrary linear combination of polynomials listed in \eqref{h1}-\eqref{h7}, as well as all their permutations, for each $\alpha$. However, now we will show that we may consider only a few special $h_{\alpha}$ and still obtain fully general wavefunction coefficients. We give an explicit argument for $\alpha=2$, but a closely analogous argument works for any $\alpha$. \\

\noindent We have already established that $h_2(k_1, k_2, k_3) = \sum_{a} n_a k_a^2 + \sum_{a} m_a k_a k_{a+1}$, where $n_a, m_b$ are arbitrary numerical coefficients. We then have (recall that $\tilde{\psi}_3$ is a shorthand notation for $\psi^{\rm trimmed}_3$):
\be
\begin{split} 
\frac{\psi^{+++}_3(\bfk_1,\bfk_2,\bfk_3)}{\text{SH}_{+++} } & = \sum\limits_{\sigma \in S_3} \sum\limits_a \left( n_a k_{\sigma(a)}^2 + m_a k_{\sigma(a)} k_{\sigma(a+1)} \right) (\tilde{\psi}_3 \circ \sigma) (k_{1,2,3})  \\
& = \sum\limits_a \sum\limits_{\sigma \in S_3} \left( n_{\sigma^{-1}(a)} k_a^2 +m_{\sigma^{-1}(a)} k_a k_{a+1} \right) 
(\tilde{\psi}_3  \circ \sigma) (k_{1,2,3})  \\ 
& = \sum\limits_a \left(   k_a^2 \sum\limits_{\sigma \in S_3} n_{\sigma^{-1}(a)} (\tilde{\psi}_3 \circ \sigma) (k_{1,2,3})  +  k_a k_{a+1}  \sum\limits_{\sigma \in S_3} m_{\sigma^{-1}(a)} (\tilde{\psi}_3 \circ \sigma) (k_{1,2,3}) \right)  \\
& = \sum\limits_{\text{cyclic}} k_1^2  f_{(23)}(k_1, k_2, k_3) + \sum\limits_{\text{cyclic}} k_1 k_2  g_{(12)}(k_1, k_2, k_3)\,,
\end{split}
\ee
where $f_{(23)}$ and $g_{(12)}$ are linear combinations of trimmed wavefunction coefficients, and therefore they must take the form given in \eqref{eq:trimmedansatz} and satisfy the MLT. Moreover, we employ the notation that a function of the three external energies is symmetric under the exchange of energies indicated in a subscript e.g. $f_{(23)}$ is symmetric under the exchange of $k_{2}$ and $k_{3}$, while $f_{(123)}$ would be fully symmetric. An analogous argument can be used to show that $\psi^{++-}_3(\bfk_1,\bfk_2,\bfk_3)$ can be simplified in the same way. More precisely, we have 
\begin{align} \nn
\frac{\psi^{++-}_3(\bfk_1,\bfk_2,\bfk_3)}{\text{SH}_{++-}} & =  \ \sum\limits_{\text{cyclic}} k_1^2  f_{(23)}(k_1, k_2, k_3) +  k_1 k_2 g_{(12)}(k_1, k_2, k_3)  \\
&  -  k_2 k_3 g_{(23)}(k_2, k_3, k_1) - k_3 k_1 g_{(31)}(k_3, k_1, k_2). 
\end{align}
Thus, we see that we can take $h_2(k_a)$ to be a linear combination of $k_1^2$ and $k_1 k_2$ and still get a fully general $\alpha=2$ solution. Moreover, we note that all solutions constructed from $h_2(k_a) = k_1^2$ can also be constructed using the $\alpha=0$ polarization factor $h_0(k_a) = 1$. This is because, if $ f_{(23)}(k_1, k_2, k_3)$ satisfies the MLT, then $k_1^2 f_{(23)}(k_1, k_2, k_3)$ must satisfy it too, so wavefunction coefficients of the form
\begin{align} 
\frac{\psi^{+++}_3(\bfk_1,\bfk_2,\bfk_3)}{\text{SH}_{+++} }   & = \sum\limits_{\text{cyclic}} k_1^2  f_{(23)}(k_1, k_2, k_3)\,, \\
\frac{\psi^{++-}_3(\bfk_1,\bfk_2,\bfk_3)}{\text{SH}_{++-}} & = \sum\limits_{\text{cyclic}} k_1^2  f_{(23)}(k_1, k_2, k_3)\,,
\end{align}
are already accounted for and contained in the MLT solutions for polarisation factors with $\alpha=0$. Assuming we construct solutions iteratively with increasing $\alpha$, so that $\alpha = 0$ wavefunction coefficients have already been constructed, for $\alpha = 2$ we only need to consider $h_{2}(k_1,k_2,k_3) = k_1 k_2$ to derive a complete set of such coefficients. \\

\noindent One can proceed in a similar manner at each order in $\alpha$ by studying the different allowed $h_{\alpha}$ and asking if the resulting wavefunction coefficients have already been captured by lower order solutions in $\alpha$. We find that \textit{to construct fully general wavefunction coefficients, it is sufficient to consider the following polarisation factors for the $+++$ helicity configuration:} 
\begin{align}
h_{0} &= 1\,, \label{h1a}\\
h_{1} &=i k_1 \,, \\
h_{2} &= k_2 k_3\,,\\
h_{3} &=i  I_1 I_2 I_3\,, \\
h_{4} &= I_1^2 I_2 I_3 \,,  \\
h_{5a,b} &=i  I_1^3 I_2 I_3, i  I_1 I^2_2 I^2_3\,,  \\
h_{6} &= I^2_1I^2_2I^2_3\,, \\
h_{7} &= i I^3_1I^2_2I^2_3\,. \label{h7b} 
\end{align}
Therefore, at each order in $\alpha$ we have to consider a single polarisation factor, apart from $\alpha=5$ where there are two possible structures. Note that in all cases we can write the polarisation factor in such a way that it is symmetric in the kinematical data of two out of the three external fields, which we take to be fields $2$ and $3$. We can now follow the recipe outlined above, and constrain the remaining part of the wavefunction coefficients with the MLT.


\subsection{Constraining the trimmed wavefunction}
\label{constraining_trimmed}

We now turn to the final piece of the puzzle, which requires us to solve the MLT \eqref{MLT} to constrain \eqref{eq:trimmedansatz} and therefore construct the trimmed part of the wavefunction coefficients. By writing out the allowed form of the polynomials in this ansatz we have 
\begin{align} \label{MLTansatz}
\psi^{\rm trimmed}(k_1, k_2, k_3) = \frac{1}{k_T^p} \sum\limits_{l+m+n = 3+p-\alpha} c_{lmn} k_1^l k_2^m k_3^n + \log \left(- k_T \eta_0 \right)  \sum\limits_{l+m+n = 3-\alpha} d_{lmn} k_1^l k_2^m k_3^n \nonumber \\
+\frac{1}{\eta_{0}} \sum\limits_{l+m+n = 2-\alpha} e_{lmn} k_1^l k_2^m k_3^n+\frac{1}{\eta_{0}^2} \sum\limits_{l+m+n = 1-\alpha} f_{lmn} k_1^l k_2^m k_3^n+\frac{1}{\eta_{0}^3} \sum\limits_{l+m+n = -\alpha} g_{lmn} k_1^l k_2^m k_3^n\,,
\end{align}
where $l,m,n \geq 0$ and we remind the reader that the sums are fixed by scale invariance. The following conditions are then necessary for the above ansatz to pass the MLT:
\begin{align} \label{implicitsolutions1}
d_{1,n,r-1} & = 0\,, \\
\sum_{m} {{p}\choose{n - m}} d_{0, m, 3 - \alpha - m} & = p \ c_{0,n,p+r} - c_{1, n-1, p+r} - c_{1, n, p+r-1}\,, \\ \label{implicitsolutions3}
e_{1, n, r-2}  =  f_{1, n, r-3} = g_{1, n, r-4} & = 0\,,
\end{align}
with $r \equiv 3-\alpha-n$; along with analogous conditions for all other permutations of indices. Note that the conditions that arise from the terms in the first line of \eqref{MLTansatz} decouple from those in the second line. \\

\noindent Whenever a polynomial $h_{\alpha}$ has a symmetry under interchange of external labels, the trimmed wavefunction coefficient may also be assumed to have such a symmetry without loss of generality. This is because any non-symmetric part will be cancelled out after summing over all permutations indicated in \eqref{eq:psi+++}, as we saw explicitly in the previous section in the $\alpha=2$ case. Therefore, if $h_{\alpha}(k_1,k_2, k_3) = h_{\alpha}(k_1,k_3, k_2)$, then we have
\begin{eqnarray}
c_{lmn} & = & c_{lnm}\,, \\
d_{lmn} & = & d_{lnm}\,, \\
&\vdots& \nonumber
\end{eqnarray}
Moreover, if $h_{\alpha}(k_1, k_2, k_3)$ is completely symmetric, then we have
\begin{eqnarray}
c_{lmn} & = & c_{lnm} = c_{mln}\,, \\
d_{lmn} & = & d_{lnm}  = d_{mln}\,, \\
&\vdots& \nonumber
\end{eqnarray}
As we saw above, in all cases $h_{\alpha}$ is symmetric in at least two external labels. We will now present the first few solutions for each $\alpha$, considering even and odd $\alpha$ separately.


\paragraph{Parity-even interactions} We begin with parity-even interactions which have even $\alpha$. \\

\noindent $\bm{\alpha=0}$ In this case we have $h_{0} = 1$ and so the solution to the MLT must be fully symmetric. This case is actually exactly the same as the situation for three identical scalars which was covered in \cite{MLT}. The following solutions are therefore the same as those found in that work. Given the symmetry, we present the results using the three elementary symmetric polynomials $k_{T}, e_{2}, e_{3}$. Up to $p = 3$ we have
\begin{align}
\eta_{0}^{-1}:& \quad  \frac{i(k_{T}^2 - e_{2})}{\eta_{0}}\,, \\
\eta_{0}^{-3}:& \quad \frac{i}{\eta_{0}^{3}}\,, \\
 p=0 :& \quad 4e_3 - e_2 k_T + (k_{T}^3 - 3 k_{T}e_{2} + 3 e_{3})\log (-k_T \eta_0), \quad k_{T}^3 - 3 k_{T}e_{2} + 3 e_{3}\,,  \\
 p=2 :& \quad \frac{e_2 e_3 + e_2^2 k_T - 2 e_3 k_T^2}{k_T^2}\,, \\
p=3 :& \quad \frac{e_3^2}{k_T^3}\,, \label{K3ij} \\
& \quad \vdots \nonumber
\end{align}
where, as indicated, there are two possible solutions for $  p=0 $. \\

\noindent Unitarity places the following additional constraints. The coefficients of $1 / \eta_{0}$ and $1 / \eta_{0}^3$ must be imaginary as consequence of the Cosmological Optical Theorem (COT), see Section \ref{ParitySection}. This has a nice interpretation in terms of the holographic language of (A)dS/CFT, along the lines of \cite{Maldacena:2002vr}. These two terms are bulk IR divergences and should be holographically renormalized as described in \cite{Skenderis:2002wp}. For the associated renormalization group flow to be unitary, these divergences should be imaginary, which is precisely what the COT ensures. Conversely, the COT says that the coefficient of the $1 / \eta_{0}^2$ divergence must be real. This would correspond to a counterterm with imaginary coupling constant. It is quite intriguing that the MLT forbids precisely these terms and we will discuss this elsewhere. For $p=0$ the MLT admits two solutions. The second one does not have a log and can satisfy the COT by itself with an arbitrary real coefficient. Conversely, the first solution, which contains a log, satisfies the COT only when it is combined with the second solution with a relative factor of $  i \pi/2 $ (see Section \ref{ParitySection} or \cite{COT}), namely in the combination
\begin{align}
\lambda\left[ 4e_3 - e_2 k_T + (k_{T}^3 - 3 k_{T}e_{2} + 3 e_{3})\log (-k_T \eta_0)+i \frac{\pi}{2} \left( k_{T}^3 - 3 k_{T}e_{2} + 3 e_{3} \right) \right]\,,
\end{align} 
for real $  \lambda $. \\

\noindent There are no $p=1$ solutions. This can simply be understood as follows. Recall that for cubic wavefunction coefficients, the degree $  p $ of the leading $k_{T}$ pole equals the number of derivatives. Then the absence of solutions for $  p=1 $ is related to the fact that there are no single derivative interactions one can write down (for $\alpha=0$), other than a total time derivative. Wavefunction coefficients with $  p=1 $ do arise in not-manifestly-local theories. Indeed the scalar bispectrum induced by gravity has $  p=1 $, which is consistent with the above discussion because, after integrating out the non-dynamical parts of the metric, GR displays not-manifestly-local interactions. We refer the reader to \cite{MLT} for more details on this case. Note that the COT fixes the coefficients of the terms rational in $\{ k\}$ to be real. \\

\noindent $\bm{\alpha=2}$ In this case, we have $h_{2}(k_1,k_2,k_3) = k_2 k_3$ without loss of generality, so we take the ansatz to also be symmetric in $k_{2}$ and $k_{3}$. The leading solutions are then
\begin{align}
\eta_{0}^{-1}:& \quad  \frac{i}{\eta_{0}}\,, \\
 p=2 :& \quad  \frac{e_3 + e_2 k_T - k_T^3}{k_T^2} \,, \\
 p=3 :& \quad \frac{k_1^2 \left( k_1^2 + 3 k_1 k_{23} + 2 (k_{23}^2 + k_2 k_3) \right)}{k_T^3}\,, \\
p=4 :& \quad \frac{k_2^2 k_3^2 \left( k_T + 3 k_1 \right)}{k_T^4}\, , \\
& \quad \vdots \nonumber
\end{align}
We see that only a simple $\eta_0=0$ pole is allowed with a constant and imaginary residue, by unitarity. In terms of total-energy poles, the leading solution has a degree two pole which is related to the fact that such wavefunction coefficients arise from bulk vertices with at least two derivatives. Again, the COT demands that the coefficients of these rational terms are real. \\

\noindent $\bm{\alpha=4.}$ Here we again have a single choice for the polarisation factor which is $h_{4}(k_1,k_2,k_3) = I_1^2 I_2 I_3$. This must be combined with an $\alpha=4$ solution to the MLT, which is symmetric in $k_{2}$ and $k_{3}$. Clearly no $\eta_{0} = 0$ poles are allowed, and the leading solutions are  
\begin{align}
 p=4 :& \quad  \frac{3 e_3 +  k_T e_2 + k_T^3}{k_T^4} \, , \\
 p=5 :& \quad \frac{k_1^2 \left( k_1^2 + 5 k_1 k_{23} + 4 (k_{23}^2 + 3 k_2 k_3) \right)}{k_T^5}\,, \\
& \quad \vdots \nonumber
\end{align}
Again we see that the lowest possible total energy pole has degree $4$.\\
%

\noindent $\bm{\alpha=6.}$ Finally, we have $h_{6} =  I_1^2 I_2^2 I_3^2$. This is fully symmetric, so we can present solutions to the MLT using the elementary symmetric polynomials. There are no $\eta_{0} = 0$ poles and the leading solutions are
\begin{align}
 p=6 :& \quad  \frac{15 e_3 + 3 k_T e_2 + k_T^3}{k_T^6}\, , \\
  p=8 :&  \quad \frac{7 e_2 e_3 + k_T e_2^2 - 2 k_T^2 e_3 }{k_T^8} \, , \\
& \quad \vdots \nonumber
\end{align}

\noindent In each case, solutions with higher-order $k_{T}$ poles can be easily computed. We see that an IR-divergent logarithm is only permitted for $\alpha = 0$, while IR-divergences in the form of $\eta_{0} = 0$ poles can only arise for $\alpha=0,2$ and they always come with imaginary coefficients. In Section \ref{FinalForm} we will use these solutions to write down the final form of the leading $+++$ and $++-$ bispectra.

\paragraph{Parity odd interactions} We now turn to odd $\alpha$ which correspond to parity-odd interactions. \\

\noindent $\bm{\alpha=1}$ In this case we have $h_{1}(k_1,k_2,k_3) = k_1$. This must be combined with an $\alpha=1$ solution to the MLT, symmetric in $k_{2}$ and $k_{3}$. The leading solutions are 
\begin{align}
\eta_{0}^{-2}:& \quad  \frac{1}{\eta_{0}^2}\,, \\
 p=0 :& \quad k_1^2, ~  k_T^2 - 2 e_2,  \\
p=1 :&  \quad  \frac{2 e_3 - e_2 k_T}{k_T} + (k_{T}^2-2e_{2}) \log(-k_T \eta_0)\,, \\
 p=2 :& \quad \frac{k_1^2 \left( k_T(k_2 + k_3) + k_2 k_3 \right)}{k_T^2}  - k_1^2 \log(-k_T \eta_0)\,, \\
p=3 :& \quad  \frac{-2 e_3 k_T^2 + 2 e_3 e_2 + k_T e_2^2}{k_T^3}\, , \\
& \quad \vdots \nonumber
\end{align}
where, as indicated, there are two possible solutions for $  p=0 $. We see that the only allowed $\eta_{0}=0$ pole is of degree two, as it should be for $  \alpha=1 $ because of scale invariance. Interestingly, we also see that IR-divergent logarithms are also permitted but \textit{only} in combination with total-energy poles. This is in contrast to even $\alpha$ where logarithms could contribute as the only singular term. The solutions with higher total-energy poles that are not shown here do not have logarithms. \\

\noindent Unitarity places the following additional constraints. All terms without logs can appear with real coefficients. The two solutions containing a log, namely $  p=1 $ and $  p=2 $, solve the Cosmological Optical Theorem (COT) only when accompanied by a corresponding $  p=0 $ solution with a relative coefficient of $  i \pi/2 $, namely in the combinations
\begin{align}
\frac{2 e_3 - e_2 k_T}{k_T} + (k_{T}^2-2e_{2}) \left[ \log(-k_T \eta_0)+i \frac{\pi}{2} \right] &\\
\frac{k_1^2 \left( k_T(k_2 + k_3) + k_2 k_3 \right)}{k_T^2}  - k_1^2 \left[ \log(-k_T \eta_0)+i \frac{\pi}{2} \right]& \,,
\end{align}
with real overall coefficients. Notice that, since we are considering parity-odd interactions, it is only the imaginary part of these trimmed wavefunction coefficients, namely that proportional to $i \pi /2 $, that contributes to the bispectrum. \\


\noindent $\bm{\alpha=3}$ Here we can choose $h_{3}(k_1,k_2,k_3) = I_1 I_2 I_3$ and the solution to the MLT may be assumed to be fully symmetric. No $\eta_{0} = 0$ poles are allowed, and the leading solutions are 
\begin{align}
 p=0 :& \quad 1\,,  \\
p=3 :&  \quad   \frac{2 e_3 + e_2 k_T}{k_T^3} - \log(-k_T \eta_0)\,, \\
 p=5 :& \quad  \frac{4 e_2 e_3 + e^2_2 k_T - 2 e_3 k_T^2}{k_T^5} \,, \\
& \quad \vdots \nonumber
\end{align}
Again the higher order solutions do not contain logarithms, so only a single solution with such a IR-divergence is allowed in this case. As above, unitarity in the form of the Cosmological Optical Theorem (COT) requires that the $ p=3  $ term, which contains a log, must appear together with the (trivial) $  p=0 $ solution in the combination
\begin{align}
\frac{2 e_3 + e_2 k_T}{k_T^3} - \left[ \log(-k_T \eta_0)+i \frac{\pi}{2} \right]\,,
\end{align}
with a real overall coefficient. \\

\noindent $\bm{\alpha=5.}$ In this penultimate case there are two choices for $h_{5}$: $h_{5}(k_1,k_2,k_3) = I_1^3 I_2 I_3$ and $h_{5}(k_1, k_2, k_3) =  I_1 I^2_2 I^2_3$. Both must be multiplied by a solution to the MLT that is symmetric in $k_2$ and $k_3$. No $\eta_{0} = 0$ poles or logarithmic terms are allowed, and the leading solutions are 
\begin{align}
 p=5 :& \quad  \frac{8 e_3 + 2 k_T e_2 + k_T^3}{k_T^5}  \,,  \\
p=6 :&  \quad    \frac{k_1^2 \left( k_1^2 + 6 k_1 k_{23} + 5 (k_{23}^2 + 4 k_2 k_3) \right)}{k_T^6}\,, \\
& \quad \vdots \nonumber
\end{align}
both with real coefficients by unitarity. \\

\noindent $\bm{\alpha=7.}$ In this final case, we have $h_{7}(k_1,k_2,k_3) = I^3_1I^2_2I^2_3$ and so the solution to the MLT needs to be symmetric $k_2$ and $k_3$. The leading solutions are 
\begin{align}
 p=7 :& \quad \frac{24 e_3 + 4 k_T e_2 + k_T^3}{k_T^7}\,,  \\
 p=8 :& \quad \frac{k_1^2 \left( k_1^2 + 8 k_1 (k_2 + k_3) + 7 ((k_2+k_3)^2 + 6 k_2 k_3 ) \right)}{k_T^8} \,, \\
& \quad \vdots \nonumber
\end{align}

\noindent Again, higher order solutions are easily found. As we have emphasised a number of times, only the coefficients of the logarithms contribute to the final bispectra for these parity-odd interactions. We have found only three solutions with logarithms which are also required to come alongside total-energy poles which will ultimately drop out from the correlator. It is important to stress that the fact we only have three logarithmic terms is true to all orders in derivatives. Indeed, all remaining solutions not explicitly shown above are purely rational. We can therefore extract the full form of parity-odd graviton bispectra, to all orders in derivatives, from these MLT solutions. Given that there is only a single polarisation structure for $\alpha=1,3$, there are only three independent parity-odd graviton bispectra. We will discuss this further in Section \ref{FinalForm} where we construct the final form of the correlators. 

\paragraph{Contact reconstruction formula} In this section we have derived wavefunction coefficients for graviton interactions without any reference to flat space. However, there also exists a well-defined relationship between wavefunction coefficients in de Sitter and scattering amplitudes in flat space: the residue of the leading total-energy pole of a wavefunction coefficient contains the flat space amplitude (see also \cite{Benincasa:2018ssx,Baumann:2021fxj,Hillman:2021bnk} for additional relations between correlators and amplitudes). This was first noticed in \cite{Maldacena:2011nz,Raju:2012zr} and then an explicit formula was derived in \cite{COT}. For $n$ external fields the relationship is 
\begin{align}
\psi_{n} = (p-1)! (i H)^{p-n-1} \frac{e_{n} A_{n}^{(p-n+3)}}{k_{T}^p} + \ldots\,,
\end{align}
where $\smash{e_{n} = \prod_{a=1}^{n}k_{a}}$ is a product of the $n$ energies and here we have re-inserted the factors of Hubble. The ellipsis denote terms with subleading total-energy poles and $A_{n}^{(p-n+3)}$ is the part of the corresponding scattering amplitude that contains the largest scaling in energy and momentum, which is of order $p-n+3$. For $n=3$ which is the primary focus of this work, this leading total-energy pole picks out that part of the amplitude that comes from operators with $p$ derivatives. \\

\noindent One may go a step further and hope that with the knowledge of the scattering amplitude, as well as the form of the de Sitter mode functions, the full de Sitter wavefunction coefficient could be produced since it is the same bulk interaction vertex that gives rise to the amplitude and the wavefunction. As we have seen above, some knowledge of the de Sitter mode functions is contained in the MLT and indeed in a recent paper \cite{Bonifacio:2021azc} solutions to the MLT were used to convert a contact flat space amplitude into a contact de Sitter wavefunction via a \textit{contact reconstruction formula}:
\begin{align} \label{eq:ampToWFnpts}
\psi_n  = (p-1)! (iH)^{p-n-1}\sum_{m=0}^{n}  \sum_{\pi \in S_n} \frac{ A_n^{(p-n+3)}\big|_{ \{ k_{\pi(j)} =0\}_{j=n-m+1}^n } \prod_{i=1}^{n-m} k_{\pi(i)} }{m! (n-m)! k_T^{p-m} \prod_{l=1}^m (p-l) }\,,
\end{align}
where the sum $\sum_{\pi \in S_n}$ runs over the $n!$ permutations $\pi$ of $\{1,2,\ldots,n \}$. For $n=3$ this reconstruction formula takes the following form
\begin{align} \label{eq:reconstruct-n=3}
\psi_3  & = (p-1)! i^p H^{p-4} \bigg[ \frac{ A_3^{(p)} k_1 k_2 k_3 }{ k_T^{p} } + \frac{   A_3^{(p)}\big|_{k_1 =0} k_2 k_3+ A_3^{(p)}\big|_{k_2 =0} k_1 k_3  + A_3^{(p)}\big|_{k_3 =0} k_1 k_2  }{ k_T^{p-1} (p-1) } \nonumber \\
& + \frac{  A_3^{(p)}\big|_{ k_2=k_3=0} k_1 + A_3^{(p)}\big|_{ k_1=k_3=0} k_2+ A_3^{(p)}\big|_{ k_1=k_2=0} k_3}{k_T^{p-2}  (p-1)(p-2) } + \frac{A_3^{(p)}\big|_{ k_1=k_2=k_3=0 } }{k_T^{p-3}  (p-1)(p-2)(p-3) }\bigg]\,.
\end{align}
This formula is valid for $p \geq 4$ where the time integrals in the bulk computation of these wavefunction coefficients do not produce logarithms or purely analytic terms. In this case \eqref{eq:reconstruct-n=3} yields the full wavefunction. For $p \leq 3$ the time integrals can yield such logarithms or analytic terms which are not captured, but in those cases the total-energy poles can still be computed using this formula; then one would need to write down an ansatz for the MLT solution and fix the additional terms that are ultimately required to satisfy the MLT. For more details we refer the reader to \cite{Bonifacio:2021azc}. \\

\noindent Instead of taking the route outlined in this paper, one could in principle use \eqref{eq:reconstruct-n=3} to construct graviton bispectra. The $p$-derivative amplitude that we must input is simply given by taking one of the polarisation factors we classified in Appendix \ref{app:contractions}, multiplying this $SO(3)$ invariant object by a polynomial in the energies of degree $(p-\alpha) > 0$, followed by summing over permutations \cite{PSS}. The final sum over permutations is crucial since as can be seen from \eqref{eq:reconstruct-n=3}, the wavefunction coefficient will only have the correct Bose symmetry if the amplitude does. For $p \geq 4$ this procedure will generate all possible bispectra. Note that here we are advocating to use this contact reconstruction formula using polarisation tensors rather than the spinor helicity formalism since in $A_{3}^{(p)}$ the energy dependence needs to be from bulk time derivatives \textit{only}. When the amplitude is written in terms of spinors, there is an energy dependence that has arisen from the polarisation factor itself rather than from bulk time derivatives, as we explained above. With the final result computed from \eqref{eq:reconstruct-n=3}, one can convert this expression into the spinor helicity formalism using the expressions given in Appendix \ref{app:contractions}. Above we have presented the leading order MLT solutions for each $\alpha$, one can in principle use this reconstruction formula to generate \textit{all} higher-order solutions.


\subsection{The final form of graviton bispectra} \label{FinalForm}
With all of the ingredients at hand, we can now write down the final form of the wavefunction coefficients and extract the corresponding correlators. We will concentrate on the $+++$ and $++-$ helicity configurations (since the other two are easily obtained from those by a parity transformation, with an extra $-$ sign for odd $\alpha$) and again work at each order in $\alpha$ treating the even and odd cases separately. Note that we classify the final form of the bispectra in terms of the leading pole of the MLT solutions presented in the previous subsection. Once we sum over permutations there can be cancellations meaning that the final form has a lower order pole. However, it is the solution to the MLT whose leading degree pole is generically equal to the number of derivatives in a corresponding bulk vertex. Each of the bispectra below can be multiplied by a real coupling which we denote as $g_{\alpha,p}$, and we absorb all $O(1)$ factors that appear when we go from a wavefunction to correlator (c.f. \eqref{WFtoCorrelator}) into these couplings.

\paragraph{Parity-even interactions} We begin with even $\alpha$ where both the rational parts and the logarithmic parts contribute to the correlator, as shown in Section \ref{ParitySection}. \\

\noindent $\bm{\alpha=0}$ Since in this case we have $h_{\alpha} = 1$, both the final $+++$ and $++-$ bispectra are easily read off from the solutions to the MLT given above. We simply take the spinor helicity factors, multiply them by the MLT solutions and then divide by the power spectrum of each external field which contributes a factor of $1 / e_{3}^3$. We have
\begin{align}\label{prima}
 p=0 :& \quad e_{3}^3 B^{+++}_{3} = g_{0,0} \text{SH}_{+++}[4e_3 - e_2 k_T + (k_{T}^3 - 3 k_{T}e_{2} + 3 e_{3})(\log (-k_T \eta_0 / \mu)]\,, \\
&  \quad e_{3}^3 B^{++-}_{3} = g_{0,0}\text{SH}_{++-}[4e_3 - e_2 k_T + (k_{T}^3 - 3 k_{T}e_{2} + 3 e_{3})\log (-k_T \eta_0 / \mu)]\,, \\
 p=2 :& \quad e_{3}^3 B^{+++}_{3} = g_{0,2} \text{SH}_{+++} \frac{e_2 e_3 + e_2^2 k_T - 2 e_3 k_T^2}{k_T^2}\,, \\
 & \quad e_{3}^3 B^{++-}_{3} = g_{0,2} \text{SH}_{++-} \frac{e_2 e_3 + e_2^2 k_T - 2 e_3 k_T^2}{k_T^2}\,,\\
  p=3 :& \quad e_{3}^3 B^{+++}_{3} = g_{0,3} \text{SH}_{+++} \frac{e_3^2}{k_T^3}\,, \\
 & \quad e_{3}^3 B^{++-}_{3} = g_{0,3} \text{SH}_{++-}  \frac{e_3^2}{k_T^3}\,,\\
& \quad \vdots \nonumber
\end{align}
This $p=0$ bispectrum corresponds to a combination of a potential term in the bulk of the form $\gamma_{ij}^3$ and the contribution $k_{T}^3 - 3 k_{T}e_{2} + 3 e_{3}$ which is the graviton version of the well studied local non-Gaussianity \cite{LNG}. The local shape arises from taking the free theory for the massless graviton and performing a field redefinition $\gamma_{ij} \rightarrow \gamma_{ij} + \gamma_{ik}\gamma_{kj}$. Such a redefinition does not alter the $S$-matrix and so its contribution to the wavefunction must be regular as $k_{T} \rightarrow 0$, which it is. The log piece is produced by the $\gamma_{ij}^3$ vertex which appears in Solid Inflation \cite{SolidInflation} and in the slow-roll limit it is the leading contribution from this interaction. The $p=3$ bispectrum corresponds to that of a $\dot{\gamma}_{ij}^3$ vertex in the bulk which appears in the Effective Field Theory of Inflation (EFToI) \cite{Cheung:2007st}, without corrections to the two-point function and with an independent coefficient \cite{CabassBordin}. We provide more details about these examples in Section \ref{SymmetryBreaking}. \\

\noindent $\bm{\alpha=2}$ In this case the polarisation factor is not fully symmetric, so after we multiply it by a solution to the MLT, we need to symmetrize the result. We find 
\begin{align}
 p=2 :& \quad e_{3}^3 B^{+++}_{3} = g_{2,2} \text{SH}_{+++}\frac{e_2 (e_3 + e_2 k_T - k_T^3)}{k_T^2}\,, \\
&  \quad e_{3}^3 B^{++-}_{3} = g_{2,2} \text{SH}_{++-}\frac{(k_1k_2-k_2k_3-k_3k_1)(e_3 + e_2 k_T - k_T^3)}{k_T^2}\,, \\
 p=3 :& \quad e_{3}^3 B^{+++}_{3} = g_{2,3} \text{SH}_{+++} \frac{e_3 (6 e_3 + 2 e_2 k_T + k_T^3)}{k_T^3}\,, \\
 & \quad e_{3}^3 B^{++-}_{3} = g_{2,3} \text{SH}_{++-} \frac{-e_3 (4 e_3 + k_T (4 e_2 + I_3^2 + 2 I_3 k_T - k_T^2))}{2 k_T^3}\,,\\
& \quad \vdots \nonumber
\end{align}
Since GR is a two-derivative, parity-even theory, its bispectrum in de Sitter space must be contained within the solutions we have written up to this point. Indeed, if we first take $\mu = - k_{T} \eta_{0} e^{-\tilde{g}_{0,0} / g_{0,0}}$, and then
\begin{align}
g_{0,2} = 2 \tilde{g}_{0,0} = -g_{2,2}, \qquad g_{0,0} = 0, \qquad \text{(GR tuning)}
\end{align}
then both the $+++$ and $++-$ wavefunction coefficients are those of GR \cite{Maldacena:2011nz}. We remind the reader that on the total-energy poles we recover the amplitude, and in GR the $+++$ amplitude vanishes while the $++-$ amplitude does not. This tells us that in GR the $+++$ bispectrum should not have such a pole while the $++-$ one should have a degree-$2$ pole. If we take an arbitrary linear combination of these bispectra and demand that the $+++$ wavefunction does not have a total-energy pole, while the $++-$ has a non-zero total-energy pole, then the result is a linear combination of GR and the local non-Gaussianity. In \cite{Maldacena:2011nz} these conditions along with full de Sitter symmetry was enough to uniquely pick out GR. Without some additional symmetry principle, we cannot set the coefficient of the local non-Gaussianity coupling to zero. Interestingly, this GR bispectrum is the leading order one in the EFToI \cite{Creminelli:2014wna,CabassBordin,Bordin:2017hal}: the breaking of boosts is only felt at higher-order in derivatives. \\

\noindent $\bm{\alpha=4}$ Again in this case the polarisation factor is not fully symmetric, so we take the solutions to the MLT and then symmetrise appropriately. We find 
\begin{align}
 p=4 :& \quad e_{3}^3 B^{+++}_{3} = g_{4,4} \text{SH}_{+++} I_1 I_2 I_3 \frac{3 e_3 + e_2 k_T + k_T^3}{k_T^3}\,, \\
&  \quad e_{3}^3 B^{++-}_{3} = g_{4,4} \text{SH}_{++-}  I_1 I_2 I_3 \frac{3 e_3 + e_2 k_T + k_T^3}{k_T^3}\,, \\
 p=5 :& \quad e_{3}^3 B^{+++}_{3} = g_{4,5} \text{SH}_{+++} I_1 I_2 I_3 \frac{24 e_2 e_3 + 6 e_2^2 k_T - 9 e_3 k_T^2 + e_2 k_T^3}{k_T^5}\,, \\ \nn
 & \quad e_{3}^3 B^{++-}_{3} = g_{4,5} \text{SH}_{++-} I_1 I_2 (2 k_T)^{-4} \Big[  12 {e_2}^2 {k_T}+48 {e_2} {e_3}+{e_2} {k_T} \left(3 {I_3}^2-2 {I_3} {k_T}+{k_T}^2\right) \\
 & \quad \quad \quad \quad \quad \quad \quad \quad \quad \quad \quad \quad \quad \quad \quad \quad\,\,
  + 2 {e_3} \left(6 {I_3}^2-6 {I_3} {k_T}-7 {k_T}^2\right)+{I_3}^2 {k_T}^3-{k_T}^5 \Big]\,,\\
& \quad \vdots \nonumber
\end{align}

\noindent $\bm{\alpha=6}$ Here we have $h_{\alpha} = I_1^2 I_2^2 I_3^2$, which is fully symmetric and no symmetrization is necessary when constructing the full bispectra. We have:
\begin{align}
 p=6:& \quad e_{3}^3 B^{+++}_{3} = g_{6,6} \text{SH}_{+++} I_1^2 I_2^2 I_3^2  \frac{15 e_3 + 3 k_T e_2 + k_T^3}{k_T^6}  \, , \\
&  \quad e_{3}^3 B^{++-}_{3} = g_{6,6} \text{SH}_{++-} I_1^2 I_2^2  \frac{15 e_3 + 3 k_T e_2 + k_T^3}{k_T^4}  \,  , \\
 p=8:& \quad e_{3}^3 B^{+++}_{3} = g_{6,8} \text{SH}_{+++}I_1^2 I_2^2 I_3^2 \frac{7 e_2 e_3 + k_T e_2^2 - 2 k_T^2 e_3 }{k_T^8} \, , \\
 & \quad e_{3}^3 B^{++-}_{3} = g_{6,8} \text{SH}_{++-} I_1^2 I_2^2  \frac{7 e_2 e_3 + k_T e_2^2 - 2 k_T^2 e_3 }{k_T^6}\,, \label{ultima}\\
& \quad \vdots \nonumber
\end{align}

\paragraph{Parity-odd interactions} We now turn to odd $\alpha$ where only the coefficient of the logarithm can contribute to the correlator. In all cases it must be multiplied by $i \pi /2$. We absorb the $\pi / 2$ factor into the overall coupling $g_{\alpha,p}$ and then the additional factor of $i$ combines with the factors of $i$ appearing in each $h_{\alpha}$, c.f. \eqref{h7b}, to give real coefficients. As we showed above, to all orders in derivatives logarithms can only appear for $\alpha=1,3$ and give rise to a total of three solutions. Since in each case the logarithmic solutions to the MLT must always come with total-energy poles, we still classify these solutions by the corresponding $p$. \\

\noindent $\bm{\alpha=1}$ In this case we need to multiply $h_{1}(k_{1},k_{2},k_{3}) = k_{1}$ by the appropriate solutions to the MLT and then symmetrise. We find
\begin{align}\label{96}
p=1: & \quad e_{3}^3 B^{+++}_{3} = g_{1,1} \text{SH}_{+++} k_T \left( k_T^2 - 2 e_2 \right) \,, \\ \label{96a}
&  \quad e_{3}^3 B^{++-}_{3} = g_{1,1} \text{SH}_{++-}  I_3 \left( k_T^2 - 2 e_2 \right)\,, \\ \label{96b1}
p=2: & \quad e_{3}^3 B^{+++}_{3} = g_{1,2} \text{SH}_{+++} \left(-3 e_3 + k_T e_2 \right)\,, \\ \label{96b}
 & \quad e_{3}^3 B^{++-}_{3} = g_{1,2} \text{SH}_{++-} \left(k_1 (k_2^2 + k_3^2) + k_2(k_1^2 + k_3^2) - k_3 (k_1^2 + k_2^2) \right)\,.
\end{align}
Possible operators that generate these bispectra are, respectively (up to constant factors),
\begin{align}\label{eq:ParityOddOperator1}
& a(\eta)^{-1} g_{1,1} \eps_{ijk} \gamma_{il} \gamma_{lm} \partial_j \gamma_{km}  , \\
& a(\eta)^{-2} g_{1,2}  \eps_{ijk} \gamma'_{il} \gamma_{lm} \partial_j \gamma_{km} \label{eq:ParityOddOperator2}
\,.
\end{align}

\noindent $\bm{\alpha=3}$ In this case we have $h_{3}(k_{1},k_{2},k_{3}) = I_{1}I_{2}I_{3}$ which is already symmetric. We then have a unique $\log$ term yielding
\begin{align}
p=3: & \quad e_{3}^3 B^{+++}_{3} = g_{3,3} \text{SH}_{+++} I_1 I_2 I_3 =g_{3,3} \text{SH}_{+++} \left(  -8 e_{3} + 4 e_{2} k_{T} - k_{T}^3\right) \,, \\
&  \quad e_{3}^3 B^{++-}_{3} = g_{3,3} \text{SH}_{++-}  I_1 I_2 k_T\,.\label{101}
\end{align}
This can be generated, up to a constant factor, by the operator
\begin{equation}\label{eq:ParityOddOperator3}
a(\eta)^{-3} g_{3,3} \eps_{ijk} \partial_l \gamma_{i n} \partial_m \gamma_{j l} \partial_n \gamma_{k m}
\,.
\end{equation}

\noindent These bispectra correspond to those in computed in \cite{CabassBordin} where the three couplings were tuned as dictated by the symmetries of the EFToI. Although in that work the bispectra were presented using polarisation tensors, we have checked that they are all indeed captured by our expressions and provide details in Section \ref{EFToI}. Such parity-odd interactions do not appear on their own in EFToI; rather, they come with a correction to the two-point function \cite{CabassBordin}. We also discuss this further in Section \ref{EFToI}. It is also worth pointing out that our results tell us that for bulk vertices with more than three derivatives, and therefore $ p \geqslant 4$ degree poles in the solutions to the MLT, there are no contributions to the bispectra. Indeed, IR-divergent logarithms can only appear when 
\begin{align}
\text{Condition for IR-divergent logs:} ~~~~ 2 n_{\partial_{\eta}} + n_{\partial_{i}} \leqslant 3 \,,
\end{align}
where $  n_{\partial_{\eta}} $ and $  n_{\partial_{i}} $ are respectively the number of time and space derivatives in the parity-odd interaction. Note that here we assume that each field in the cubic vertex contains at most one time derivative which can always be guaranteed by using the equations of motion. This offers a complementary proof that the parity-odd $\text{Weyl}^3$ vertex in de Sitter space leads to a vanishing bispectrum. Indeed, this is a six derivative vertex and therefore the corresponding wavefunction does not have logarithms and therefore the correlator vanishes. See \cite{Maldacena:2011nz,Shiraishi:2011st,Soda:2011am} for further discussions. 

\paragraph{Discussion} So far in this paper we have bootstrapped three-point wavefunction coefficients arising from tree-level and manifestly local bulk graviton self-interactions. We have made very minimal assumptions. We assumed the usual massless de Sitter mode functions, and assumed that the vertices are $SO(3)$ and scale invariant. With this full catalogue at hand, one can now search for interesting subsets. Indeed, given a particular symmetry breaking pattern for inflation, as recently classified in \cite{ZoologyNG}, only some of these bispectra will be permitted and non-linear realisations of the broken symmetries could result in relations among the couplings $g_{\alpha,p}$, as is the case for GR. One could attempt to perform the classification at the level of the Lagrangian using an effective field theory approach, however the main message of the bootstrap approach is that the Lagrangian route might not be the most efficient. Rather, one would like to take this full catalogue of bispectra and use soft theorems to classify consistent subsets, or even better to use these objects as the building blocks of higher-point functions. We have learned from the $S$-matrix programme that gluing together three-point amplitudes to form consistent four-point ones can be very constraining \cite{Benincasa:2007xk,PSS}. We expect this gluing procedure to also be very constraining for cosmology and plan to explore this in future work. For parity-even vertices in the EFToI, there is only a single operator at both cubic and quartic order in derivatives that does not modify the two-point function \cite{CabassBordin}. It would be interesting to rederive this result directly using bootstrap methods. In any case, in Section \ref{SymmetryBreaking} we provide a discussion of how these bispectra could be classified by the EFToI \cite{Cheung:2007st} or as the leading contributions to the bispectra of Solid Inflation \cite{SolidInflation}, following the approach of \cite{ZoologyNG}.


\subsection{Parity-odd bispectra involving gravitons \textit{and} scalars} \label{AddingScalars}

\noindent Given that we have discovered very few possible parity-odd bispectra for three gravitons, let us provide a more complete analysis by also considering bispectra involving a scalar. As is well-known, the bispectrum for three scalars cannot break parity. This is easily seen given that there is no non-zero way to contract two independent momenta with an epsilon tensor, and so there are simply no parity-odd tensor structures in the absence of polarisation tensors. Let us therefore concentrate on scalar-scalar-graviton ($B^{00+}_{3}$) and scalar-graviton-graviton ($B^{0++}_{3}$) bispectra. As always, other helicity configurations can be extracted from these as explained in Section \ref{ToRule}. \\

\noindent As we have done throughout this work, we concentrate on manifestly local interactions and so the solutions to the MLT that we have classified previously in this section can be used to construct trimmed wavefunctions when we also have scalars: the MLT applies to scalars and gravitons alike. Our results of Section \ref{ParitySection} also apply and so for contact interactions there can be no singularities in the parity-odd bispectra. Our job to classify $B^{00+}_{3}$ and $B^{0++}_{3}$ is then a simple one: we first write down all possible parity-odd tensor structures, and then multiply these by a solution to the MLT. Here we will concentrate on the contributions to the bispectrum rather than the wavefunction and so the relevant part of the solution to the MLT is the coefficient of a logarithm, as we have explained in detail above. These logs can only occur for $\alpha=1,3$ and so we only need to consider these tensor structures. 

\paragraph{Scalar-scalar-graviton} First consider $B^{00+}_{3}$. In this case there is only a single parity-odd tensor structure, which has $\alpha=3$: 
\begin{align} \label{PPG1}
\epsilon_{ijk}  e^{h_{3}}_{im}(\bfk_{3}) k_{1}^{j} k_{2}^{k} k_{1}^{m}\,,
\end{align} 
and permutations. The relevant solution to the MLT contains a log with a $k$-independent coefficient just as was the case for $\alpha=3$ with three gravitons. If we multiply \eqref{PPG1} by the appropriate MLT solution and sum over permutations, then the contribution to the bispectrum written in terms of spinor helicity variables is
\begin{align} \label{SSTbispectrum}
e_{3}^3B_{3}^{00+} = h_{3,3} \frac{[13]^2 [23]^2}{k_{3}^2 [12]^2} I_{3}^2 k_{3}\,.
\end{align} 
Regardless of the form of the symmetry breaking pattern, this is the \textit{only} such parity-odd bispectrum when the bulk interactions are manifestly local, to all orders in derivatives.

\paragraph{Scalar-graviton-graviton} Moving onto $B^{0++}_{3}$ we find a single tensor structure for $\alpha=1$ and two for $\alpha=3$. Up to permutations we have
\begin{align}
\alpha=1&: \epsilon_{ijk}  e^{h_{2}}_{im}(\bfk_{2}) e^{h_{3}}_{jm}(\bfk_{3})  k_{k}^{2}\,,  \\
\alpha=3&:  \epsilon_{ijk}  e^{h_{2}}_{im}(\bfk_{2}) e^{h_{3}}_{ml}(\bfk_{3})  k_{j}^{2}k_{k}^{3}k_{l}^{1} \quad \text{and} \quad \epsilon_{ijk}  e^{h_{2}}_{il}(\bfk_{2}) e^{h_{3}}_{jm}(\bfk_{3})  k_{k}^{2}k_{l}^{1}k_{m}^{1}\,.
\end{align}
Now for $\alpha=1$ we need to multiply this tensor structure by a degree-$2$ polynomial that arises from the coefficient of a log in a solution to the MLT. We find three such solutions: this tensor structure can be multiplied by $k_{1}^2, k_{2}^2$ or $k_{3}^2$ with arbitrary coefficients. Each of the appropriate MLT solutions also have rational contributions with $k_{T}$ poles: one solution has a simple pole while the other two solutions have $k_{T}^{-2}$ poles. A complete basis is
\begin{align} \label{Alpha1}
e_{3}^3 B_{3}^{0++} = \frac{[23]^4}{k_{2}^2k_{3}^2} \left[q_{1,1}(k_{2}+k_{3})k_{1}^2 + q_{1,2,a}(k_{2}^3+k_{3}^3)+q_{1,2,b}(k_{2}k_{3}^2+k_{3}k_{2}^2) \right]\,.
\end{align}
Now for $\alpha=3$ we find that when converted to spinor helicity variables, the two $\alpha=3$ structures are equivalent and since they already scale as $\sim k^3$, the relevant part of the MLT solution is simply a constant multiplied by a log. We therefore have a single solution for the $\alpha=3$ $B_{3}^{0++}$ bispectrum which turns out to be a linear combination of those from $\alpha=1$ in \eqref{Alpha1}. It follows that \eqref{Alpha1} is a complete list, to all orders in derivatives. From these bispectra we can also extract those for $B_{3}^{0+-}$. We have
\begin{align} 
e_{3}^3 B_{3}^{0+-} = \frac{I_{2}^4}{k_{2}^2k_{3}^2} \frac{[12]^4}{[31]^4}  \left[q_{1,1}(k_{2}-k_{3})k_{1}^2 + q_{1,2,a}(k_{2}^3-k_{3}^3)+q_{1,2,b}(k_{2}k_{3}^2-k_{3}k_{2}^2) \right]\,,
\end{align}
where the overall factor is a necessary consequence of helicity scaling and the absence of divergences. \\

\noindent One of our main messages in this paper is that parity-odd contact bispectra, arising from manifestly local cubic interactions, are small in number. In Table \ref{NumberOfCouplings} we summarise the number of independent couplings associated with tree-level parity-odd bispectra of manifestly local scalars and gravitons, to all orders in derivatives, and with exact scale invariance. In inflationary models, we would expect additional bispectra that mix scalars and gravitons and violate manifest locality. These will arise when we integrate out the non-dynamical modes. We still expect the shapes of such correlators to be heavily constrained given our discussion in Section \ref{ParitySection}. The primary difference is that in those cases the logs that appear in the wavefunction can in principle be multiplied by poles as one of the external energies is taken soft. In any case, in Section \ref{SymmetryBreaking} we comment on when the above bispectra appear in the effective field theory of inflation and solid inflation.
\begin{table}[h!]
\begin{center}
\begin{tabular}{c c c c c c}
Parity-odd bispectra  &  SSS &  SST &  STT  &  TTT  \\
\hline
no. of couplings  & 0  & 1 & 3 & 3  
\end{tabular}
\caption{For manifestly local and scale invariant theories, the table specifies the number of independent parity-odd tree-level bispectra for all possible combinations of scalars (``S'') and gravitons (``T'') to all orders in derivatives.}
\label{NumberOfCouplings}
\end{center}
\end{table}


\section{Graviton bispectra and symmetry breaking patterns} \label{SymmetryBreaking}

In this section, we want to study which of the graviton and graviton-scalar three-point functions we have discussed so far can arise during inflation depending on the particular way in which de Sitter boosts are broken. We consider the 
Effective Field Theory of Inflation (EFToI) \cite{Cheung:2007st} and the symmetry breaking pattern of a solid, i.e.~Solid 
Inflation \cite{SolidInflation}. 


\subsection{Effective field theory of inflation} \label{EFToI}
Let us begin with the EFToI which is the most well-studied symmetry breaking pattern for inflation. Here the symmetry breaking is driven by a single scalar that acquires a time dependent vev. The background homogeneity and isotropy is then manifest and an approximate shift-symmetry for the resulting Goldstone mode ensures approximately scale invariant primordial correlators. In the decoupling limit and on subhorizon scales, where we can neglect gravity and the expansion of the universe, the Goldstone theory is that of a superfluid \cite{Son:2002zn}. \\

\noindent First, let us stress that at tree-level all cubic graviton interactions are manifestly local: one does not need to worry about non-manifestly local interactions coming from solving the Hamiltonian and momentum constraints in GR. The reason is that for the three-point function it is sufficient to solve the constraints at linear order\footnote{More generally, the solution of the constraints to order $ n  $ is sufficient to write down the action to order $ (2n + 1)  $ or less \cite{Pajer:2016ieg}.} \cite{Maldacena:2002vr} and at this order a two-tensor cannot mix with the scalars and transverse vector in $  g^{0\mu} $. Hence, the Manifestly Local Test (MLT) we used throughout this paper does indeed capture graviton bispectra in the EFToI. Now, what are the building blocks for the graviton operators? Initially consider operators that give rise to non-trivial cubic graviton self-interactions, but \textit{do not} alter the graviton's quadratic action with respect to the GR contribution. This case corresponds to the setup in this paper: standard dS mode functions for the massless graviton plus bispectra arising from manifestly-local cubic self-interactions. To find these building blocks we can stop the expansion of all geometric objects constructed from the foliation at leading order in perturbations. 
We can either use $\dot{\gamma}_{ij}$, which is $2\,\delta K^i_{\hphantom{i}j}$ 
at leading order in perturbations ($\delta K_{\mu\nu}$ being the fluctuation in the extrinsic curvature of constant-time hypersurfaces), or 
$a^{-2}\partial_k\partial_l{\gamma}_{ij}$. The indices $ijkl$ cannot be, however, chosen arbitrarily. We can either have the combination\footnote{The square brackets on a pair of indices denote anti-symmetrization with weight one, $  A_{[ij]}\equiv \left( A_{ij}-A_{ji} \right)/2 $.}
\be
a^{-2}\big(\partial_k\partial_{[i}\gamma_{j]l} - \partial_l\partial_{[i}\gamma_{j]k}\big)\,,  
\ee
corresponding to the Riemann tensor ${^{(3)}}\!R^{ij}_{\hphantom{ij}kl}$ on constant-time hypersurfaces, or 
\be
a^{-2}\partial^2\gamma_{ij}\,, 
\ee
corresponding to the Ricci tensor ${^{(3)}}\!R^{i}_{\hphantom{i}j}$. 
We can then freely take further time derivatives or spatial derivatives of these building blocks, since we can project derivatives either 
parallel or orthogonal to $n^\mu$, the normal four-vector to constant-time hypersurfaces.

\paragraph{Parity even} Let us consider a few parity-even examples (\emph{beyond} the bispectrum of GR) before moving to the parity-odd case. 
The constraints on the building blocks forbid us from having $p< 3$ for $\alpha=0$ and $p<5$ for $\alpha=2$ and $\alpha=4$. Two examples are the following: we have the dimension-$6$ and dimension-$7$ operators 
\be\label{ints}
\text{$\int d \eta d^3 x \,a(\eta)\,\gamma'_{ij} \gamma'_{jk} \gamma'_{ki}$\quad and \quad $\int d \eta d^3 x \, \gamma'_{ij} \gamma'_{jk}\partial^2\gamma_{ki}\,.$}
\ee 
Since no spatial derivative is contracted with the indices of $  \gamma_{ij} $, both have $\alpha=0$ and both give the same trimmed wavefunction $\psi^{\rm trimmed}(k_1, k_2, k_3)$ which may be assumed to be symmetric since the polarisation factors are. We find
\be
\psi^{\rm trimmed}(k_1, k_2, k_3) = \frac{e^2_3}{k^3_T}\,,
\ee 
which is our $\alpha=0$, $p=3$ solution from Section \ref{sec:bispectra}. In general we expect that the order of the total-energy pole is given by \cite{BBBB}
\begin{align}\label{general}
p=1+\sum_{A} (\Delta_{A}-4)\,,
\end{align}
where the sum is over all vertices $  A $ with mass dimension $  \Delta_{A} $. For the tree-level bispectrum of gravitons and scalars this tells us that $  p $ is the total number of spatial and time derivatives. Indeed, for the first interaction in \eqref{ints}, we get a $k_{T}^{-3}$ pole as expected. For the second interaction in \eqref{ints}, we naively expect a $k_{T}^{-4}$ pole. However, the amplitude corresponding to this interaction vanishes, so the residue of the $k_{T}^{-4}$ pole is zero\footnote{On-shell we can replace the $\partial^2$ with two time derivatives and then it is clear that this interaction is a total time derivative and does therefore not contribute to the energy conserving $S$-matrix.}. A similar observation was made for the DBI limit of the EFToI in \cite{GJS}. Another example is the dimension-$7$ operator 
\be
\int d \eta d^3 x \, \gamma'_{jk} \gamma'_{il}\big(\partial_k\partial_{(i}\gamma_{j)l} - \partial_l\partial_{(i}\gamma_{j)k}\big) \,, 
\ee 
which has $\alpha=2$ because two spatial derivatives are contracted with the indices of $\gamma_{ij}$. The trimmed wavefunction coefficient is
\be 
\psi^{\rm trimmed}(k_1, k_2, k_3) = \frac{k_{1}^2 k_{2}^2}{k_{T}^4}(k_{1}+k_{2}+4k_{3})\,, 
\ee 
i.e.~$p=4$, as expected from \eqref{general}. Note that this is the trimmed wavefunction for one of the permutations where the third leg in the diagram is not differentiated with respect to time. One would need to follow the rules outlined in Section \ref{sec:bispectra} to find the final expressions with the correct symmetries. Despite appearances, this trimmed wavefunction does indeed satisfy the MLT for each leg.

\begin{figure}
  \centering
  \includegraphics[width=.8\linewidth]{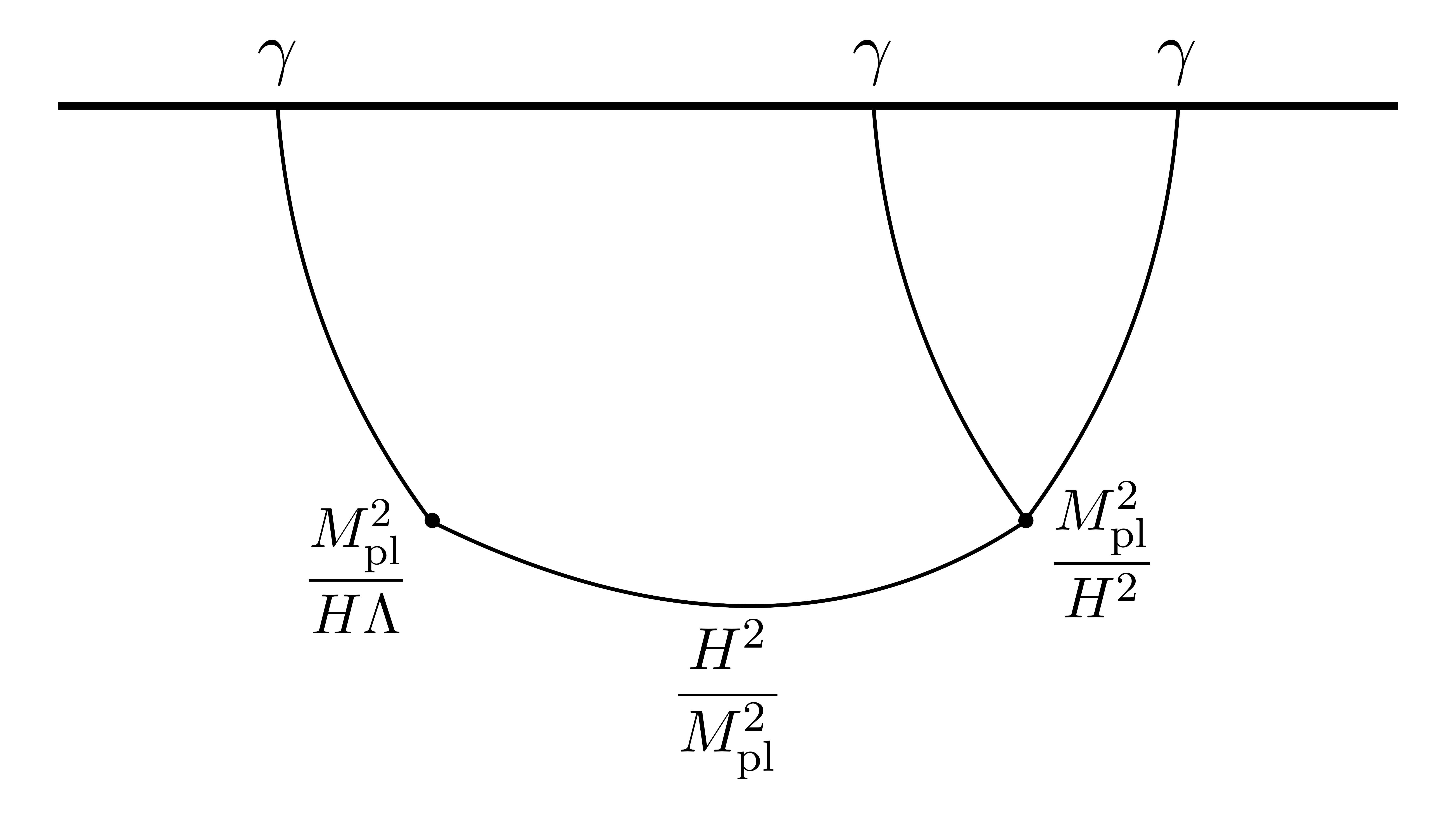}  
  \caption{Three-point exchange diagram in the parity-odd sector of the EFT of Inflation}
  \label{fig:3ptexchangediagram}
\end{figure}

\paragraph{Parity odd} As we have seen already in the previous sections, the parity-odd case is much more constrained. While it is possible to write infinitely many parity-odd high-dimension operators, only one contributes to the three-point function. This is the three-dimensional Chern-Simons term, 
i.e. 
\be \label{CSterm}
\frac{M_{\text{pl}}^2}{\Lambda}\int  d \eta d^3 x \,a(\eta) \epsilon_{ijk} \,\Bigg[\frac{{^{(3)}\Gamma}^l_{im}\partial_j{^{(3)}\Gamma}^m_{kl}}{2} + \frac{{^{(3)}\Gamma}^l_{im}{^{(3)}\Gamma}^m_{jn}{^{(3)}\Gamma}^n_{kl}}{3}\Bigg]\,, 
\ee 
where ${^{(3)}\Gamma}^k_{ij}$ are the Christoffel symbols for the covariant derivative on hypersurfaces of constant time and we have introduced a new scale $\Lambda$ on dimensional grounds. 
It is possible to show (see e.g.~\cite{CabassBordin}, Appendix B), that this term is equal to the $W \tilde{W}$ combination (where $W$ is shorthand for the Weyl tensor and $\tilde{W}$ for its dual) multiplied by $f(\phi)\propto \phi$ where $\phi$ is the inflaton, up to a boundary term\footnote{Given the high number of derivatives, 
such boundary term vanishes very fast at late times and does not contribute to the three-point function.} 
and an operator that has $\alpha=1$ and $p=3$ (which does not, then, contribute to correlators as we showed in Section \ref{ParitySection}). 
At cubic order in perturbations, this operator is equal to \cite{CabassBordin}
\be
\label{3dcs_action}
\begin{split}
\frac{M_{\text{pl}}^2}{\Lambda}\int  d \eta d^3 x \,a(\eta) \,\bigg[&\frac{1}{4}\epsilon_{ijk}\gamma_{kn}\partial_l\gamma_{nm}\partial_j\partial_l\gamma_{im} 
+ \frac{1}{4}\epsilon_{ijk}\gamma_{ln}\partial_n\gamma_{im}\partial_j\partial_l\gamma_{km} 
- \frac{1}{4}\epsilon_{ijk}\partial_m\gamma_{nj}\partial_n\gamma_{lm}\partial_i\gamma_{lk} \\
& - \frac{1}{12}\epsilon_{ijk}\partial_m\gamma_{lj}\partial_n\gamma_{mi}\partial_l\gamma_{nk} 
+ \frac{1}{4}\epsilon_{ijk}\partial_n\gamma_{lm}\partial_k\gamma_{ln}\partial_j\gamma_{mi} 
+ \frac{1}{4}\epsilon_{ijk}\partial_m\gamma_{nj}\partial_m\gamma_{lk}\partial_l\gamma_{ni} \\ 
& + \frac{1}{4}\epsilon_{ijk}\partial_n\gamma_{mk}\partial_l\gamma_{mn}\partial_j\gamma_{li}\bigg]\,. 
\end{split}
\ee
We recognize the tensor structures summarized in Appendix~\ref{app:contractions}. The first and sixth terms in 
the above equation are the first and second tensor structures in Eq.~\eqref{eq:g1}. Indeed they only have one spatial derivative contracted with the graviton's indices and have $\alpha=1$. The other five terms are found in Eqs.~\eqref{eq:g3-zero_k}, \eqref{eq:g3-one_k}, \eqref{eq:g3-two_k} and have $\alpha=3$. If we take the mode functions to be the usual massless ones in dS (see below for a discussion on this point) then the bispectrum coming from this sum of interactions is given by a linear sum of the parity-odd bispectra in Section \ref{FinalForm}. The relevant constraints are
\begin{align} \label{tunings}
g_{1,1} = -2g_{1,2} = - \frac{1}{6}g_{3,3}\,,
\end{align}  
and by fixing $g_{1,1}$ in terms of $M_{\text{pl}}^2 / (H \Lambda)$, and reinserting the factor of $\pi$, we find for the $+++$ configuration 
\begin{align} 
e_{3}^3B_{\text{CS,contact}}^{+++} &= - \frac{\pi}{4}\frac{H}{\Lambda}\frac{H^4}{M_{\text{pl}}^4} \text{SH}_{+++}(99e_{3} - 53 k_{T}e_{2}+14k_{T}^3) \,, 
\end{align} 
where the subscript CS stands for ``Chern-Simons''. If we rewrite this using the tunings of Eq.~\eqref{tunings}, we find 
\begin{align} 
e_{3}^3B_{\text{CS,contact}}^{+++} &= {- \frac{\pi}{4}\frac{H}{\Lambda}\frac{H^4}{M_{\text{pl}}^4}} \text{SH}_{+++} \Big(2k_T(k_T^2-2e_2) - (-3e_3+k_Te_2) -12I_1I_2I_3\Big) \,, \label{EFToIContact+++} \\
e_{3}^3B_{\text{CS,contact}}^{++-} &=  {- \frac{\pi}{4}\frac{H}{\Lambda}\frac{H^4}{M_{\text{pl}}^4}} \text{SH}_{++-} \Big(2I_3(k_T^2-2e_2)- \big(k_1 (k_2^2 + k_3^2) + k_2(k_1^2 + k_3^2) - k_3 (k_1^2 + k_2^2)\big) \nonumber \\
&\hphantom{{- \frac{\pi}{4}\frac{H}{\Lambda}\frac{H^4}{M_{\text{pl}}^4}} \text{SH}_{++-} \Big(\;\;\;\;} -12I_1I_2k_T\Big)\,. \label{EFToIContact++-}
\end{align} 
The non-linear realization of boosts not only forces the different operators to appear with tuned coefficients, it 
also forces a contribution to the quadratic graviton action. One can see how this is necessary from the fact that 
the bispectrum from Eq.~\eqref{3dcs_action}, by itself, gives a contribution $\sim q^{-3}$ in the soft $q$ limit\footnote{This 
comes from the first two terms in Eq.~\eqref{3dcs_action}.},
spoiling the consistency relation of GR. The modification to the quadratic action is 
\begin{equation}
\label{eq:3dcs_quadratic_action}
-\frac{M_{\text{pl}}^2}{4 \Lambda}\int  d \eta d^3 x \,a(\eta) \, \epsilon_{ijk} \partial_{i} \partial_{m} \gamma_{jl} \partial_{m} \gamma_{lk}\,,
\end{equation} 
and gives rise to the new three-point exchange diagram shown in Figure \ref{fig:3ptexchangediagram} where the cubic vertex is the one from GR.
In this paper we have not discussed such exchange diagrams. However, given that such an operator is the only source of parity violation in the EFToI, we find it interesting to consider the wavefunction coefficient and resulting bispectrum from this diagram which must appear in addition to the contact ones we just derived. Note that we are treating the correction to the two-point function perturbatively which, as shown in \cite{Creminelli:2014wna}, is valid as long as the approximately constant Hubble scale during inflation is smaller than the scale $\Lambda$. Indeed, the correction to the late-time power spectrum $\braket{\gamma^{\pm}_{\mathbf{k}}\gamma^{\pm}_{\mathbf{k}'}}'=P^\pm(k)$ of the graviton is \cite{Creminelli:2014wna}
\be
\delta P^{\pm}(k) = \pm\pi\,\frac{H}{\Lambda} \frac{H^2}{M_{\text{pl}}^2k^3}\,, 
\ee
where the factor of $\pi$ is enforced by unitarity, as explained in Section \ref{ParitySection}, and the $\pm$ indicates that the helicities have been split by this parity-odd correction. To compute this correction to the power spectrum the relation
\begin{align}
i \epsilon_{ijk}k_{j}e_{km}^{h}(\bfk) = \lambda_{h} k\, e^{h}_{im}(\bfk) 
\end{align}
proves useful, where $\lambda_{\pm} = \pm 1$. By considering this correction perturbatively, we can use the usual bulk-to-boundary and bulk-to-bulk propagators arising from the massless mode functions, as we have done to compute \eqref{EFToIContact+++} and \eqref{EFToIContact++-}. It is worth noting that to get a parity-odd bispectrum one could use any parity-even vertex for the right-hand sub-diagram but this one is of the same order as the contact contributions arising from \eqref{3dcs_action}. The contribution to the wavefunction from the contact interactions \eqref{3dcs_action} is $M_{\text{pl}}^2 / (H \Lambda)$. For the diagram in Figure \ref{fig:3ptexchangediagram}, the GR vertex contributes a factor of $M_{\text{pl}}^2 / H^2$, the quadratic mixing contributes a factor of $M_{\text{pl}}^2 / (H \Lambda)$, while the bulk-bulk propagator scales in the same way as the graviton power spectrum and so contributes a factor of $H^2 / M_{\text{pl}}^2$. Multiplying these factors together shows that the contact diagram and this exchange diagram contain the same dependence on $M_{\text{pl}}, H$  and $\Lambda$, as expected from the consistency relations of the EFToI \cite{CabassBordin}. If in Figure \ref{fig:3ptexchangediagram} we took the cubic vertex to be given by the sum in \eqref{3dcs_action} then this (parity-even) contribution would scale as $M_{\text{pl}}^2 / (H \Lambda) \times H / \Lambda$ and for $H < \Lambda$ such a diagram is suppressed. \\

\noindent  Now, up to cubic order the GR action is 
\begin{align}
S_{GR} = \frac{M_{\text{pl}}^2}{8} \int  d \eta d^3 x \,a(\eta)^2 [\gamma_{ij}' \gamma_{ij}' - \partial_{l}\gamma_{ij}\partial_{l}\gamma_{ij}+(2 \gamma_{ik}\gamma_{jl} - \gamma_{ij}\gamma_{kl})\partial_{k}\partial_{l}\gamma_{ij} + \mathcal{O}(\gamma^4)]\,,
\end{align}
and given that both the GR vertex and the new quadratic term only have spatial derivatives, none of the propagators in the bulk time integrands are differentiated, so the relevant integrals are the same for all permutations. The only integral we need to compute for this exchange diagram is
\begin{align}
- i \int d \eta \int d \eta' a(\eta)  a(\eta')^2  K(k_{1},\eta) G(k_{1}, \eta, \eta') K(k_{2},\eta') K(k_{3},\eta')\,,
\end{align} 
where we have used momentum conservation to write the internal energy as $k_{1}$, and have included an overall $-i$ as dictated by the Feynman rules. We will use (see e.g.~\cite{MLT})
\begin{align}
K(k,\eta) &= (1 - ik \eta)e^{i k \eta}\,\,,\\
G(k, \eta, \eta') &= 2 P(k)[\theta(\eta - \eta') K(k,\eta') \text{Im} K(k,\eta) + (\eta \leftrightarrow \eta')]\,,
\end{align}
where $P(k)$ is the GR power spectrum arising from the usual massless mode functions. We can compute this integral exactly and while the full expression is not very illuminating, we find that the result is \textit{purely imaginary}. The result of this integral is then multiplied by a polarisation factor which is \textit{purely real}, as we discussed in Section \ref{ParitySection}, so it follows that the contribution to the wavefunction from this digram is a pure phase, so it does not contribute to the bispectrum. It is interesting to note that this diagram contributes logarithms to the wavefunction which by unitarity have to come with $i \pi$ contributions too \cite{COT}.  As we explained in Section \ref{ParitySection}, for contact diagrams the only logarithmic divergences are of the form $\log(-k_{T}\eta_{0})$, so $i \pi$ contributions are inevitable. However, for this three-point exchange diagram we find a number of logarithmic terms with different arguments and it turns out that the coefficients of these logs are such that all $i \pi$ contributions cancel out! This observation clearly deserves further attention and we plan to come back it in the near future. \\ 

\noindent Although the contribution of Figure \ref{fig:3ptexchangediagram} to $\psi_{3}$ is pure phase, this modification of the quadratic action still leads to a parity-odd correction to the bispectrum, as we will now show\footnote{We thank Aaron Hillman for discussions about this point.}. First, we write the full $\psi_2$ as $\psi_2^{(0)}+ \delta \psi_2$, i.e. the leading part from GR plus a small correction due to the parity-odd quadratic term. To linear order in $1  / \Lambda$, the relevant contributions to the wavefunction for computing the bispectra are 
\begin{equation}
\begin{split}
\Psi[\gamma, \eta_0] = \exp \bigg\{& {-\dfrac{1}{2}} \int_{\bfk} \sum_{\lambda} \left( \psi_2^{(0)}(k)  +  \delta \psi_2^{\lambda}(k)   \right) \gamma^{\lambda}_{\k} \gamma^{\lambda}_{-\k} \\
&-\dfrac{1}{3!} \int_{\bfk_1,\bfk_2,\bfk_3} (2 \pi)^3 \delta^{(3)} \left(\sum \k_i \right)  \sum\limits_{\{\mu_i\}} \psi_3^{\mu_1 \mu_2 \mu_3}(\{ k_i \} , \{ \k_i \} )  \gamma^{\mu_1}_{\k_1} \gamma^{\mu_2}_{\k_2} \gamma^{\mu_3}_{\k_3}
+ \ldots \Big\}\,,
\end{split}
\end{equation}
where $\psi_{3}$ contains all contributions from GR and our parity-breaking CS term, and we 
have used the fact that due to $SO(3)$ invariance helicities do not mix at quadratic order. To linear order in $1 / \Lambda$ we then have
\be
\label{FullBispectrum}
\begin{split}
B_{3} = \frac{1}{\prod\limits_{i=1}^3\mathcal{P}_2^{(0)}(k_i)} \left( - \mathcal{P}_3^{ \{ \lambda_i \}}( \{ \k_i \})  +  \mathcal{P}_3^{\{ \lambda_i \}}(\{ \k_i \})\bigg(\frac{ \delta \mathcal{P}_2^{\lambda_1}(k_1) }{\mathcal{P}_2^{(0)}(k_1)}
 + 2 \  \text{permutations}\bigg)   \right)\,, 
\end{split}
\ee
where the permutations are of both momenta and helicity labels, and we have defined 
\begin{equation}
\mathcal{P}_n^{ \{ \mu_i \} }(\{ k_i \}, \{ \bfk_i \}) = \psi_n^{\{ \mu_i \}}(\{ k_i \}, \{ \bfk_i \})  + \psi_n^{\{ \mu_i \}}(\{ k_i \}, \{ - \bfk_i \})^*, 
\end{equation} 
for $n\geq 3$, while for $n=2$ we use $SO(3)$ invariance to simplify the definition of $\mathcal{P}^\lambda_2(k)$ as $\psi_2^{\lambda}(k) + \psi_2^{\lambda}(k)^\ast$. 
As we explained above, the contribution to $\psi_{3}$ from Figure \ref{fig:3ptexchangediagram} drops out of $\mathcal{P}_3$; thus the only parity-odd contribution to $\mathcal{P}_3$ is fixed by the contact interactions in \eqref{3dcs_action} and the contributions of these interactions are given by \eqref{EFToIContact+++} and \eqref{EFToIContact++-}. However, $\delta \mathcal{P}_2$ is non-zero. The expressions that we now need to compute the full bispectra are 
\be
\mathcal{P}_2^{\pm}(k) = \frac{M_{\text{pl}}^2}{H^2}k^3\bigg(1\mp\frac{\pi H}{\Lambda}\bigg) 
\ee
and
\begin{align}
\mathcal{P}_{3,\text{GR}}^{+++}( \{ \bfk_i \}) &=  \frac{M_{\text{pl}}^2}{4 H^2} \text{SH}_{+++}(e_{3}+k_{T}e_{2}-k_{T}^3)\,, \\
\mathcal{P}_{3, \text{GR}}^{++-}( \{ \bfk_i \}) &= \frac{M_{\text{pl}}^2}{4 H^2} \text{SH}_{++-} \frac{ I_{3}^2}{k_{T}^2}(e_{3}+k_{T}e_{2}-k_{T}^3)\,,
\end{align}
which we have computed from the cubic Einstein-Hilbert action using the bulk formalism. 
It follows that the \textit{full} parity-odd contributions to the bispecta at $\mathcal{O}(1 / \Lambda)$, by summing all terms in \eqref{FullBispectrum} with 
those in Eqs.~\eqref{EFToIContact+++}, \eqref{EFToIContact++-}, are given by
\begin{align} 
e_{3}^3B_{\text{CS,total}}^{+++} &= {- \frac{\pi}{4}\frac{H}{\Lambda}\frac{H^4}{M_{\text{pl}}^4}} \text{SH}_{+++} \Big(2k_T(k_T^2-2e_2) - (-3e_3+k_Te_2) -12I_1I_2I_3 
- 3(k_T^3 - e_2k_T - e_3)\Big) \,, \label{EFToIFULL+++} \\
e_{3}^3B_{\text{CS,total}}^{++-} &=  {- \frac{\pi}{4}\frac{H}{\Lambda}\frac{H^4}{M_{\text{pl}}^4}} \text{SH}_{++-} \bigg(2I_3(k_T^2-2e_2)- \big(k_1 (k_2^2 + k_3^2) + k_2(k_1^2 + k_3^2) - k_3 (k_1^2 + k_2^2)\big) \nonumber \\
&\hphantom{{- \frac{\pi}{4}\frac{H}{\Lambda}\frac{H^4}{M_{\text{pl}}^4}} \text{SH}_{++-} \Big(\;\;\;\;} -12I_1I_2k_T + \frac{ I_{3}^2}{k_{T}^2}(e_{3}+k_{T}e_{2}-k_{T}^3)\bigg)\,. \label{EFToIFULL++-}
\end{align} 
Again, we see that these parity-odd corrections are suppressed by $H / \Lambda$ compared to the GR contributions. 
Using the relation $P^{\pm}(k) = 1/{\cal P}^{\pm}(k)$, it is straightforward to check that these bispectra satisfy the consistency condition for large wavelength gravitons, i.e.~ $\smash{\braket{\gamma^{h_S}_{\mathbf{k}-\mathbf{q}/2}\gamma^{h_S}_{-\mathbf{k}-\mathbf{q}/2}\gamma^{h_L}_{\mathbf{q}}}' \sim \frac{3}{2}P^{h_L}(q)P^{h_S}(k)e^{h_L}_{ij}(\mathbf{q})\hat{k}^i\hat{k}^j}$ for $q/k \to 0$. \\

\noindent To conclude, many of the bispectra we have computed in Section \ref{sec:bispectra} do indeed arise in the EFToI, without corrections to the quadratic theory. The parity-odd contact bispectra, however, necessarily come with a correction to the two-point function that can be treated perturbatively and results in a \textit{total} parity-odd contribution to the bispectrum in the EFToI given in \eqref{EFToIFULL+++} and \eqref{EFToIFULL++-}. Although it is very interesting that all parity-odd corrections can be computed, they are suppressed relative to the GR contribution and will therefore be very difficult to detect observationally. This suppression was also noted in \cite{Bartolo:2017szm}. In this paper we have restricted ourselves to exact scale invariance, and away from this limit other shapes are possible \cite{Bartolo:2020gsh}.  As we mentioned above, it would be very interesting to pick out this EFToI subset of our full catalogue directly at the level of the correlator rather than going back to the Lagrangian. We hope to return to this in the future. \\

\noindent In Section \ref{AddingScalars} we showed that \eqref{SSTbispectrum} is the unique scale invariant, manifestly-local and parity-odd bispectrum of two scalars and a graviton. Here we will argue that this bispectrum \textit{does not} appear in the EFToI. For the reader's convenience the structure of this bispectrum is 
\begin{align} 
e_{3}^3B_{3}^{00+} = h_{3,3} \frac{[13]^2 [23]^2}{k_{3}^2 [12]^2} I_{3}^2 k_{3}\,,
\end{align} 
and the relevant polarisation factor is 
\begin{align} \label{SSTPF}
\epsilon_{ijk}  e^{h_{3}}_{im}(\bfk_{3}) k_{1}^{j} k_{2}^{k} k_{1}^{m}\,,
\end{align} 
up to permutations. Let's first ask if such a tensor structure can appear in the EFToI without a correction to the graviton's two-point function. In this case the corresponding operator must be built out of the building blocks we listed above. Since such an interaction must have three spatial derivatives (simply from the form of the polarisation factor) and no time derivatives (to ensure $  2 n_{\partial_{\eta}} + n_{\partial_{i}} \leq 3 $ and therefore a non-vanishing bispectrum) the only possible building blocks contain two derivatives acting on the graviton. However, it is easy to see from the structure of \eqref{SSTPF} that there is no way for two of the $k's$ to correspond to the graviton. This implies that if such a bispectrum is to appear in the EFToI, it should come with a correction to the graviton's two-point function. \\

\noindent The leading correction to the graviton's two-point function is the one we discussed above, namely \eqref{eq:3dcs_quadratic_action} which, as dictated by symmetries, appears in the EFToI in the form of the Chern-Simons term \eqref{CSterm}, see also \cite{Creminelli:2014wna}. We have checked that to leading order in slow-roll this Chern-Simons term does not contain a three-derivative scalar-scalar-graviton interaction and therefore cannot produce this bispectrum. We therefore conclude that \eqref{SSTbispectrum} does not appear in the EFToI. We note that in \cite{Bartolo:2017szm} a scalar-scalar-graviton interaction with four-derivatives was derived from this Chern-Simons term. This interaction is slow-roll suppressed, but given our discussion in Section \ref{ParitySection} it also has too many derivatives to produce a log in the corresponding wavefunction and so the corresponding bispectrum is zero (rather than simply small).

\subsection{Solid inflation} \label{SolidInflation}

Let us now switch to the symmetry breaking pattern of solid inflation \cite{SolidInflation}. Here the symmetry breaking is driven by a multiplet of scalar fields that pick up \textit{spatial} vevs. Internal symmetries of the scalars then ensure that the background geometry is homogeneous and isotropic. In stark contrast to the EFToI, the fluctuations in solid inflation break spatial diffeomorphisms, and yield the following effective field theory description \cite{ZoologyNG}. \\

\noindent The building blocks in unitary gauge 
are constructed from the one-forms $\partial_\mu x^i$, with the requirement that latin indices are contracted in an $SO(3)$-invariant way. 
An important role is played by the trace 
\be
\label{X_def}
X=g^{ii}\,,
\ee
which is a proxy for time, and by the four-vector 
\begin{equation}
\label{eq:solid-1}
O^\mu = \frac{\bm{e}^{\mu\nu\rho\sigma}\epsilon_{ijk}\partial_\nu x^i\partial_\rho x^j\partial_\sigma x^k}{6\sqrt{\det(g^{mn})}}\,, 
\end{equation} 
which we can use to take time derivatives of diffeomorphism scalars (via $O^\mu\nabla_\mu$). Then, we can take spatial derivatives of diffeomorphism scalars via 
\begin{equation}
\label{eq:solid-9}
D^\perp_i = \frac{g^{i\mu}\nabla_\mu}{\sqrt{X/3}}\,. 
\end{equation}
This reduces to $\partial_i/a$ at zeroth order in perturbations. The last ingredient is the $SO(3)$ tensor 
\be
\label{eq:solid-8}
\Gamma^{ij} = \delta_{ij} - \frac{3g^{ij}}{X}\,,
\ee
which is equal to $\gamma_{ij}$ at leading order in perturbations. \\

\noindent With these building blocks it is then possible to write all possible manifestly-local cubic operators involving three gravitons. These are always built from an object of the form 
\begin{equation} 
\label{eq:solid-14} 
({D^\perp_{i_1}\cdots D^\perp_{i_{\alpha_1}}\Gamma^{il}}) ({D^\perp_{j_1}\cdots D^\perp_{j_{\alpha_2}}\Gamma^{jm}}) ({D^\perp_{k_1}\cdots D^\perp_{k_{\alpha_3}}\Gamma^{kn}})\,, 
\end{equation} 
where indices are contracted with $\delta_{ij}$ or $\epsilon_{ijk}$. 
Contracting the indices $i_1,\dots i_{\alpha_1}$, $j_1,\dots j_{\alpha_1}$, $k_1,\dots k_{\alpha_1}$ allows us to obtain all the tensor structures 
discussed in Section~\ref{subsec:tensor_structures} and Appendix~\ref{app:contractions}. 
Adding time derivatives to $\Gamma^{il}$, $\Gamma^{jm}$, $\Gamma^{kn}$ only changes $\psi^{\rm trimmed}_3(k_1,k_2,k_3)$ and is always allowed in this solid inflation EFT. So \textit{all of the interactions we have considered in Section \ref{sec:bispectra} can arise in solid inflation}. As it is clear from our bootstrap approach, there are a number of degeneracies at the level of the action that do not appear when working directly with observables. However, as an example, one possible set of interactions in solid inflation that can give rise to the parity-odd bispectra that we derived in Section \ref{FinalForm} are
\begin{align}\label{10}
g_{1,1} &:\quad \int d^4x\,\sqrt{-g}\,\epsilon_{ijk}\Gamma^{kn}\Gamma^{nm}D^\perp_j\Gamma^{im}\,, \\
g_{1,2} &:\quad \int d^4x\,\sqrt{-g}\,\epsilon_{ijk}\Gamma^{kn}O^\mu\nabla_\mu{\Gamma}^{nm}D^\perp_j\Gamma^{im}\,, \label{with_time_derivative} \\ 
g_{3,3} &:\quad \int d^4x\,\sqrt{-g}\,\epsilon_{ijk}D^\perp_m\Gamma^{lj}D^\perp_n\Gamma^{mi}D^\perp_l\Gamma^{nk} \,,\label{12}
\end{align}  
where $O^\mu$ is defined in Eq.~\eqref{eq:solid-1}. In addition to parity-odd graviton bispectra, each of these operators also generates mixed bispectra containing gravitons and scalars, such as for example scalar-scalar-graviton and scalar-graviton-graviton bispectra. No purely scalar bispectra are generated because scalar bispectra cannot be parity-odd. To see why the scalars must enter the game, notice that to leading order in scalar and tensor perturbations we have
\begin{align}\label{nml}
\Gamma^{ij}=\gamma_{ij}+2\zeta \delta_{ij}-6\frac{\partial_{i}\partial_{j}}{\partial^{2}}\zeta\,,
\end{align}
where $ \zeta $ are curvature perturbations on constant-energy time slices. Furthermore, this expression makes it clear that the interactions involving the scalar $  \zeta $ are \textit{not} manifestly local due to the appearance of the inverse Laplacian in the last term in \eqref{nml}, a fact that we have verified with an explicit computation. Hence, the mixed scalar-scalar-graviton and scalar-graviton-graviton bispectra generated by these operators are not the ones we derived in Section \ref{AddingScalars}, where we used the Manifestly Local Test (MLT) and therefore described only manifestly local interactions. In Section \ref{ssec:S2N} we will see that, in solid inflation, the signal-to-noise ratio for the bispectra involving one or more scalar is always larger than that for the purely graviton bispectrum. Therefore, if the parity-odd graviton bispectra derived here were to be detected, then either one should also see the corresponding parity-odd mixed bispectra or one would conclude that the symmetry breaking pattern during inflation is different from that assumed in solid inflation (and the EFT of inflation). \\

\noindent Since in Section~\ref{AddingScalars} we have bootstrapped \emph{only} the manifestly local parity-odd scalar-scalar-graviton bispectrum, it is natural to ask whether that mixed bispectrum can be generated in solid inflation. It is straightforward to see that the answer is yes. 
Let us consider the operator 
\be\label{SOoperator}
\int d^4x\,\sqrt{-g}\,X^{-2}\epsilon_{ijk}D^\perp_lXD^\perp_kXD^\perp_j\Gamma^{li}\,,
\ee
which starts at cubic order in perturbations. Using $X = 3a^{-2}(1+2\zeta)$, 
together with Eq.~\eqref{nml}, we see that to lowest order it is equal to 
\be
\label{00plusoperator}
4\int d \eta d^3 x \,a(\eta) \, \epsilon_{ijk} \partial_l\zeta\partial_k\zeta\partial_j\gamma_{li}\,.
\ee
Furthermore, the operator in \eqref{SOoperator} does not introduce any other mixed interaction. In particular, it does not generate terms with two gravitons and one $\zeta$, with three gravitons (given that $X$ does not contain $\gamma_{ij}$), or with three scalars (they all vanish by integration by parts). The interaction in Eq.~\eqref{00plusoperator} gives the bispectrum of Eq.~\eqref{SSTbispectrum}. Since the coupling constant of this operator can be large, we conclude that the parity-odd scalar-scalar-graviton bispectrum that we have bootstrapped can indeed arise and be large in solid inflation. We plot its shape in Figure \ref{fig:SST}.


\paragraph{Mode functions} Let us now discuss our assumption that the mode functions for the graviton and scalar are the usual massless de Sitter ones. Let us start with the graviton. As one may have anticipated from the breaking of spatial diffeomorphisms, in solid inflation the graviton acquires a mass. Indeed, in unitary gauge solid inflation can be thought of as a theory of Lorentz-violating massive gravity \cite{Phases}. On the surface this seems problematic for our assumptions, but it turns out that this mass is slow-roll suppressed. Let us quickly review how this happens (we refer the reader to \cite{SolidInflation} for more details.) We write the non-Einstein-Hilbert part of the action $S$ as 
\begin{equation}
\label{eq:solid-2}
S = \int d^4x\,\sqrt{-g}\,\big\{{\cal L}_0(X) + M^4(X,\delta Y,\delta Z)\big\}\,,
\end{equation} 
where we defined 
\label{eq:solid-3}
\begin{align}
\delta Y &= Y - \frac{1}{3} = \frac{g^{ij}g^{ji}}{X^2} - \frac{1}{3}\,, \label{eq:solid-3-2} \\ 
\delta Z &= Z - \frac{1}{9} = \frac{g^{ij}g^{jk}g^{ki}}{X^3} - \frac{1}{9}\,, \label{eq:solid-3-3}
\end{align}
and $X = g^{ii}$ was defined in Eq.~\eqref{X_def}. Without loss of generality, $M^4$ can be written as an expansion in powers of $\delta Y$ and $\delta Z$, each multiplied by a function of $X$. 
This mimics the expansion in powers of $g^{00}+1$ (with time-dependent coefficients) of the EFToI 
action at zeroth order in derivatives, thanks to the fact that $\delta Y$ and $\delta Z$ start at second order 
in perturbations around a FLRW background. Consequently, it is only ${\cal L}_0$ whose 
dependence is fixed by the background evolution: we have 
\label{eq:solid-4}
\begin{align}
3M^2_{\rm pl}H^2 &= {-{\cal L}_0}\,, \label{eq:solid-4-1} \\
\varepsilon\equiv{-\frac{\dot{H}}{H^2}} &= \frac{d\log{\cal L}_0}{d\log X}\,, \label{eq:solid-4-2}
\end{align}
where we recognize the slow-roll parameter $\varepsilon$ on the left-hand side of \eqref{eq:solid-4-2}. What are the contributions of \EQ{solid-2} to the graviton action (which are added to those from Einstein gravity)? It is straightforward to see that 
\begin{equation}
\label{eq:solid-5}
S_{\gamma\gamma} = \int{d}^4x\,a^3\bigg(\frac{1}{6}\frac{{d}{\cal L}_0}{{d}\log X} 
+ \frac{1}{9}\frac{\partial M^4}{\partial Y} + \frac{1}{9}\frac{\partial M^4}{\partial Z}\bigg)\gamma_{ij}\gamma_{ji}\,.
\end{equation}
Using \EQ{solid-4-2}, together with the fact that the propagation speed $c^2_{\rm T}$ 
of the transverse part of the Goldstone modes $\pi^i$ is \cite{SolidInflation} 
\begin{equation}
\label{eq:solid-5-bis}
c^2_{\rm T} = 1 + \frac{2}{3}\frac{M^4_{,Y} + M^4_{,Z}}{X{\cal L}_{0,X}}\,,
\end{equation} 
we see that the graviton has a small mass given by
\begin{equation}
\label{eq:solid-6}
m^2_\gamma = {-2}\dot{H}c^2_{\rm T}=2H^{2}\varepsilon c^2_{\rm T} \,. 
\end{equation} 
One might ask what happens if other (higher-derivative) operators
are turned on. No other operator 
can contribute to the mass term aside from Eq.~\eqref{eq:solid-5}. They could, however, modify Eq.~\eqref{eq:solid-4-2} 
and therefore modify how one converts from Eq.~\eqref{eq:solid-5} to Eq.~\eqref{eq:solid-6} via the definition of the 
speed of sound $c^2_{\rm T}$. As long as these operators are only small perturbative corrections to the solid inflation 
action of Eq.~\eqref{eq:solid-2}, the graviton mass will still be given by Eq.~\eqref{eq:solid-6} at leading order, and thus it will be slow-roll-suppressed. Therefore, to leading order in slow-roll one can use the massless dS mode functions to compute the bispectra, as we have been doing and as was done in \cite{ZoologyNG,Endlich:2013jia}. In the scale invariant limit, it then follows that \textit{all} of the parity-even and parity-odd bispectra we have constructed in this paper using the MLT are the leading contributions to the graviton bispectra in solid inflation.  \\

\noindent Let us now turn to the scalar's mode functions. We have shown that the unique parity-odd $B_{3}^{00+}$ can indeed arise from an interaction in solid inflation under the assumptions we have made in this paper which translate into constraints from the COT, MLT and boostless bootstrap rules that enable us to write down an ansatz for the wavefunction. It turns out that in solid inflation the mode functions for $\zeta$ are not the usual ones for a massless scalar in dS \cite{SolidInflation}, yet the COT, MLT and our ansatz still apply and so this unique bispectrum can indeed be generated when using the corrected mode functions for $\zeta$. \\

\noindent In more detail, it was shown in \cite{SolidInflation} that in the slow-roll limit the $\zeta$ mode functions contain an extra term relative to the usual ones for a massless scalar in dS which corrects the $\zeta$ bulk-boundary propagator to
\begin{align} \label{zetaB2B}
K_{\zeta}(k, \eta) \sim (1 - i c_{L} k \eta + c_{L}^2 k^2 \eta^2 / 3 )e^{i c_{L} k \eta}\,,
\end{align}
where we have omitted overall constant factors\footnote{We note that $\zeta$ and $\gamma$ are not conserved in solid inflation which induces time dependence in both the power spectra and bispectra of these modes. However, in the slow-roll limit this time dependence is small and to capture it one needs to keep slow-roll corrections in the mode functions. Importantly for us, the shapes of solid inflation bispectra are unaffected by this mild time dependence. We refer the reader to \cite{SolidInflation} for details.}. Now, one can easily verify that this new bulk-boundary propagator satisfies \eqref{BulkBoundaryCOT} and any contact diagrams that we derive using this propagator satisfy the contact COT \eqref{COT}. One can also see that this propagator will lead to wavefunction coefficients that satisfy the MLT. Indeed, the first derivative of \eqref{zetaB2B} vanishes at $k=0$. Finally, wavefunction coefficients due to manifestly-local interactions that are derived in the bulk formalism using this bulk-boundary propagator still take the form we assume in this paper: the energy dependence corresponds to rational functions, with only total-energy poles, with the additional possibility of $\log(- k_{T} \eta_{0})$ terms multiplied by polynomials. This can be easily seen from the fact that any integrand in the bulk formalism that is a function of \eqref{zetaB2B} and its derivatives, can also be written in terms of the usual expression for $K(k,\eta)$ and its derivatives. To capture the effects of the $k^2 \eta^2$ correction, we would need to include terms with extra time derivatives which will in turn result in wavefunction coefficients where the degree of the leading total-energy pole will not equal the number of derivatives in the corresponding bulk interaction. Indeed, we have
\begin{align} \label{fromzetatophi}
K_{\zeta}(k, \eta) = K(k, \eta) - \frac{\eta}{3} \frac{\partial K(k, \eta)}{\partial \eta}\,.
\end{align} 

\noindent For our interests in this paper, the important point is that the interaction in \eqref{00plusoperator} still generates a logarithm in the wavefunction. Indeed, given that the relevant interaction vertex has $\alpha=3$, terms coming from the second term on the RHS of \eqref{fromzetatophi} will violate $2 n_{\partial_{\eta}} + n_{\partial_{i}} \leq 3 $ and so will not alter the coefficient of the log divergence coming from three copies of $K(k, \eta)$. It follows that \eqref{SSTbispectrum} does indeed arise as a $\zeta$-correlator in solid inflation.

\subsection{Phenomenology of parity-odd interactions in solid inflation} \label{Pheno}

\begin{wrapfigure}[17]{r}{.47\linewidth}
\begin{center}
  \includegraphics[width=0.9\linewidth]{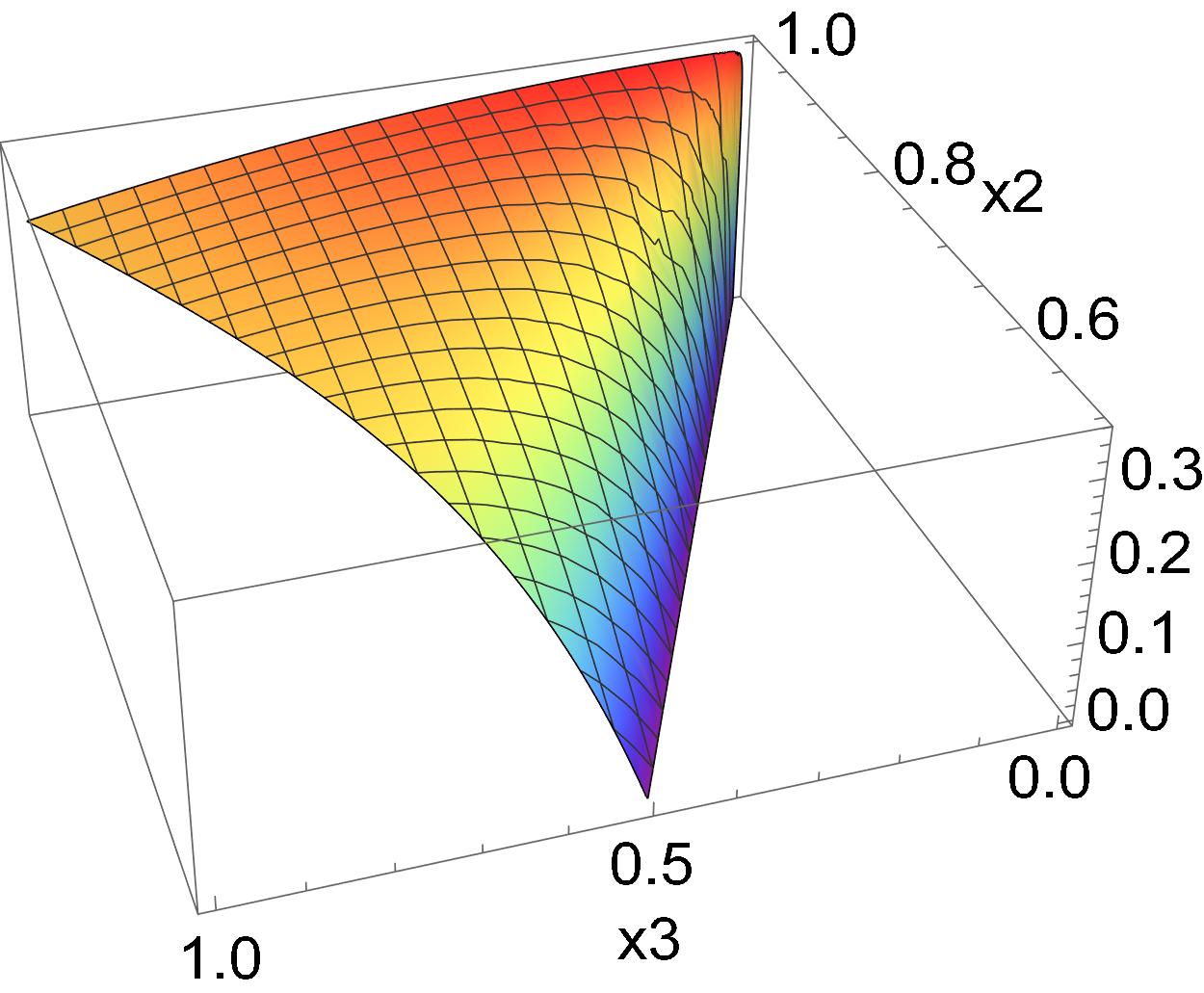}  
  \caption{The shape of the unique parity odd, manifestly local $B^{00+}_3$.}
    \label{fig:SST}
  \end{center}
\end{wrapfigure}

In the EFToI, we showed above that the parity-odd correction to the \textit{graviton} bispectrum is small relative to the GR contribution since in our analysis we took the correction to the quadratic action to be perturbative. Furthermore, the mixed scalar-scalar-graviton parity-odd bispectrum that we found in Section \ref{AddingScalars} does not arise in the EFToI. In solid inflation, however, we can choose operators that do not affect the quadratic action and can in principle give rise to large parity-odd bispectra: see \eqref{10} - \eqref{12} for gravitons (which do also introduce three-point functions involving the curvature perturbation $\zeta$) and \eqref{SOoperator} for scalar-scalar-graviton (which is the only three-point function arising from that operator). The reason we focus on these three-point functions is two-fold. On the one hand, if parity-odd non-Gaussianities involving scalars are suppressed in some model of inflation beyond the ones we discuss here, then graviton-graviton-graviton bispectra will be the leading signal. On the other hand, in the generic case where non-Gaussianities involving scalars are not suppressed (as is the case in EFToI and solid inflation), then the scalar-scalar-graviton signal will be the leading one, as we show in Section \ref{ssec:S2N}. In this short section, we therefore study the phenomenology of the parity-odd, \textit{manifestly local} graviton-graviton-graviton and scalar-scalar-graviton bispectra in more detail by plotting and commenting on the shapes of each possibility that we presented in Sections \ref{FinalForm} and \ref{AddingScalars}.  \\

\noindent Let us start with the unique scalar-scalar-graviton bispectrum, which is
\begin{align}
(\alpha=3, \  p=3): & \quad B^{00+}_{3} = h_{3,3} \frac{1}{e_3^3} \frac{[23]^2 [31]^2}{[12]^2 k_3} I_3^2 \,.
\end{align}
Taking $\bfk_3 = k_3 \hat{\bf{z}}$ without loss of generality, the bispectrum can be rewritten in terms of the graviton polarisation tensor
\begin{align}
e^{\pm}(\bfk_{3}) & = \frac{1}{\sqrt{2}} \begin{pmatrix}
0 & 0 & 0 \\
0 & 1 & \pm i \\
0 & \pm i & -1 
\end{pmatrix} \,,
\end{align}
as
\begin{align}
B^{00 h}_{3} = - 2 \lambda_h h_{3,3} \frac{1}{e_3^3}  e^h_{ij}(\k_3) k^{1}_{i} k^{2}_{j} k_3 = \lambda_h \frac{h_{3,3}}{2 \sqrt{2} e_3^3 k_{3}} k_{T}I_{1}I_{2}I_{3}  \,,
\end{align}
where $\lambda_{\pm} = \pm 1$. To see the shape of this bispectrum, in Figure \ref{fig:SST} we plot $B^{00+}_3 k_1^6$, which has a vanishing scaling dimension. The correlator vanishes in the folded limit and does not peak in the squeezed limit. \\

\noindent Let us now move to the three parity-odd graviton bispectra. For the convenience of the reader we recall that these are
\begin{align}  \nn
\alpha=1, \  p=1: & \quad B^{+++}_{3} = g_{1,1} \text{SH}_{+++} \frac{k_T \left( k_T^2 - 2 e_2 \right)}{e_3^3}\,, \\ \label{eq:ThreeOddBispectra1}
&  \quad B^{++-}_{3} = g_{1,1} \text{SH}_{++-}  \frac{I_3 \left( k_T^2 - 2 e_2 \right)}{e_3^3}\,, \\ \nn
\alpha=1, \  p=2: & \quad B^{+++}_{3} = g_{1,2} \text{SH}_{+++} \frac{-3 e_3 + k_T e_2}{e_3^3}\,, \\ 
\label{eq:ThreeOddBispectra2}
 & \quad  B^{++-}_{3} = g_{1,2} \text{SH}_{++-} \frac{k_1 (k_2^2 + k_3^2) + k_2(k_1^2 + k_3^2) - k_3 (k_1^2 + k_2^2)}{e_3^3}\,, \\ \nn
\alpha=3, \ p=3: & \quad B^{+++}_{3} = g_{3,3} \text{SH}_{+++} \frac{ I_1 I_2 I_3 }{e_3^3}\,, \\
\label{eq:ThreeOddBispectra3}
&  \quad B^{++-}_{3} = g_{3,3} \text{SH}_{++-} \frac{ I_1 I_2 k_T }{e_3^3}\,.
\end{align}

\noindent Now, for each of these bispectra the polarisation factor is unique and is fixed by the helicity transformations of the external spinors. In terms of polarisation tensors we have
\begin{align}
\text{SH}_{++ \pm} = - e^+_{ij}(\bfk_1) e^+_{jk}(\bfk_2) e^{\pm}_{ik}(\bfk_3)\,,
\end{align}
which we can express solely in terms of the energies $k_{1},k_{2},k_{3}$. Using momentum conservation and $SO(3)$ invariance, we can make each of the three external vectors lie in the $(x,y)$ plane with
\begin{align}
\bfk_{1} = k_{1}(1,0,0), \qquad \bfk_{2} = k_{2}(\cos \theta, \sin \theta ,0), \qquad \bfk_{3} = k_{3}(\cos \varphi, \sin \varphi,0),
\end{align}
where 
\begin{align}
\cos \theta = \frac{k_{3}^2 - k_{1}^2 - k_{2}^2}{2 k_{1}k_{2}}, \qquad \cos \varphi = \frac{k_{2}^2 - k_{3}^2 - k_{1}^2}{2 k_{1}k_{3}} \,.
\end{align}
The angles $\theta$ and $\varphi$ are simply those formed by $\bfk_{1}$ with $\bfk_{2}$ and $\bfk_{3}$ respectively. With this representation for $\bfk_{a}$ we can write the polarisation tensors as 
\begin{align}
e^{\pm}(\bfk_{1}) & = \frac{1}{\sqrt{2}} \begin{pmatrix}
0 & 0 & 0 \\
0 & 1 & \pm i \\
0 & \pm i & -1 
\end{pmatrix} \,, \\
e^{\pm}(\bfk_{2}) & = \frac{1}{\sqrt{2}}  \begin{pmatrix}
\sin^2 \theta & - \sin \theta \cos \theta & \mp i \sin \theta \\
- \sin \theta \cos \theta & \cos^2 \theta & \pm i \cos \theta \\
\mp i \sin \theta & \pm i \cos \theta & -1 
\end{pmatrix} \,, \\
e^{\pm}(\bfk_{3}) & = \frac{1}{\sqrt{2}}  \begin{pmatrix}
\sin^2 \varphi & - \sin \varphi \cos \varphi & \mp i \sin \varphi \\
- \sin \varphi \cos \varphi & \cos^2 \varphi & \pm i \cos \varphi \\
\mp i \sin \varphi & \pm i \cos \varphi & -1 
\end{pmatrix} \,.
\end{align}
It is then straightforward to see that 
\begin{align}
\text{SH}_{+++} =  -\frac{k_T^3 \left( 8 e_3 - 4 k_T e_2+ k_T^3 \right) }{16 \sqrt{2} e_3^2}\,, \\
\text{SH}_{++-} =  -\frac{I_3^3 \left( - 8 e_3 - 4 I_3 e'_2+ I_3^3 \right) }{16 \sqrt{2} e_3^2}\,.
\end{align}
Note that, perhaps surprisingly, these expressions are purely rational. Here we have defined $e'_{2}$ which is simply $e_{2}$ with the sign of $k_{3}$ flipped i.e. $e'_{2} = k_{1}k_{2} - (k_{1}+k_{2})k_{3}$. \\

\begin{figure}
\centering
\begin{subfigure}{.48\textwidth}
  \centering
  \includegraphics[width=.8\linewidth]{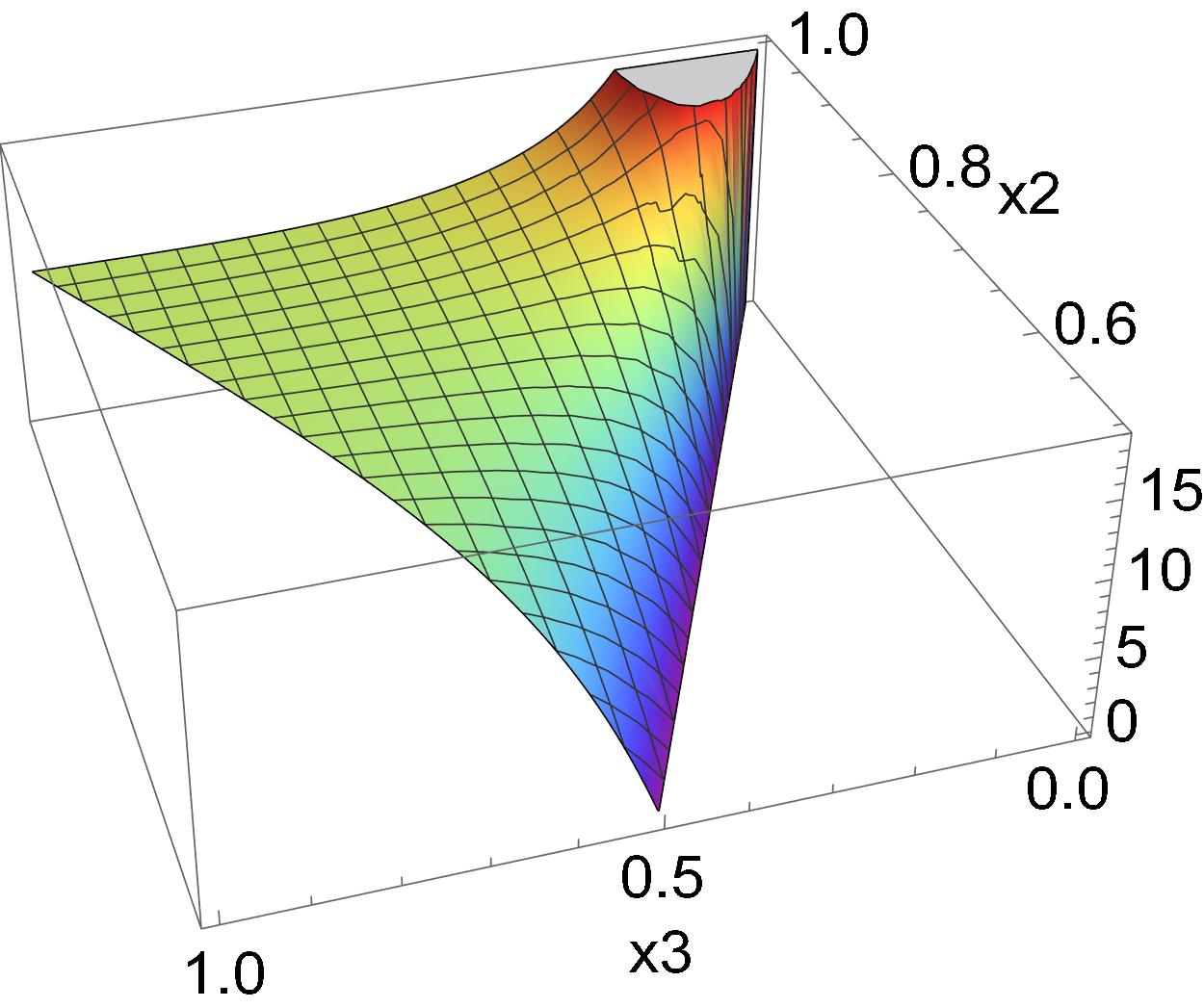}  
  \caption{The shape of $B^{+++}_3$ for $\alpha=1$, $p=1$.}
  \label{fig:sub-first}
\end{subfigure}
\begin{subfigure}{.48\textwidth}
  \centering
  \includegraphics[width=.8\linewidth]{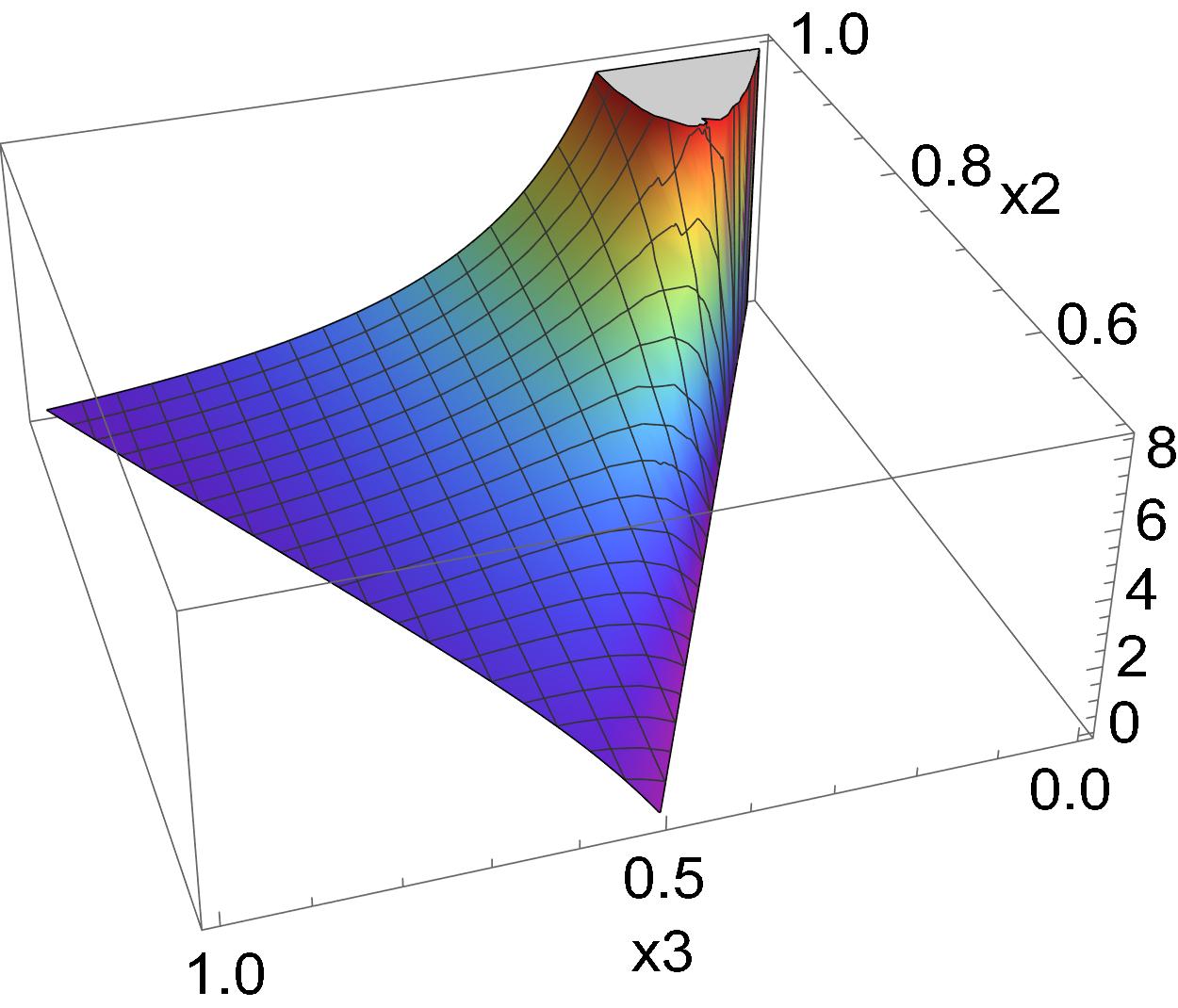}  
  \caption{The shape of $B^{++-}_3$ for $\alpha=1$, $p=1$.}
  \label{fig:sub-third}
\end{subfigure}  \\
\begin{subfigure}{.48\textwidth}
  \centering
  \includegraphics[width=.8\linewidth]{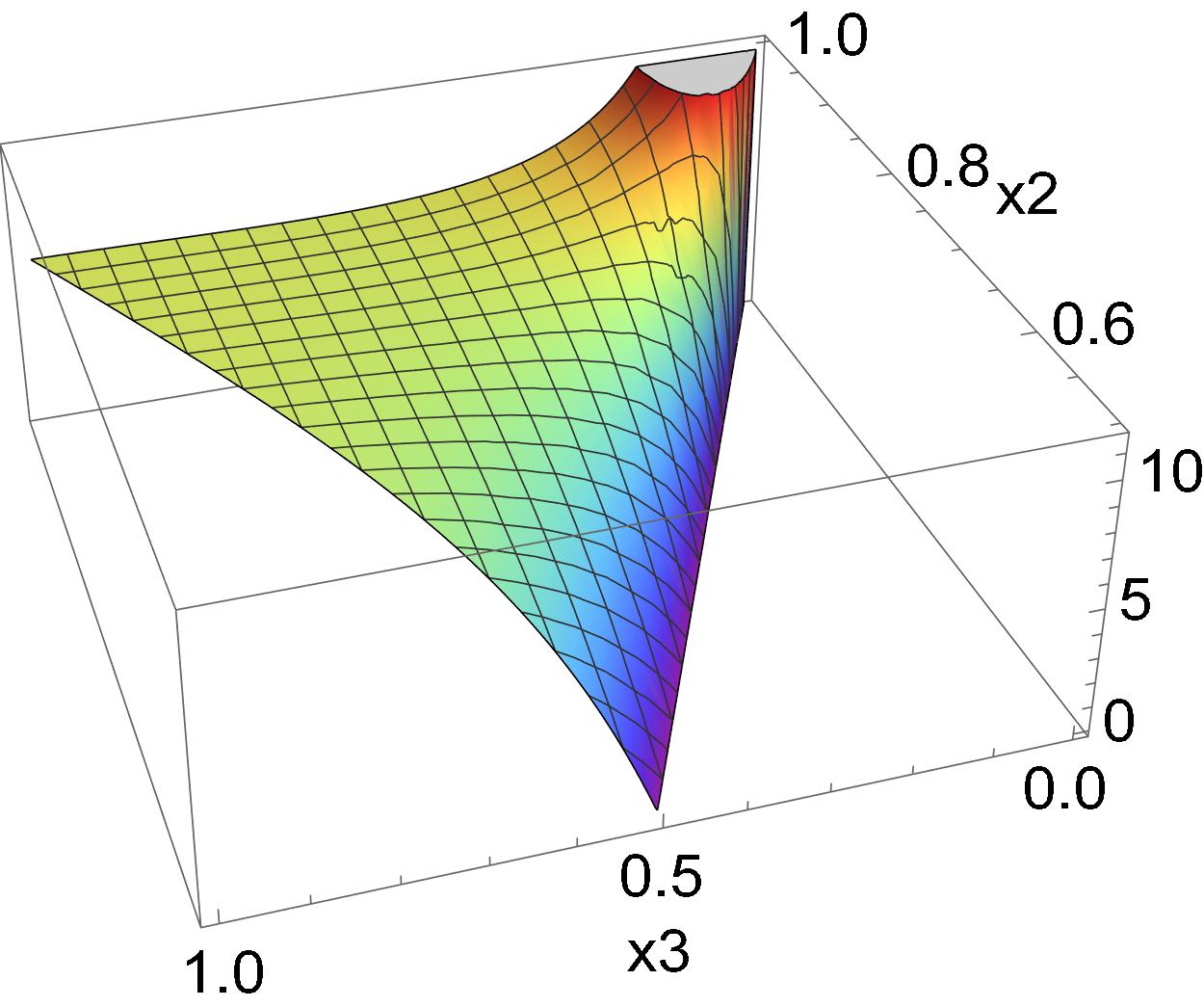}  
  \caption{The shape of $B^{+++}_3$ for $\alpha=1$, $p=2$.}
  \label{fig:sub-second}
\end{subfigure}
\begin{subfigure}{.48\textwidth}
  \centering
  \includegraphics[width=.8\linewidth]{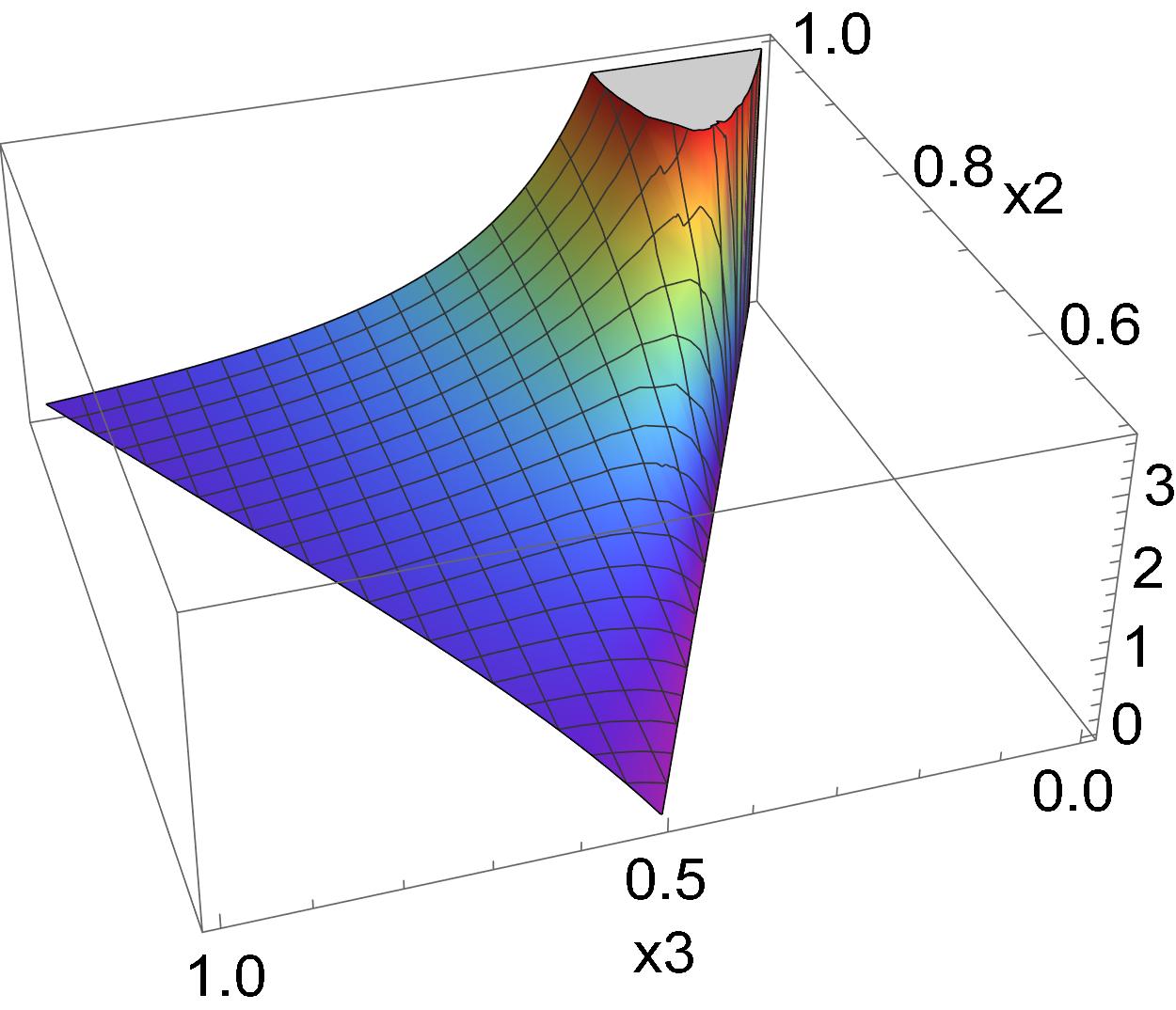}   
   \caption{The shape of $B^{++-}_3$ for $\alpha=1$, $p=2$.}
  \label{fig:sub-fourth}
\end{subfigure} \\
\begin{subfigure}{.48\textwidth}
  \centering
  \includegraphics[width=.8\linewidth]{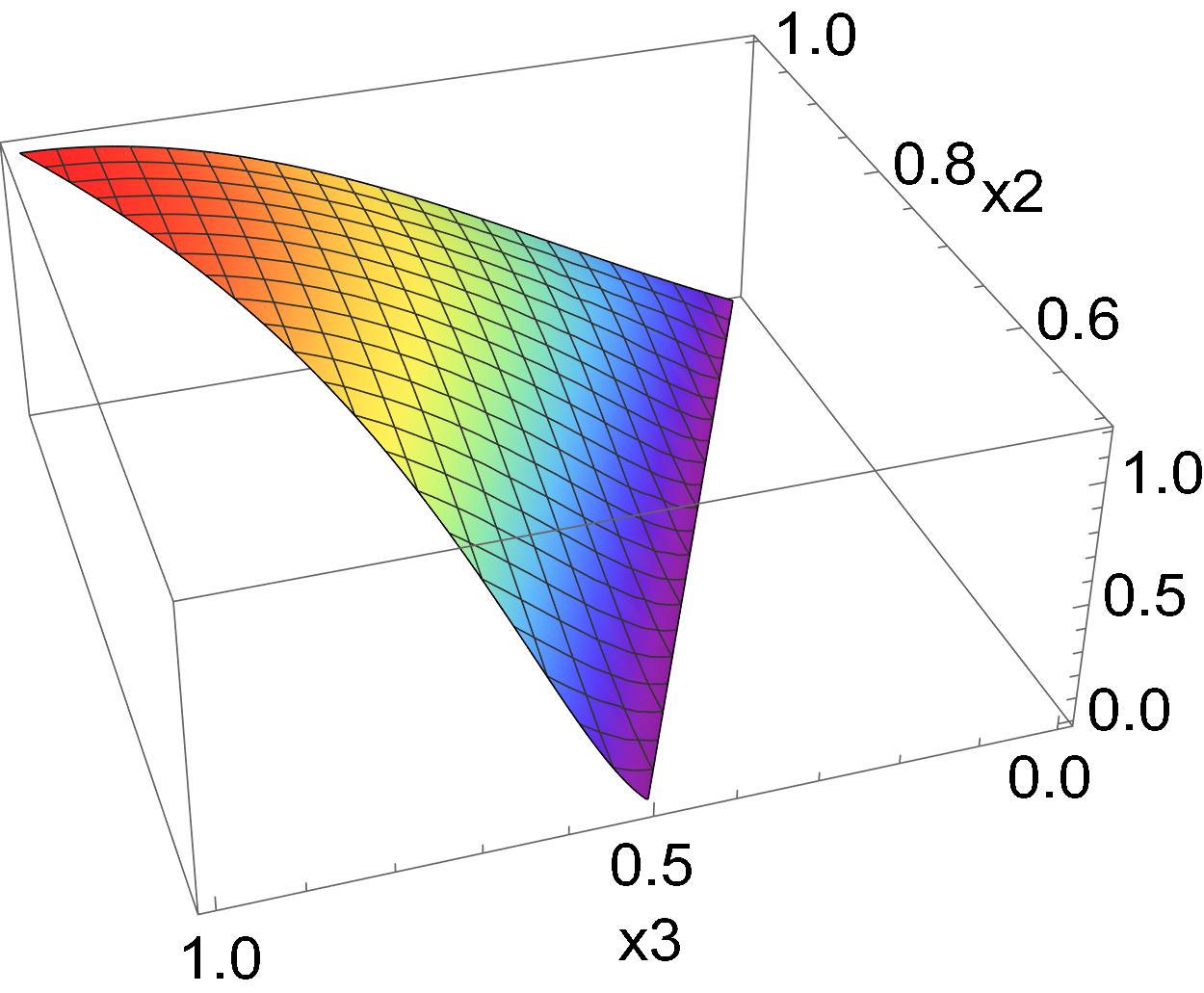}  
  \caption{The shape of $B^{+++}_3$ for $\alpha=3$, $p=3$.}
  \label{fig:sub-second}
\end{subfigure}
\begin{subfigure}{.48\textwidth}
  \centering
  \includegraphics[width=.8\linewidth]{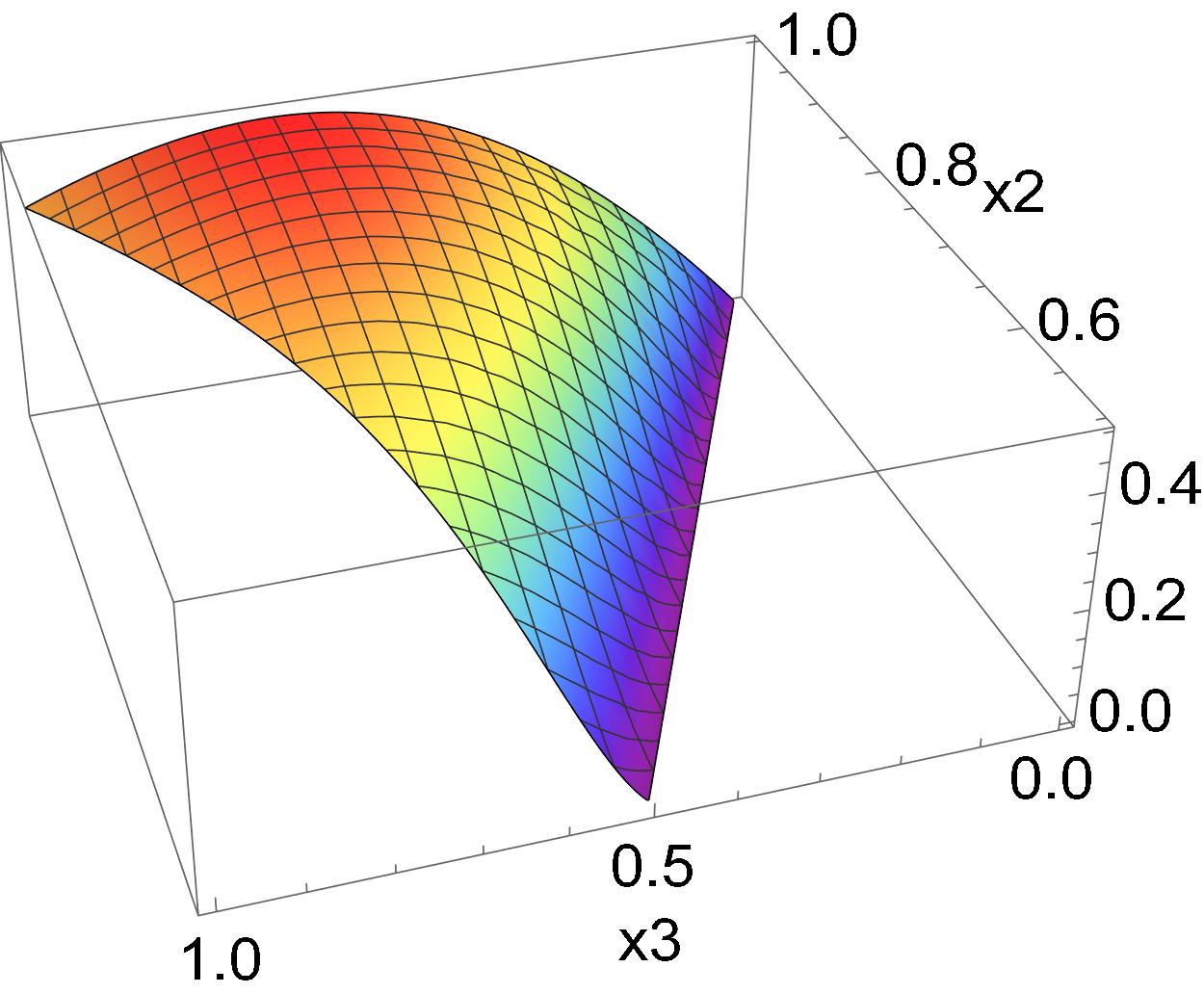}  
  \caption{The shape of $B^{++-}_3$ for $\alpha=3$, $p=3$.}
  \label{fig:sub-second}
\end{subfigure}
\caption{Shapes of each of the three tree-level, contact parity-odd bispectra consistent with the MLT.}
\label{fig:B3Shapes}
\end{figure}

\noindent To see the behaviour of these different shapes we plot $B^{++ \pm}_3 k_1^6$ for each of the three couplings. These combinations have vanishing scaling weight and can be written as  functions of the dimensionless parameters 
\begin{align}
x_2 \equiv \frac{k_2}{k_1}, \qquad x_3 \equiv \frac{k_3}{k_1} \,.
\end{align}
The shapes can be found in Figure \ref{fig:B3Shapes}. We see that both $\alpha=1$ parity-odd bispectra peak in the squeezed limit for all helicities, but have an angular dependence which causes them to vanish when all spatial momenta are parallel. More specifically, in the squeezed limit $k_3 \ll k_1, k_2$, all $\alpha=1$ bispectra are proportional to $\sin^2 \left( \measuredangle (\bfk_1, \bfk_3) \right)$. By contrast, the $\alpha=3$ parity-odd bispectrum vanishes in the squeezed limit for all helicities, and is large in the equilateral configuration. \\

\subsection{On the detectability of graviton and scalar bispectra}\label{ssec:S2N}

In this subsection we discuss the signal-to-noise ratio $  S/N $ for general bispectra. We use this analysis to argue that, since $  \left( S/N \right)^{2} $ scales with the power spectrum of the fields involved, it is larger for bispectra that contain more scalar fields, all other things being equal.  \\

\noindent So far we have seen that the three parity-odd graviton bispectra that we have bootstrapped to all orders in derivatives can indeed arise in solid inflation. Since there cannot be any parity-odd scalar bispectra, the graviton bispectra do not have any counterpart in the purely scalar sector and are therefore unconstrained by current data. In solid inflation they can appear with a large coefficient and should therefore be considered an important observational target for observations of the polarization of the Cosmic Microwave Background (CMB). It would be interesting search for these parity-odd graviton bispectra also with gravitational wave interferometers. Since both ground and space based interferometers probe scales that are very much shorter than cosmological scales, the possibility to detect a primordial stochastic background of gravitational waves in the conceivable future hinges on having a blue tilt in the tensor power spectrum. It is worth keeping in mind that such a blue tilt is at odds with the assumption of scale invariance that we have used extensively in this work.\\

\noindent The operators in \eqref{10}-\eqref{12} generate a parity-odd graviton bispectrum, but also scalar-scalar-graviton and scalar-graviton-graviton bispectra. It is therefore interesting to ask which of these signals can be seen first. To assess the theoretical detectability of a bispectrum we look at the signal-to-noise ratio. For concreteness and convenience, we assume that we can access the full three-dimensional distribution of the fields within a volume $  V  $ and up to a resolution of order $  k_{\text{max}}^{-1} $. Let us consider the following action for three massless fields $  \varphi_{a} $ with $  a=1,2,3 $, which can be scalars or gravitons,
\begin{align}\label{ints-bis}
S=\int d^{3}x d\eta \, a^{4} \left[ \sum_{a=1}^{3} \frac{\Delta_{a}^{2}}{2}\left( \partial_{\mu}\phi_{a} \right)^{2}+ g a^{-p} \partial^{p} \phi_{1}\phi_{2}\phi_{3}\right]\,,
\end{align}
where $  g $ is a coupling constant, $   \Delta_{a}$ is an arbitrary normalization, and we have schematically indicated that the interaction has $  p $ derivatives and therefore comes with the appropriate power of the scale factor required by scale invariance. The indices of the spatial derivatives can be contracted with the indices of the gravitons, with $  \delta_{ij} $ or with the anti-symmetric Levi-Civita symbol $  \e_{ijk} $, so that this discussion captures parity-odd interactions as well. For example, for the graviton $   \gamma_{ij} $ we would have $   \Delta_{\gamma}=\Mpl/2$ and for curvature perturbations $  \Delta_{\zeta}=\Mpl \sqrt{2\e} \ll \Delta_{\gamma} $. The power spectra are found to be
\begin{align}
\ex{\phi_{a}\phi_{a}}'=\frac{H^{2}}{2\Delta_{a}^{2}}\frac{1}{k^{3}}\equiv \frac{A_{a}}{k^{3}}\,,
\end{align}
where the prime indicates that we are dropping the factor $  (2\pi)^{3}\delta^{3}(\bfk) $. The bispectrum induced by the interactions in \eqref{ints-bis} in the in-in formalism takes the following schematic form
\begin{align}\label{Bqui}
\ex{\phi_{1}\phi_{2}\phi_{3}}'=B_{123}&=\int d\eta \ex{[H_{\text{int}},\phi_{1}\phi_{2}\phi_{3}]}\\
&\sim g H^{p-4} \left( \prod_{a=1}^{3}\frac{A_{a}}{k_{a}^{3}} \right) R_{3}(k_{1},k_{2},k_{3})\,,
\end{align}
where $  R_{3} $ is a rational function of the momenta that scales as $  k^{3} $, up to possible logarithmic terms. For parity-even interactions we expect $ R_{3}\sim \text{Poly}_{p+3}/ k_{T}^{p}$, while for parity-odd interactions we have proven that no $  k_{T} $ pole can arise and so $ R_{3}\sim \text{Poly}_{3} $. Notice that the bispectrum therefore scales as the power spectrum of each of the fields. Then, we define the dimensionless signal-to-noise ratio $  S/N $ as (see e.g. \cite{Behbahani:2011it})
\begin{align}
\left(  \frac{S}{N}\right)^{2}&=V^{3}\int_{\bfk_{1}\bfk_{2}\bfk_{3} }  \frac{\ex{\phi_{1}(\bfk_{1})\phi_{2}(\bfk_{2})\phi_{3}(\bfk_{3})}\ex{\phi_{1}(\bfk_{1})\phi_{2}(\bfk_{2})\phi_{3}(\bfk_{3})}}{\ex{\phi_{1}(\bfk_{1})\phi_{2}(\bfk_{2})\phi_{3}(\bfk_{3})\phi_{1}(\bfk_{1})\phi_{2}(\bfk_{2})\phi_{3}(\bfk_{3})}}\\
&=V^{3}\int_{\bfk_{1}\bfk_{2}\bfk_{3} } (2\pi)^{3}\delta\left(  \sum_{a=1}^{3}\bfk_{a}\right) \frac{B_{123}(k_{1},k_{2},k_{3})^{2} \times (2\pi)^{3}\delta(\mathbf{0})}{\prod_{a=1}^{3}(2\pi)^{3}\delta(\mathbf{0})P_{a}}\,,
\end{align}
where we estimated the denominator, i.e. the noise, in the Gaussian theory since we have in mind non-Gaussianities that are perturbatively close to the Gaussian theory. For a finite-volume survey we substitute $  (2\pi)^{3}\delta(\mathbf{0})=V $ and use \eqref{Bqui} to find
\begin{align}
\left(  \frac{S}{N}\right)^{2}&=V\int_{\bfk_{1}\bfk_{2}} \frac{\left(  gH^{p-4}R_{3} \prod_{a=1}^{3}A_{a}\right)^{2}}{e_{3}^{3}\prod_{a=1}^{3}A_{a}}\\
&=g^{2} H^{2p-8}\,\left( \prod_{a=1}^{3}A_{a} \right)\, \left(  Vk_{\max}^{3}\right)\,,
\end{align}
where we estimated the momentum integrals with dimensional analysis\footnote{Here we focus our attention on the parametric scaling of $  S/N $. The reader should be mindful that this discussion neglects the fact that different bispectra might have very different shapes and so momentum integrals might give rise to large numerical factors that are not captured by dimensional analysis. This is not the case for the parity-odd bispectra we are considering in this work.}. Since we can write $  V\sim k_{\text{min}}^{-3} $ and the number of independent data points is $  N_{\text{data}}\sim (k_{\max}/k_{\text{min}})^{3} $, the last factor confirms the intuition that our ability to detect a signal scales as $ S/N \sim N_{\text{data}}^{1/2}  $. From the above expression we deduce that if two interactions have the same coupling constant $  g $, then the interaction involving fields with the largest power spectrum has the most signal-to-noise ratio and therefore should be the main observational target. \\

\noindent If we apply this result to the parity-odd bispectra generated in solid inflation by the operators \eqref{10}-\eqref{12} we conclude that \textit{the scalar-scalar-graviton bispectrum is expected to have an $  S/N $ larger than the graviton bispectrum by a factor of $  \epsilon^{-1} $}, which is the inverse of the small tensor-to-scalar ratio. (See \cite{Duivenvoorden:2019ses} for a detailed analysis of efficient CMB estimators of this signal.) To summarize, we briefly discuss some possible scenarios in which the manifestly local parity-odd bispectra that we computed in this work can be the leading observational signals:
\begin{itemize}
\item The manifestly local, parity-odd scalar-scalar-graviton bispectrum that we computed in \eqref{PPG1} and which is generated in solid inflation by the interaction in \eqref{SOoperator} does not have a purely scalar counterpart because of symmetry, and therefore can be the leading observational signal in solid inflation or in other non-minimal symmetry breaking patterns.
\item If one has access only to the gravitational sector, as it is the case for example if one considers only interferometric and pulsar observations of gravitational waves, then the parity-odd graviton bispectra in \eqref{96}-\eqref{101} can be the leading observational signals in solid inflation. A detection of these signals would rule out the effective field theory of inflation.
\item A detection of the parity odd graviton bispectra in \eqref{96}-\eqref{101} that is not accompanied by correlated parity-odd scalar-scalar-graviton and scalar-graviton-graviton bispectra with a much higher signal-to-noise ratio would rule out both the effective field theory of inflation and solid inflation. It would be interesting to investigate what symmetry breaking pattern could be consistent with this possibility.
\end{itemize}

\subsection{Perturbativity, naturalness and strong coupling} \label{sec:PaN}
Since we have claimed at the beginning of this paper that the bispectra we study, in particular the parity-odd bispectra of Section \ref{Pheno} can be \textit{large}, we need to verify how large they can be without compromising the validity of our analysis. One might worry that loop corrections could spoil our conclusions. Such corrections can come in a number of forms. Loops could introduce brand new shapes coming from performing new bulk time integrals coming from loop diagrams. These will introduce more complicated shapes that we have not considered here, but these will always be suppressed relative to the ones we have computed as long as we work below the strong coupling scale which we estimate below. Loops could also alter the quadratic action which we have assumed takes on the GR form. Such corrections could be in the form of operators with three or more derivatives that introduce new diagrams that contribute to the bispectrum. We will show below that these corrections are always small if we work below the strong coupling scale. Corrections to the quadratic action could also arise in the form of a large mass correction to the graviton. In Solid Inflation, where our large parity-odd bispectra can arise, the graviton is massive but the mass is very small and in this section we pay special attention to the issue of large mass corrections within the context of naturalness. A reader not interested in the details of the calculation may skip to the end of this section, where we summarize our findings. \\

\noindent We can write a general Lagrangian up to cubic order as
\be
\mathcal{L}[\gamma_c] = \mathcal{L}_{GR}[\gamma_c] + \sum\limits_i f^{(i)}_m \frac{(H \eta)^{m-4}}{\Lambda^{m-2}} \partial^m \gamma_c^2 + \sum\limits_i g^{(i)}_n \frac{(H \eta)^{n-4}}{M_{\rm pl} \Lambda^{n-2}} \partial^n \gamma_c^3   + \mathcal{O} \left( \gamma_c^4 \right) \, ,
\ee
where $\gamma_c := M_{\rm pl} \gamma$ is the canonically normalized field, $\partial^n \gamma_c^m$ is a shorthand notation for an $n-$derivative operator and $f^{(i)}_n$ and $g^{(i)}_n$ are the dimensionless coupling constants. We have included all powers of $\eta$ that are required by scale invariance. The parity-odd interactions that contribute to bispectra have $n=1,2$ and $3$, and their dimensionless coupling constants are denoted by $g_1, g_2, g_3$, respectively. At tree-level these operators do not correct the quadratic action which allows us to conclude that they can yield a large contribution to the graviton non-Gaussianity relative to the GR contribution if 
\be
(\mathcal{L}_3)_{\rm new} \gg (\mathcal{L}_3)_{\rm GR} \sim \frac{H}{M_{\rm P}} \mathcal{L}_2\,\,.
\ee
In the above, $\mathcal{L}_2$ is simply the GR quadratic Lagrangian, as we have assumed throughout the paper. Ideally, we want a stronger notion of a large non-Gaussianity, namely that the signal-to-noise (S-to-N) in the $3-$point function is close to that of the power spectrum. This would mean
\be
f_{\rm NL} ~ \gamma \lesssim \mathcal{O}(1).
\ee
Crucially the non-Gaussian contributions need to be smaller than the vacuum one to remain within a perturbative analysis.
\noindent These two conditions entail, respectively, (at horizon crossing)
\begin{align} \label{eq:TwoEstimates}
g_n \frac{H^{n-2}}{\Lambda^{n-2}} \gg 1, \\
g_n \frac{H^{n-2}}{\Lambda^{n-2}} \lesssim \frac{M_{\rm pl}}{H}\,,
\end{align}
and these would need to be satisfied for $H \ll M_{\rm pl}$ again so that the vacuum contribution dominates over the GR cubic contribution. We see that it is possible to have large non-Gaussianities relative to GR, while remaining perturbative. However, we must remember that the tree-level bispectra derived in this work are good approximations only if the loop contributions to the bispectra are suppressed, while the GR quadratic Lagrangian assumed throughout is only natural if loop corrections to it are insignificant. Let us therefore estimate (i) the size of quantum corrections to the quadratic Lagrangian and (ii) the size of loop contributions to the bispectra. \\

\noindent We start by estimating the UV cutoff scale $\Lambda_c$. This can be done by deriving the scale $\Lambda_*$ at which the theory becomes strongly coupled, since at that energy new physics is expected to be important \cite{Baumann:2011su}. This corresponds to a limiting scenario where loop corrections are the largest, although it is still possible that the cutoff lies much below the strong coupling scale, which would correspond to a weakly coupled UV completion which we will comment on later. Now consider a general cubic operator with $n$ derivatives. A rough estimate of the strong coupling scale can be derived by examining the breakdown of perturbative unitarity in flat-space i.e. by asking when the $\gamma \gamma \to \gamma \gamma$ scattering amplitude is of order $1$. We work in flat space as this is a good approximation for energy scales well above the Hubble scale and indeed we want the theory to be valid in such a regime. A back-of-the-envelope estimate in the flat space limit yields
\be
\mathcal{A}_4 \sim g_n^2 \frac{E^n}{M_{\rm pl} \Lambda^{n-2}} \frac{1}{E^2} \frac{E^n}{M_{\rm pl} \Lambda^{n-2}} = \left( g_n \frac{E^{n-1}}{M_{\rm pl} \Lambda^{n-2}} \right)^2 ,
\ee
implying that the strong coupling scale $\Lambda_*$ associated to an $n-$derivative operator is
\be
\Lambda_* \sim \left( \frac{1}{g_n} M_{\rm pl} \Lambda^{n-2} \right)^{\frac{1}{n-1}} .
\ee
(For $n=1$, we take $\Lambda_* \sim M_{\rm pl}$, since we expect new physics to be relevant at $M_{\rm pl}$, if not earlier.) For the EFT to be useful, we should require that it is valid at least up to Hubble scale, so that $\Lambda_* > H$. Thanks to the presence of the $M_{\rm pl}$ factor, this is consistent with the above estimate of $\Lambda_*$, as well as with (\ref{eq:TwoEstimates}). \\

\noindent Let's now estimate the size of the loop corrections to the quadratic Lagrangian. First we focus our attention on a particular $n$-derivative operator and cut off the loop momentum at the relevant $\Lambda_*$ given above. In the absence of a symmetry that would protect a small value of a given coupling, the radiative correction to the coefficient of a $(\partial^a \gamma_c)^2$ operator due to a loop with two $n-$derivative vertices is of the order
\be
\delta \mathcal{L}_{(\partial^a \gamma_c)^2} \sim \frac{1}{M_{\rm pl} \Lambda^{n-2}} \frac{1}{M_{\rm pl} \Lambda^{n-2}} \int^{\Lambda_*} d^4 p g_n^2 \frac{p^{n-a} p^{n-a}}{p^4} \sim g_n^2 \frac{\Lambda_*^{2n - 2a}}{M_{\rm pl}^2 \Lambda^{2n - 4}} \sim \Lambda_*^{2 - 2a}.
\ee
The ratio of the loop contribution $(\mathcal{L}_2)_{\rm new}$ to the GR contribution at $E \sim H$ is of the order
\be
\frac{(\mathcal{L}_2)_{\rm new}}{(\mathcal{L}_2)_{\rm GR}} \sim \left( \frac{\Lambda_*}{H} \right)^{2 - 2a}.
\ee
Now for $a > 1$ this is a small contribution since we take $\Lambda_* > H$. For $a=1$ we would have a correction to the two derivative GR action but such corrections are harmless since we can always do field redefinitions that bring the quadratic action into the canonical form \cite{Creminelli:2014wna}. However, we see that the mass term ($a=0$) could receive a large quantum correction. An important exception is for $n=3$ where we have a shift symmetry. In this case a small graviton mass is protected by the shift symmetry of the interaction (\ref{eq:ParityOddOperator3}). For the other two operators (\ref{eq:ParityOddOperator1})-(\ref{eq:ParityOddOperator2}) it looks like a large mass could be generated, but before jumping to such a conclusion one would need to perform a fully fledged computation to check if once all polarisation sums are included such a correction is still non-zero and large. We leave such an analysis for future work. \\

\noindent In the above we have assumed that there is only one cubic operator which not only generates radiative corrections, but also defines the cutoff scale. Suppose, however, that we have multiple cubic operators $\mathcal{O}_1, \ldots, \mathcal{O}_k$. If the $g_n$ couplings do not differ by too many orders of magnitude, then the cutoff scale $\Lambda_c$ is the one associated to the highest-dimension operator and this can alter our conclusions about large corrections to the mass. In the case of our three parity-odd interactions (\ref{eq:ParityOddOperator1})-(\ref{eq:ParityOddOperator2}), (\ref{eq:ParityOddOperator3}) the lowest cutoff is $\Lambda_* = \sqrt{\frac{M_{\rm P} \Lambda}{g_3}}$. The radiative correction to the coefficient at $(\partial^a \gamma)^2$ due to a loop with two $n-$derivative vertices is then of the order
\be
\delta \mathcal{L}_{(\partial^a \gamma_c)^2}  \sim \frac{1}{M_{\rm pl} \Lambda^{n-2}} \frac{1}{M_{\rm pl} \Lambda^{n-2}} \int^{\Lambda_*} d^4 p g_n^2 \frac{p^{n-a} p^{n-a}}{p^4} \sim g_n^2 \frac{\Lambda_*^{2n - 2a}}{M_{\rm pl}^2 \Lambda^{2n - 4}} \sim \frac{g_n^2}{g_3^{n-a}} M_{\rm pl}^{n-a-2} \Lambda^{4 - n -a} .
\ee
Comparing this with the GR contribution at $E \sim H$, we have
\be
\frac{(\mathcal{L}_2)_{\rm new}}{(\mathcal{L}_2)_{\rm GR}} \sim  \frac{g_n^2}{g_3^{n-a}}  \left( \frac{H^2}{M_{\rm P} \Lambda} \right)^{a -1} \left( \frac{\Lambda}{M_{\rm P}} \right)^{3-n} .
\ee
For $a \geqslant 1$ the corrections are small. For $a=0$, only $n=1,2$ contribute due to the shift symmetry for $n=3$. We have
\be
n =  1:  \frac{(\mathcal{L}_2)_{m^2}}{(\mathcal{L}_2)_{\rm GR}} \sim \frac{g_1^2}{g_3}\frac{\Lambda^3}{M_{\rm P}H^2},
\quad n =  2:  \frac{(\mathcal{L}_2)_{m^2}}{(\mathcal{L}_2)_{\rm GR}}  \sim \frac{g_2^2}{g_3^2}\frac{\Lambda^2}{H^2}.
\ee 
We see that the $g_1$ could dominate the GR contribution ($f_{NL} \gg 1$, but $f_{NL} \ll M_{\rm P}/H$) while keeping $\delta m^2$ small. This applies as long as the cutoff scale is dictated by the $n=3$ operator which, as we have shown before, could be very large ($f_{NL} \sim M_{\rm P}/H$) since its loops do not correct the mass. On the other hand, we need a hierarchy $g_3 \gg g_2 \frac{\Lambda}{H}$ if the radiative corrections to $m^2$ from the $n=2$ parity-odd operator are supposed to be small. In this case it is difficult to keep non-Gaussianity from $g_2$ larger than the GR non-Gaussianity, while keeping loop corrections under control. \\

\noindent Let us now study loop contributions to the parity-odd tree-level shapes we have computed in this paper. We are interested mostly in the regime where the energy in the loop is large (close to $\Lambda_*$), so the loop is effectively deep inside the horizon. We can therefore again work in flat-space and estimate the size of the loop corrections to the three-particle \textit{amplitude} and compare it with the tree-level result. We should be careful, however, to only put derivatives on the external legs in such a way that we reproduce the structure of one of our parity-odd operators, since otherwise the loop diagram will not contribute to the parity-odd bispectrum as we have shown in Section \ref{FinalForm}. We can therefore put $m=1,2$ or $3$ derivatives on the external legs. An estimate of such a loop diagram proceeds similarly as before. For loop diagrams with three instances of the same $n$-derivative parity-odd operator, assuming the cutoff $\Lambda_* = \left( \frac{1}{g_n} M_{\rm P} \Lambda^{n-2} \right)^{1/(n-1)}$ is dictated by that operator, we find
\be \label{eq:RatioLT}
\frac{\mathcal{A}^{\rm 1-loop}_3}{\mathcal{A}^{\rm tree}_3} \sim g_n^{\frac{m-1}{n-1}}g_m^{-1} \left( \frac{M_{\rm P}}{\Lambda} \right)^{\frac{n-m}{n-1}}.
\ee
\begin{itemize}
\item If $n=3$, then shift symmetry must be preserved at any loop order, which means that loop diagrams cannot generate the shapes (\ref{96})-(\ref{96b}), but only the three-derivative shape. Thus, it suffices to consider $m=3$. It seems that the ratio (\ref{eq:RatioLT}) is equal to $1$, although in our estimate we neglected combinatorial factors as well as factors of $(2 \pi)$. In any case, identifying the cutoff with the perturbative unitarity breakdown scale is supposed to only give us an order-of-magnitude estimate, and it is not unnatural to have a slightly lower cutoff which would further suppress loops which scale as $\Lambda_c^4$.
\item If $n=2$, then loop contributions are small for $m=3$, again $\mathcal{O}(1)$ for $m=2$, while they are large for $m=1$. However, if the $n=3$ interaction is also present and dominates the signal, then the cutoff is lowered from $M_{\rm pl}/g_2$ to $\sqrt{M_{\rm pl} \Lambda / g_3}$, and all the loop contributions from $n=2$ are small.
\item For $n=1$, high energies are suppressed and we don't observe any UV divergences in the loop constructed out of three copies of the $n=1$ operator. Instead, we ought to consider the loop constructed out of two GR vertices and one $n=1$ operator. Regardless of the structure of derivatives on the external legs, this loop diagram is suppressed, relative to tree level, by $M_{\rm pl}^{-2}$, but has at most two factors of $\Lambda_{*}$ since at most four derivatives can be put on the internal legs. Therefore, loop contributions due to $n=1$ are small.
\end{itemize}

Let us conclude this section by summarising our findings:
\begin{itemize}
\item If we consider the parity-odd operators \textit{individually}, then $g_3$ (and only $g_3$) can be so large that the (\ref{101}) bispectrum has a S-to-N ratio comparable to that of the power spectrum, without the need for fine-tuning. This is because of shift symmetry of (\ref{eq:ParityOddOperator3}), which protects a small mass from receiving quantum corrections. Meanwhile, the other two parity-odd operators do contribute to the mass via radiative corrections, and natural values of $g_1$ and $g_2$ must be very small, meaning that the associated signals are weaker than the GR bispectra.
\item By identifying the cutoff scale of the theory with the scale at which the three-derivative parity-odd interactions given by (\ref{eq:ParityOddOperator3}) become strongly coupled, $\Lambda_c = \sqrt{M_{\rm P} \Lambda / g_3}$, we can consider a more general case in which we have multiple parity-odd operators. In this case, non-Gaussianities generated by (\ref{eq:ParityOddOperator1}) may be larger than GR non-Gaussianities (but with S-to-N smaller than in the power spectrum) while $g_1$ remains natural. However, the coefficient of (\ref{eq:ParityOddOperator2}) is bounded by $g_2 \ll g_3 \frac{H}{\Lambda}$, implying that the region of parameter space where (\ref{96b1})-(\ref{96b}) are larger than GR is very limited.
\item Tree level calculations are a good approximation for the three-derivative parity-odd interaction: loop contributions to the bispectrum can be suppressed without the need for fine-tuning. Overall, the $g_{3}$ operator is best placed to give large non-Gaussianities, both in the sense of being large compared to GR but also with a sizeable S-to-N, while keeping loop corrections under control. 
\end{itemize}




\section{Summary and future directions}\label{summary}

In this work we have, for the first time, bootstrapped tree-level inflationary graviton bispectra to all orders in derivatives. Under a minimal set of assumptions, we have detailed how one can write down these bispectra without working with a concrete inflationary model. We used spatial translations, spatial rotations and scale invariance to write down a general ansatz for the corresponding wavefunction of the universe. Assuming that the mode functions are the usual ones of a massless graviton with Bunch-Davies initial conditions, we used locality and unitarity to constrain the wavefunction coefficients. We considered all possible tree-level contributions, including IR-divergences at future infinity, $\eta_{0} \to 0$. We imposed locality by demanding that the wavefunction coefficients satisfy the Manifestly Local Test (MLT) introduced in \cite{MLT} which is a simple differential constraint that all $n$-point functions of massless gravitons should satisfy. Solutions to the MLT replace solutions to the time integrals that one is required to calculate in the bulk formalism. The beauty of the MLT is that it allows us to compute non-Gaussian shapes without having to consider the unobservable bulk time evolution. We imposed bulk unitarity using the Cosmological Optical Theorem (COT) \cite{COT}. We presented our results succinctly in Section \ref{sec:bispectra} using the cosmological spinor helicity formalism of \cite{Maldacena:2011nz}, and we computed all bispectra for both parity-even and parity-odd interactions.  \\

\noindent In Section \ref{ParitySection}, we showed which part of the wavefunction contributes to the correlator, for contact diagrams. We concentrated on contact diagrams since our focus in this paper is on tree-level bispectra but many of our results in that section hold for any tree-level $n$-point function. We showed that only the part of the wavefuction that breaks the $\{k\} \rightarrow \{-k\}$ symmetry, where $\{ k \}$ are the external energies, can contribute to the correlator. This is a direct consequence of bulk unitarity and can be easily derived from the COT for contact diagrams. For graviton bispectra, this tells us that for parity-even interactions both the rational part and the log part of the wavefunction can appear in the correlator, whereas for parity-odd interactions the only allowed contributions are regular at both $  \eta_{0}\to 0 $ and $  k_{T}\to 0 $. Indeed, unitarity in the form of the COT tells us that the log must always appear in the combination $\log(-k_{T} \eta_{0}) + \frac{i \pi}{2}$ and for parity-odd interactions it is the $\frac{i \pi}{2}$ piece that contributes to the correlator. This allowed us to show that, to all orders in derivatives, for parity-odd graviton self-interactions there are only three independent couplings that contribute to the bispectrum. This is not evident when using concrete Lagrangians and the in-in formalism and therefore offers a neat example of where the bootstrap approach can be very advantageous.  \\

\noindent In Section \ref{SymmetryBreaking}, we showed that our parity-breaking graviton bispectra appear in both the Effective Field Theory of Inflation (EFToI), and in solid inflation. For the former, a correction to the two-point function is forced by the non-linearly realized symmetries. By accounting for this correction, we computed the full parity-odd contribution to the graviton bispectrum. The associated non-Gaussianity is too small to be detected observationally in any conceivable future. Conversely, for solid inflation there is no symmetry that forces a correction to the two-point function, so the three parity-odd bispectra we have computed can indeed arise with arbitrary coefficients. Given that such operators do not contribute to the bispectrum of curvature perturbations, which cannot violate parity, there are no strong observational bounds on the size of these non-Gaussianities. We plotted the associated shapes in Figure \ref{fig:B3Shapes}. \\

\noindent With this catalogue of graviton non-Gaussianities at hand, we outline here a few directions for future work
\begin{itemize}
\item To derive our catalogue of graviton bispectra we did not assume any particular symmetry breaking pattern for the inflationary dynamics. Indeed, we have captured all scale invariant contributions, assuming the usual massless mode functions. It would be very interesting to develop further criteria to identify those non-Gaussianites that are consistent only in the presence of additional degrees of freedom. For example, we expect that only some couplings can appear in the EFToI and in future work we plan to use soft theorems/consistency relations to extract this subset. It would also be very interesting to take these three-point building blocks and to glue them together to form four-point functions. By demanding that the full four-point function satisfies some consistency constraints, we will also be able to pick out interesting subsets of our full catalogue. This approach would be very similar to that used to constrain cubic interactions in flat space with $S$-matrix consistency conditions \cite{Benincasa:2007xk,PSS}, and in \cite{Baumann:2020dch} assuming invariance under de Sitter boosts. Deriving this full catalogue is the first step towards distinguishing between different symmetry breaking patterns for inflation directly at the level of the observable. This will complement the recent Lagrangian analysis of \cite{ZoologyNG} and ultimately lead to a more efficient way of ``simplifying" inflationary predictions \cite{Bordin:2017hal}. For example, we expect there to be only a single three-derivative correction to the graviton bispectrum in the EFToI \cite{CabassBordin}, and we plan to develop bootstrap techniques that enables us to efficiently extract this result without having to use the Lagrangian or bulk time evolution.
\item Given the small number of possible parity-odd graviton bispectra in solid inflation, it would be interesting to study the associated bulk operators. In particular one would like to know when those same operators give also rise to interactions between the graviton and curvature perturbations. It is also very important to study the quantum stability of these operators and possible perturbative unitarity bounds on their size. 
\item Finally, we notice that Ref.~\cite{Liu:2019fag} showed that for manifestly-local interactions, all parity-odd scalar correlation functions vanish at tree level. It would be interesting to see if their result can be generalized to spinning particles. 
\end{itemize}

\noindent Our understanding of physical observables in nature becomes increasingly more opaque as we approach the real world. In anti-de Space (AdS) we have the gauge-gravity duality that provides us with a good understanding of the structure of boundary observables. In flat-space, the object of interest is the $S$-matrix. The $S$-matrix bootstrap has lead to a good understanding of the tree-level properties of amplitudes, with progress now being made on the analytic structure at loop level. Finally, we have de Sitter space, which appears to describe the early and late phases of our universe very well. We are only now starting to understand the general structure of cosmological correlators in de Sitter, both at tree and loop level. We hope that our results will contribute to broadening this understanding and to provide theoretical guidance on the physical modeling of inflation. 

\paragraph{Acknowledgements} We thank Maria Gutierrez Guillen, Sadra Jazayeri and Jo{\~a}o Melo for collaboration during the early phases of this work. We further thank Daniel Baumann, James Bonifacio, Carlos Duaso Pueyo, Harry Goodhew, Aaron Hillman, Austin Joyce, Gui Pimentel and Jacopo Salvalaggio for useful discussions. E.P. and D.S. have been supported in part by the research program VIDI with Project No. 680-47-535, which is (partly) financed by the Netherlands Organisation for Scientific Research (NWO). G.C. acknowledges support from the Institute for Advanced Study. J.S. has been supported by a scholarship from STFC.


\appendix


\section{From polarisations to spinors} \label{app:contractions} 

In this appendix we construct all possible polarisation factors for three gravitons and explain how one can convert these into spinor expressions using the spinor helicity formalism. We consider parity-even and parity-odd structures separately. Throughout we suppress the momentum dependence of the polarisation tensors. Note that throughout we only contract momenta with polarisation tensors as any pair of contracted momenta can be written in terms of the energies (norms) which we include in the trimmed part of the wavefunction c.f. \eqref{GeneralWFC}. Indeed, we have 
\begin{align}
\bfk_{a} \cdot \bfk_{b} = \frac{1}{2}(k_{c}^2 - k_{a}^2 - k_{b}^2)\,, \qquad a \neq b \neq c\,.
\end{align}
When we convert the following expressions into spinors, their symmetry properties will become manifest. 
\subsubsection*{Parity-even tensor structures}
For parity-even structures we need to contract spatial momenta with
\begin{equation}
e^{h_1}_{i_{1}i_{2}}e^{h_2}_{i_{3}i_{4}}e^{h_3}_{i_{5}i_{6}}\,,
\end{equation}
using $\delta_{ij}$. We work order by order in the total number of derivatives $\alpha$. \\

\noindent $\bm{\alpha=0}$ In this case there is clearly only a single structure which is given by
\begin{equation}
e^{h_1}_{ij}e^{h_2}_{jk}e^{h_3}_{ki}\,.
\end{equation}
This structure is fully symmetric and when converted to spinors this contraction simply yields
\begin{align}
\text{SH}_{+++}\,,
\end{align}
for the all-plus configuration. \\

\noindent $\bm{\alpha=2}$ In this case we have two possibilities. For the first we contract the two momenta with the same polarisation tensor and for the second we contract each momentum with different polarisations. Using momentum conservation and the transversality of the polarisation tensors, there is then a single option for the labels of the momenta, up to permutations. We have
\begin{equation}
\label{eq:g2}
\text{$e^{h_1}_{lm}e^{h_2}_{lm}e^{h_3}_{ij}k^i_1k^j_2$\quad and\quad $e^{h_1}_{lm}e^{h_2}_{il}e^{h_3}_{jm}k^i_1k^j_1$}\,. 
\end{equation}
These two structures appear in GR with tuned coefficients. The first structure is symmetric in labels $1$ and $2$ while the second is symmetric in $2$ and $3$. If we sum over permutations and convert to spinors then we have 
 \begin{align}
\text{SH}_{+++}\times \text{Poly}_{2}= \text{SH}_{+++}\left(a_{0} e_{2}+a_{2} k_{T}^2 \right)\,.
\end{align}

\noindent $\bm{\alpha=4}$ In this case we have a single option. All momenta need to be contracted with polarisation tensors and then using the fact that the polarisations are traceless yields a single possibility. Again, momentum conservation and transversality yields a single possibility for the labels, up to permutations. We have
\begin{equation}
\label{eq:g4}
e^{h_1}_{lk}e^{h_2}_{mk}e^{h_3}_{ij}k^i_1k^j_2k^l_3k^m_3\,.
\end{equation} 
This structure is symmetric in $2$ and $3$ and when we sum over permutations and convert to spinors we have 
 \begin{align}
\text{SH}_{+++}\times \text{Poly}_{4}= \text{SH}_{+++}\left(k_{T}^4 - k_{T}^2 e_{2} + 8 k_{T}e_{3} \right)\,.
\end{align}
\noindent $\bm{\alpha=6}$ Finally, in this case there is a single option with all polarisation tensor indices contracted with momenta. We have
\begin{equation}
\label{eq:g6}
e^{h_1}_{il}e^{h_2}_{jm}e^{h_3}_{kn}k^i_2k^l_3 k^j_3k^m_1 k^k_1k^n_2\,. 
\end{equation}
This structure is fully symmetric and yields
 \begin{align}
\text{SH}_{+++}\times \text{Poly}_{6}= \text{SH}_{+++}\left(k_{T}^6 - 8 k_{T}^4 e_{2} + 16 k_{T}^3 e_{3} + 16 k_{T}^2 e_{2}^2 -64 k_{T} e_{2}e_{3} + 64 e_{3}^2 \right)\,.
\end{align}
when we convert to spinors.
\subsubsection*{Parity-odd tensor structures}
We now turn to parity-odd structures where we need to contract momenta with 
\begin{equation}
\epsilon_{i_{1}i_{2}i_{3}}e^{h_1}_{i_{4}i_{5}}e^{h_2}_{i_{6}i_{7}}e^{h_3}_{i_{8}i_{9}}.
\end{equation}
As above, in all cases there is a single option for the labels, up to permutations. \\

\noindent $\bm{\alpha=1}$ In this case there are two possible structures with the single momentum either contracted with a polarisation tensor or with the epsilon tensor. We have  
\begin{equation}
\label{eq:g1}
\text{$\epsilon_{ijk}e^{h_1}_{il}e^{h_2}_{lm}e^{h_3}_{km}k^j_3$ \quad and \quad $\epsilon_{ijk}e^{h_1}_{il}e^{h_2}_{jm}e^{h_3}_{kl}k^m_3$}\,. 
\end{equation}
The first of these is symmetric under the exchange $ 1 \leftrightarrow 2  $, while the second is symmetric under $ 1 \leftrightarrow 3  $. When symmetrized over all possible permutations of the three energies, these two contractions coincide up to a minus sign. This fact can be checked using explicit expressions for the polarization tensors, but it is not at all obvious. Conversely, it is easy to see in the spinor helicity formalism where both contractions must take the form 
\begin{align}
\text{SH}_{+++} \times \text{Poly}_{1}\,,
\end{align}
where the only permutation-invariant linear symmetric polynomial is $  \text{Poly}_{1}=k_{T} $. \\
 
\noindent $\bm{\alpha=3}$ In this case we have six possibilities and we classify them according to how many momenta are contracted with the epsilon tensor. First consider the case where none of the momenta are contracted with the epsilon tensor. Given the properties of the polarisation tensors, we then have a single possibility given by 
\begin{equation}
\label{eq:g3-zero_k}
\epsilon_{ijk}e^{h_1}_{il}e^{h_2}_{jm}e^{h_3}_{kn}k^l_3k^m_1k^n_2\,.
\end{equation}
Now when one of the momenta is contracted with the epsilon tensor we have two possibilities since the remaining two momenta can be contracted with the same polarisation tensor or with two different ones. We have 
\begin{equation}
\label{eq:g3-one_k}
\text{$\epsilon_{ijk}e^{h_1}_{nl}e^{h_2}_{jm}e^{h_3}_{km}k^n_2k^i_3k^l_3$ \, , \quad 
$\epsilon_{ijk}e^{h_1}_{jl}e^{h_2}_{nm}e^{h_3}_{kn}k^m_1k^l_2k^i_1$ \quad and \quad 
$\epsilon_{ijk}e^{h_1}_{jl}e^{h_2}_{nm}e^{h_3}_{kn}k^m_1k^l_2k^i_3$}\,.
\end{equation} 
Finally, we can contract two momenta with the epsilon tensor. There are then two possibilities: the third momentum must be contracted with a polarisation tensor, and the other index of this polarisation can be contracted with the epsilon tensor or another polarisation. We have
\begin{equation}
\label{eq:g3-two_k}
\text{$\epsilon_{ijk}e^{h_1}_{nl}e^{h_2}_{nm}e^{h_3}_{il}k^m_1k^j_2k^k_3$\quad and\quad $\epsilon_{ijk}e^{h_1}_{im}e^{h_2}_{ln}e^{h_3}_{ln}k^j_1k^k_2k^m_3$}\,. 
\end{equation} 
Upon symmetrization over all possible permutations of the three energies, only three of the above six contractions are linearly independent (for example \eqref{eq:g3-zero_k} and the first two in \eqref{eq:g3-one_k}). To see this with explicit polarization tensors requires a laborious calculation. Conversely, this can be easily seen using spinor helicity variables, where the most generic $  \alpha=3 $ (symmetrized) contraction must take the form
\begin{align}
\text{SH}_{+++}\times \text{Poly}_{3}= \text{SH}_{+++}\left(  a_{0} e_{3}+a_{1} k_{T}e_{2}+a_{3}k_{T}^{3}\right)\,,
\end{align}
which has indeed three free coefficients $  a_{0,1,3} $. \\

\noindent $\bm{\alpha=5}$ In this case we have a total of three possibilities. One of them corresponds to having only one momentum contracted with the epsilon tensor, while for the others two of the momenta are contracted with the epsilon tensor. We have
\begin{equation}
\label{eq:g5}
\text{
$\epsilon_{ijk}e^{h_1}_{mq}e^{h_2}_{nq}e^{h_3}_{lk}k^i_1k^j_2k^m_2k^n_1k^l_1\,,$\quad 
$\epsilon_{ijk}e^{h_1}_{mn}e^{h_2}_{lq}e^{h_3}_{qk}k^i_1k^j_2k^m_2k^n_2k^l_1\,,$ 
and\quad 
$\epsilon_{ijk}e^{h_1}_{mn}e^{h_2}_{qj}e^{h_3}_{lk}k^i_2k^m_3k^q_1k^l_1k^n_3$}\,.
\end{equation} 
When we sum over permutations and convert to spinors we have only two structures:
\begin{align}
\text{SH}_{+++} \text{Poly}_{5} = ~ & a ~ \text{SH}_{+++} (-3k_{T}^5+20k_{T}^3 e_{2} -24 k_{T}^2 e_{3}-32 k_{T}e_{2}^2 + 64 e_{2}e_{3}) \\
+ ~ & b ~ \text{SH}_{+++} (k_{T}^5-8k_{T}^3 e_{2} +8 k_{T}^2 e_{3}+16 k_{T}e_{2}^2 -32 e_{2}e_{3})\,.
\end{align}
\noindent $\bm{\alpha=7}$ Finally, in this last case we have a single possibility given by 
\begin{equation}
\label{eq:g7}
\epsilon_{ijk}e^{h_1}_{mn}e^{h_2}_{qp}e^{h_3}_{lk}k^i_1k^j_2k^m_2k^n_2k^q_1k^p_1k^l_1\,,
\end{equation}
and once we sum over permutations and convert to spinors we have
\begin{align}
\text{SH}_{+++} \text{Poly}_{7} = \text{SH}_{+++}(k_{T}^7 - 8 k_{T}^5 e_{2} + 16 k_{T}^4 e_{3} + 16 k_{T}^3 e_{2}^2 -64 k_{T}^2 e_{2}e_{3} + 16 k_{T}e_{3}^2)\,.
\end{align}
Note that in the above we have used the fact that three momenta cannot be contracted with the epsilon tensor due to momentum conservation.
\subsubsection*{Converting to spinors}

Now that we have all of the possible polarisation factors, we can convert them into spinor expressions using the spinor helicity formalism. As we explained in detail in Section \ref{sec:bispectra}, given the form of the $+++$ polarisation factor, one can easily construct the ones for the other helicity configurations. The following expressions hold for three-point kinematics only. In the parity-even case the only expressions we need are 
\begin{subequations}
\label{eq:app_contractions-1}
\begin{align}
e^{a+} \cdot  e^{b+}  &= -\frac{[ab]^2}{k_a k_b}\,, \label{eq:app_contractions-1-1} \\
p^{a} \cdot e^{b+}  &= \frac{(ab)[ab]}{\sqrt{2} k_b} \,, \label{eq:app_contractions-1-2}
\end{align}
\end{subequations} 
where we have used the relations presented in Section \ref{SHF}. For parity-odd structures we use the general expression 
\begin{align}
\epsilon_{ijk}V^{a}_{i}V^{b}_{j}V^{c}_{k} = \frac{i}{4}(\langle ab \rangle [ab] (cc) + \langle ab \rangle [ca] (cb) + \langle bc \rangle [ab] (ac))\,,
\end{align}
where each $SO(3)$ vector contains the spatial parts of a null four-vector $V_{\mu}$ which is converted to spinors using the standard expressions
\begin{align}
V^{\mu} = \frac{1}{2} (\bar{\sigma}^{\mu})^{\dot{\alpha}\alpha}V_{\alpha \dot{\alpha}}, \qquad V_{\alpha \dot{\alpha}} = V^{\mu}(\sigma_{\mu})_{\alpha \dot{\alpha}}, \qquad V_{\alpha \dot{\alpha}} = \lambda_{\alpha} \tilde{\lambda}_{\dot{\alpha}}\,.
\end{align}
The expressions that we need are then 
\begin{subequations}
\label{eq:app_contractions-2}
\begin{align}
\epsilon_{ijk}\,e^{a+}_ie^{b+}_je^{c+}_k &= 
- i \sqrt{2} \frac{[ab][bc][ca]}{k_a k_b k_c}\,, \label{eq:app_contractions-2-1} \\
\epsilon_{ijk}\,p^{a}_ie^{a+}_je^{b+}_k &= 
- i \frac{[ab]^2}{k_b}\,, \label{eq:app_contractions-2-2} \\
\epsilon_{ijk}\,p^{a}_ip^{b}_je^{a+}_k &= -\frac{i}{\sqrt{2}}[ab](ba)\,. \label{eq:app_contractions-2-3}
\end{align}
\end{subequations}

\noindent Note that by momentum conservation, we only need to consider cases where one of the momenta has the same label as one of the polarisation tensors. We used these relations to derive the list of possible $h_{\alpha}(k_{1},k_{2},k_{3})$ in \eqref{h1} to \eqref{h7}. Notice that for some $\alpha$ there are fewer choices for $h_{\alpha}$ compared to the polarisation structures above.

\bibliographystyle{JHEP}
\bibliography{refs}

\end{document}